\PassOptionsToPackage{dvipsnames,prologue}{xcolor}

\documentclass[manuscript,screen]{acmart}

\usepackage{acro}
\usepackage{multirow}
\usepackage{makecell}
\usepackage{amsmath}
\usepackage{lineno,hyperref}
\usepackage{adjustbox}
\usepackage{graphicx}
\usepackage{xltabular}
\usepackage[most]{tcolorbox}
\usepackage[normalem]{ulem}
\usepackage{xcolor}

\newcommand{\greentriangleup}{$\color{ForestGreen} \blacktriangle \;$}
\newcommand{\yellowcircle}{$\color{YellowOrange} \bullet \;$}

\AtBeginDocument{%
  }

\setcopyright{acmlicensed}
\copyrightyear{2018}
\acmYear{2018}
\acmDOI{XXXXXXX.XXXXXXX}

\acmJournal{TORS}
\acmVolume{37}
\acmNumber{4}
\acmArticle{111}
\acmMonth{8}

\DeclareAcronym{RS}{
    short = RS,
    long  = Recommender System
}

\DeclareAcronym{ML}{
    short = ML,
    long  = Machine Learning
}

\DeclareAcronym{XAI}{
    short = XAI,
    long = Explainable AI
}

\DeclareAcronym{LLM}{
    short = LLM,
    long = Large Language Models
}

\DeclareAcronym{KG}{
    short = KG,
    long = Knowledge Graph
}

\DeclareAcronym{NLP}{
    short = NLP,
    long  = Natural Language Processing
}

\DeclareAcronym{BLEU}{
    short = BLEU,
    long = Bilingual Evaluation Understudy
}

\DeclareAcronym{ROUGE}{
    short = ROUGE,
    long = Recall-Oriented Understudy for Gisting Evaluation
}

\DeclareAcronym{RQ}{
    short = RQ,
    long = Research Question
}

\DeclareAcronym{SHAP}{
    short = SHAP,
    long = Shapley Additive exPlanations
}

\DeclareAcronym{LIME}{
    short = LIME,
    long = Local Interpretable Model-agnostic Explanations
}

\DeclareAcronym{TF-IDF}{
    short = TF-IDF,
    long = Term Frequency–Inverse Document Frequency
}

\DeclareAcronym{PEM}{
    short = PEM,
    long =  Property-based Explanation Model
}

\DeclareAcronym{LOD}{
    short = LOD,
    long = Linked Open Data
}

\DeclareAcronym{NDCG}{
    short = NDCG,
    long = Normalized Discounted Cumulative Gain
}

\DeclareAcronym{MovieLens}{
    short = MovieLens,
    long = MovieLens Latest Small
}

\DeclareAcronym{LastFM}{
    short = LastFM,
    long = Hetrec2011 Lastfm 2k
}
\begin{document}

\title[A Comparative Survey Analysis of Offline Explanation Metrics in Recommender Systems]{Can Offline Metrics Measure Explanation Goals? A Comparative Survey Analysis of Offline Explanation Metrics in Recommender Systems}

\author{André Levi Zanon}
\email{andre.zanon@insight-centre.org}
\orcid{0000-0003-0526-2678}
\affiliation{%
  \institution{Insight Centre for Data Analytics, School of Computer Science and IT, University College Cork}
  \streetaddress{University College Cork}
  \city{Cork}
  \state{Cork}
  \country{Ireland}
  \postcode{T12 YN60}
}

\author{Leonardo Chaves Dutra da Rocha}
\email{lcrocha@ufsj.edu.br}
\orcid{0000-0002-4913-4902}
\affiliation{%
  \institution{Departamento de Ciência da Computação, Universidade Federal de São João del-Rei}
  \streetaddress{Praça Frei Orlando, 170}
  \city{São João del Rei}
  \state{Minas Gerais}
  \country{Brasil}
  \postcode{36307-352}
}

\author{Marcelo Garcia Manzato}
\email{mmanzato@icmc.usp.br}
\orcid{0000-0003-3215-6918}
\affiliation{%
  \institution{Instituto de Ciências Matemáticas e de Computação, Universidade de São Paulo}
  \streetaddress{Avenida Trabalhador São-carlense, 400 - Centro}
  \city{São Carlos}
  \state{São Paulo}
  \country{Brasil}
  \postcode{13566-590}
}

\renewcommand{\shortauthors}{Zanon, A.L., Rocha, L.C.D., Manzato, M.G}

\begin{abstract}
    In \ac{RS}, explanations help users understand why items are recommended and can enhance a system’s transparency, persuasiveness, engagement, and trust, which are known as explanation goals. However, evaluating the effectiveness of explanation algorithms offline remains challenging because explanation goals are inherently subjective. We initially conducted a rapid literature review, which revealed that algorithms are often assessed using anecdotal evidence (offering convincing examples) or using metrics that do not align with human perception. From these results, we investigated whether the selection of item attributes and interacted items affects explanation goals in explanations that generate a path connecting interacted and recommended items based on shared attributes (such as genres). We used metrics that measure the diversity and popularity of attributes and the recency of item interactions to evaluate explanations from three state-of-the-art agnostic algorithms across six recommendation systems. We then performed an online user study to compare user perceptions of explanation goals and offline metrics. Our findings indicate that engagement is sensitive to users’ perceptions of diversity in explanations, whereas transparency, trust, and persuasiveness are influenced by perceptions of both popularity and diversity. However, offline metrics require refinement to more closely align with explanation goals and user understanding.\looseness=-1
\end{abstract}



\received{}
\received[revised]{}
\received[accepted]{}

\maketitle

\section{Introduction}

\ac{RS} provide personalized suggestions based on users' past interactions and preferences. As the volume of information grows, \ac{RS} architectures have become increasingly complex in modeling the underlying relationships among users, items, and metadata~\cite{nunes2017systematic}. Consequently, explaining \ac{RS} outputs has gained significance among researchers, aiming to offer users a more personalized experience \cite{tintarev2015explaining}.\looseness=-1


Explanation algorithms in \ac{RS} generate reasons for why a particular set of recommendations is suggested \cite{zhang2020explainable}. By providing such explanations, \ac{RS}s can enhance persuasiveness, transparency, trust, and user engagement, collectively known as explanation goals, which represent key advantages of incorporating explanations \cite{balog2020measuring, tintarev2015explaining}. However, unlike item ranking, where well-established offline metrics can effectively measure an algorithm's performance, measuring explanation goals with offline metrics is challenging because they must reflect how well explanations reflect the reasoning behind a recommendation \cite{zhang2020explainable,nunes2017systematic}. Consequently, explanation goals are typically evaluated through online user experiments \cite{musto2016explod, musto2019combining, du2022post}. Table \ref{tab:def} summarizes the definitions of all explanation goals.\looseness=-1


\begin{table*}[!htbp]
\caption{Table of explanation goal definitions as in \cite{tintarev2015explaining}. Engagement is defined as in \cite{musto2016explod}.}
    \begin{tabular}{c||l}  \hline 
         \textbf{Goa}l & \textbf{Definition}  \\ \hline \hline
         Transparency  &  Explain how the system works \cite{tintarev2015explaining}   \\ \hline
         Persuasiveness  &   Convince users to interact with a recommendation \cite{tintarev2015explaining}       \\ \hline
         Trust  &   Increase the user confidence on the recommendation algorithm \cite{tintarev2015explaining}    \\ \hline
         Scrutability & Allow the user to correct the \ac{RS} \cite{tintarev2015explaining}\\ \hline
         Effectiveness & Help users to take good decisions\cite{tintarev2015explaining} \\ \hline
         Efficiency & Help users to take fast decisions \cite{tintarev2015explaining} \\ \hline
         Engagement   &  Show new information to the user about a recommendation\cite{musto2016explod} \\ \hline
    \end{tabular}
\label{tab:def}
\end{table*}

\subsection{Problem Setting}

Because explanation goals are tied to the subjective aspects of user perception and feelings, evaluating these elements is challenging and necessitates a user study for accurate assessment. Nevertheless, evaluating explanations with offline metrics can be helpful to compare different explanation algorithms and have become widely adopted in the literature \cite{10.1145/3583558}.\looseness=-1

Consequently, we conducted a rapid literature review to analyze how explanations are evaluated offline in \ac{RS}. We verified three main problems: (i) similar to the broader field of \ac{ML}, explanations in \ac{RS} are validated only by anecdotal evidence \cite{10.1145/3583558}; (ii) most of the research focusing on user studies does not evaluate explanations in offline settings; as a result, the evaluation of explanations is limited to the recruited participants, who may not represent those of the real systems \cite{10.1145/3716394}; and (iii) most of the adopted offline metrics are not correlated with explanations goals, which means that the optimization of current offline explanation metrics may not necessarily mean an improvement in user perception.\looseness=-1

Regarding this last problem, the metrics most commonly cited in our rapid literature review are \ac{NLP} metrics, such as \ac{BLEU} and \ac{ROUGE}, which evaluate the overlap of n-grams between the generated and reference texts. However, in the conversational domain, these metrics are poorly correlated with user perception \cite{10.1145/3627043.3659574}. Consequently, these challenges complicate the assessment of progress in explanation algorithms within recommendation systems and the impact of different methods on explanation goals \cite{nunes2017systematic}.\looseness=-1

Existing metrics that do not align with human perception evaluate explanations based on word co-occurrence, rather than content. To address this limitation, we investigated path metrics, an underexplored offline explanation metric that considers shared content between interacted and recommended items. We examine whether these metrics correlate with users’ perceptions of explanation goals, including transparency, persuasiveness, trust, and engagement.\looseness=-1

Explanations can show how one or more items with which a user has interacted (i.e., items in the user's profile) are connected to a recommended item through shared attributes. In an explanation such as ``Because you watched Saving Private Ryan, starring Tom Hanks, watch Forrest Gump'', the interacted item ``Saving Private Ryan'' is connected to the recommended item ``Forrest Gump'' by the actor Tom Hanks, which is the shared attribute.\looseness=-1

In this style of explanation, a \emph{path} to the recommended item is created using two main elements: (a) the interacted item, and (b) the attributes of the item. Path metrics evaluate each element from various perspectives. For interacted items, we can measure various aspects, such as the recency of the interaction and the repetition of the item across multiple explanations.\looseness=-1

\begin{figure}[!ht]
    \centering
    \includegraphics[width=.6\textwidth]{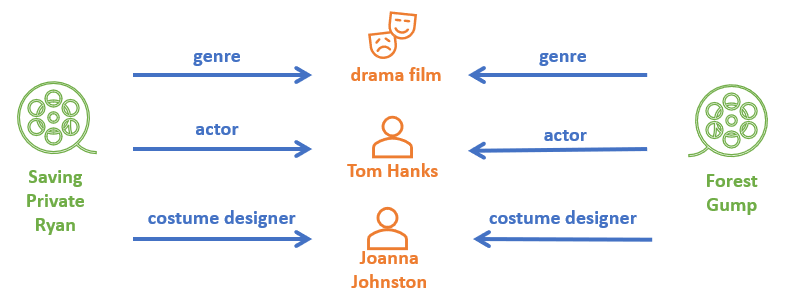}
    \caption{Example of different item's attributes for a single interacted item. Attributes are represented in \textcolor{YellowOrange}{orange} and, in \textbf{\textcolor{NavyBlue}{blue}}, the relation between the attribute and the item and in \textbf{\textcolor{YellowGreen}{green}} are items.}
    \label{fig:graph}
\end{figure}

Similarly, explanation attributes can be evaluated for their popularity across other items, which increases their familiarity with users. Figure \ref{fig:graph} illustrates the construction of explanations using different attribute types. For instance, ``drama'' is a very common attribute, as many movies belong to this genre. In contrast, ``Tom Hanks'', while a popular actor, is associated with fewer films than a genre like ``drama''. At the other extreme is the shared attribute ``Joanna Johnston'', who served as the costume designer for both films, representing a much less common attribute.\looseness=-1

\subsection{Objective and Research Questions}

Therefore, the main objective of this work is to analyze how different attributes and interacting items on paths between interacted and recommended items can impact user perception and their evaluation of explanation goals.\looseness=-1
 
The first \ac{RQ} investigated in this study is {\bf (\ac{RQ}1)}:  {\it How are explanations in \ac{RS}s evaluated with offline metrics in the literature?}. The \ac{RQ} has the objective of searching the literature for available offline explanation evaluation metrics, what aspect of the explanation it evaluates, and whether there exists support in the literature for correlation between the offline metric used and user perception of explanation goals.\looseness=-1 

Our second \ac{RQ} is {\bf (\ac{RQ}2)}: {\it In explanations that connect interacted and recommended items based on shared attributes, how does the selection of item attributes and interacted items affect explanation goals?}. The main objective of this \ac{RQ} is to verify whether a relationship exists between the construction of explanations and user perception. We hypothesize that user perception of explanation goals is tied to the different ways of selecting the elements of an explanation, particularly attributes and interacted items. Therefore, by answering this RQ, we hope to help researchers identify whether selecting attributes that are more or less popular, for instance, impacts the perception of users under explanation goals. \looseness=-1

To answer {\bf (RQ2)}, we used six path-based offline metrics to measure the recency and diversity of interacted items and the popularity and diversity of attributes in explanations for three agnostic explainable algorithms for six \ac{RS}s that create a direct state-of-the-art evolution. We also conducted an online experiment considering persuasiveness, transparency, engagement, and trust using the two algorithms that achieved the best performance on the offline metrics. Users’ opinions on the explanations generated by both algorithms were compared with the offline metrics to better understand the relationship between the online explanation goals and these offline metrics. \looseness=-1

\subsection{Overview of Main Findings and Contributions}

Considering RQ1, according to our rapid literature review, the quality of explanations is indirectly evaluated in offline experiments, as they rely on other contributions, such as improvements in ranking accuracy and diversity \cite{nunes2017systematic}. As a result, explanations are assessed using anecdotal evidence \cite{10.1145/3583558}. While some offline explanation metrics, such as NLP-based metrics, do not correlate with human perception in conversational \ac{RS} \cite{10.1145/3627043.3659574}, others, such as path-based metrics, have not been extensively evaluated in terms of user perceptions. As a result, improvements in offline metrics may not necessarily translate into enhancements in explanations as perceived by users. \looseness=-1

Regarding RQ2, our results suggest that offline measurements of attributes and interacted items reflect users’ perceptions of explanation goals to some extent. In particular, engagement appears to be sensitive to the diversity of attributes within explanations, whereas transparency, trust, and persuasiveness are influenced by users’ perceptions of both popularity and diversity. Consequently, offline explanation metrics still need to evolve to more reliably align with explanation goals and users’ understanding.\looseness=-1

The main contributions of this study are as follows:
\begin{itemize} 
    \item A survey on current metrics for evaluation of explanations in offline settings;\looseness=-1 
    \item Insights into the relation of offline path explanation metrics with explanation goals; and\looseness=-1 
    \item Guidelines for evaluating explanations in \ac{RS}s with offline experiments and open research directions in the field.\looseness=-1
\end{itemize}


\section{Definitions and Terminology}
\label{sec:def_and_term}

Before answering \ac{RQ}1 with an analysis of the rapid literature review, in this section, we introduce the different ways to generate explanations, explore fundamental concepts, and highlight methodologies associated with generating and evaluating explanations on \ac{RS}.\looseness=-1

\ac{RS} require a set of items that the user has previously interacted with to generate recommendations. These items are typically referred to as historic, interacted, or profile items\footnote{We will use the terms ``historic", ``interacted items", and ``profile items`` interchangeably}. Based on this pipeline, where \ac{RS} uses a set of items to output recommendations, explanation algorithms are divided into two main methods: agnostic and intrinsic. Agnostic methods use a separate algorithm to interpret black-box recommendations, whereas intrinsic approaches aim to generate transparent recommendations in which the resulting explanations are a direct reflection of the system's underlying reasoning \cite{zhang2020explainable,BARREDOARRIETA202082}. More recently, reordering approaches have been proposed to adjust the order of recommendations, prioritizing those with more compelling explanations \cite{balloccu2022post,ZANON2022109333}.\looseness=-1 

Agnostic methods can be integrated with any recommendation algorithm, but do not reveal the exact logic behind the recommendations \cite{rana_daddio_manzato_bridge_2022}. For this reason, model-agnostic explanations are also referred to as justifications \cite{musto2019combining}. In contrast, intrinsic methods provide more transparency because the explanations are integrated into the recommendation engine, though this can affect system latency and increase vulnerability to adversarial attacks \cite{xu2023reusable}.\looseness=-1

In addition, explanation algorithms can use various types of information to enhance explanation goals. In this context, an ``explanation style'' refers to the method or different types of data employed to explain the reason behind certain recommendations to users \cite{kouki2019personalized,bilgic2005explaining,papadimitriou2012generalized,tintarev2015explaining}. The literature also enumerates different possible explanation styles \cite{tintarev2015explaining,kouki2019personalized,bilgic2005explaining,papadimitriou2012generalized}. In this manuscript, we will use the five different explanation styles defined in \cite{kouki2019personalized}: user-based, which generates explanations based on other users (e.g., "Watch Titanic since similar users watched it as well"); Item-based, that uses similar items to justify a recommendation (e.g., "Users who watch Braveheart also watch Titanic"); Content, which uses item metadata in explanations (e.g., "Watch Titanic because you like drama movies"); Social, which uses social connections such as friends in explanations (e.g., "Watch Titanic because your friend Alice likes it"); and popularity explanations (e.g., "Titanic is highly popular among users"). Hybrid explanation styles use combinations of two or more of the previous explanation styles.\looseness=-1

After an explanation is generated based on a method and an explanation style, the evaluation can be performed in two distinct ways: with online experiments or offline experiments. We categorize online experiments with two main distinct subcategories: A\slash B testing and user trials.\looseness=-1

In A\slash B testing, the impact of explanations is assessed by comparing clicks from two groups to determine if there is improved adherence in the intervention group over the control group. In contrast, user trials do not rely on deployed systems. Participants are recruited to simulate system interactions. Unlike deployed \ac{RS}s, where the set of interacted items is updated continuously with user clicks over time, user trials involve a single session. In this session, participants are assigned or create a set of simulated interactions, receive recommendations and evaluate explanations.\looseness=-1

Although these two approaches can overlap, as it is possible to perform user trials in the form of A\slash B testing, we adopt these definitions in this work to distinguish between data-driven evaluations performed in production (or production-like) environments and controlled studies that focus on user perception.\looseness=-1

The offline evaluation of explanations varies depending on the explanation style, primarily because different styles utilize different types of information, affecting the applicability of the metrics. For instance, item-based explanation styles require metrics that evaluate the similarity between recommended items \cite{10.1145/3490099.3511104}, whereas content-based styles focus on the relevance of item attributes \cite{balloccu2022post}.\looseness=-1

Throughout this manuscript, we adopt this terminology to refer to the different aspects of explanations in a \ac{RS}. In Section \ref{sec:lib}, we use it to categorize and analyze the papers retrieved in our rapid literature review, which supports our investigation of \ac{RQ}1. This terminology also serves as the basis for our methodology, as we employ agnostic models that generate content-based style explanations to examine whether path metrics correlate with explanation goals and, in turn, address \ac{RQ}2.\looseness=-1

\section{Rapid Literature Review}
\label{sec:lib}

Considering the concepts discussed in Section \ref{sec:def_and_term}, we conducted a rapid literature review to address {\bf (RQ1)}. Unlike a systematic literature review that covers all studies on a research topic and provides an in-depth analysis of the field's past, current state, and future directions, a rapid literature review covers a shorter timeframe to analyze recent works and synthesize results and key new findings \cite{schünemann_moja_2015,smela_toumi_świerk_francois_biernikiewicz_clay_boyer_2023}. 

{In this section, we first outline the methodology of our rapid literature review. We then define the categories used to classify the selected papers. Finally, we present the findings and highlight recent research on novel offline explanation metrics for \ac{RS}.\looseness=-1

\subsection{Search and Selection}

This review was structured according to the guidelines proposed by \cite{kitchenham2009systematic} and aimed to retrieve relevant papers on explanation algorithms within the \ac{RS} community. We focused on analyzing how these explanations were generated and evaluated in the selected papers. Unlike other studies \cite{10.1145/3583558, krishnadisagreement} that surveyed and analyzed papers on explanations for \ac{ML} algorithms, our review specifically targets \ac{RS} research. Extracted papers and our analysis is available open source in a code repository\footnote{\url{https://github.com/andlzanon/offline_metrics_explanation_rec_goals/tree/main/Literature_Review}}. Figure \ref{fig:lit_review} depicts the workflow used to select the papers.\looseness=-1 

\begin{figure}[!ht]
    \centering
    \includegraphics[width=.75\textwidth]{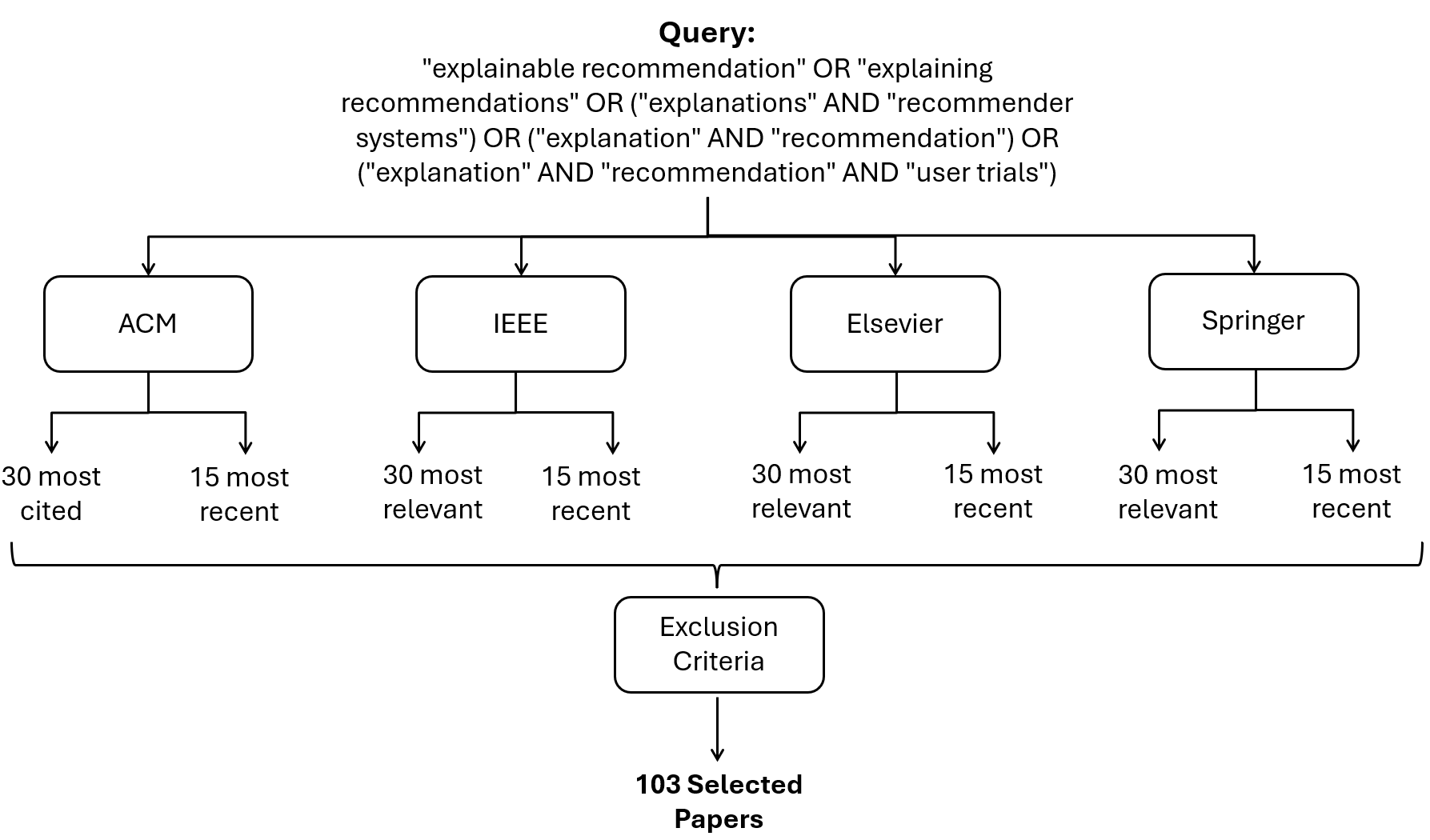}
    \caption{Workflow of the conducted rapid literature review}
    \label{fig:lit_review}
\end{figure}

Initially, to retrieve all papers that could be related to our subject, we constructed the following query: "explainable recommendation" OR "explaining recommendations" OR ("explanations" AND "recommender systems") OR ("explanation" AND "recommendation") OR ("explanation" AND "recommendation" AND "user trials"). These terms were searched for in the entire document (not just the titles) of conference papers and journal papers and were filtered by publication year ranging from 2015 to 2025.\looseness=-1

We used the search engines of the Association for Computing Machinery (ACM) Digital Library \footnote{\url{https://dl.acm.org/}}; the Institute of Electrical and Electronics Engineers (IEEE) Xplore engine \footnote{\url{https://ieeexplore.ieee.org/Xplore/home.jsp}}; Elsevier's Science Direct engine\footnote{\url{https://www.sciencedirect.com/}} and Springer Nature Link\footnote{\url{https://link.springer.com/}}. We extracted the 30 most cited papers from each search engine. On IEEE Xplore, ScienceDirect, and Springer Nature Link, we used the search engines’ ``relevance'' ranking, which considers factors such as matching of search terms with document content, the importance of the journal or conference, and the number of citations. To capture recent research and reduce bias toward older, highly cited papers, we extracted the 15 most recent papers from each search engine. This process resulted in a total of 180 papers.\looseness=-1

The search query was built with specific keywords to primarily explore how explanation algorithms were evaluated offline, regardless of whether user trials were conducted, and helped us answer \ac{RQ}1. Initially, the last term of our search query ("explanation" AND "recommendation" AND "user trials") was not included in our rapid literature review, however, after reviewing the preliminary results from the search engines, we added this term  because no papers on online studies were found. This occurs because user studies are more time-consuming than offline experiments in terms of development, as they require analyzing participant interactions with an implemented platform that generates explanations for recommendations \cite{nunes2017systematic}.\looseness=-1

After this initial search, we applied exclusion criteria and removed the following: survey papers, since we found no similar literature review paper on the offline evaluation of explanations in \ac{RS}; papers that mentioned a query term but were not about explanations for \ac{RS}; perspective papers; papers proposing new datasets for explainable recommendation; prefaces of special issues; and books. This resulted in 103 papers for the literature review analysis \looseness=-1

Figure \ref{fig:year} illustrates the distribution of articles over the years. Most of the manuscripts were published between 2023 and 2025, but the years between 2018 and 2020 are also fairly well represented. \looseness=-1

\begin{figure}[!b]
    \centering
    \includegraphics[width=.55\textwidth]{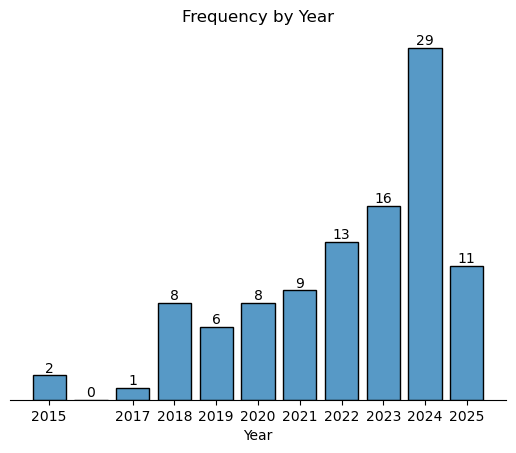}
    \caption{Distribution of the papers found by our rapid literature review by year of publication.}
    \label{fig:year}
\end{figure}

Similarly, Figure \ref{fig:sources} in Appendix \ref{apen:sources} presents the distribution of papers by journal and conference. Our search identified 56 different conferences, with the most prominent being the ACM Web Conference, ACM SIGIR Conference on Research and Development in Information Retrieval, and Conference on Information and Knowledge Management. The journals with the highest number of publications in our survey include: Knowledge-Based Systems and Neurocomputing.\looseness=-1

\subsection{Organization and Definitions}
\label{sec:org_lit}

We analyzed the papers according to the seven categories described in Section \ref{sec:def_and_term}. The classification of each paper retrieved from our search query, after applying the exclusion criteria, is presented in Table \ref{tab:lit_review} in Appendix \ref{apen:lit_review}, where each column is a category.\looseness=-1

The first categoric is \textbf{Style} and refers to the explanation style used to generate explanations and their \textbf{Method}, the second category, characterizes how the explanations are generated on the recommendation process.\looseness=-1

Multiple categorizations exist for explanation styles \cite{tintarev2015explaining,kouki2019personalized,bilgic2005explaining,papadimitriou2012generalized}. We adopted the one proposed in \cite{kouki2019personalized}, detailed in Section \ref{sec:def_and_term}, as it covers most of the studies from our rapid literature review. However, with the growing body of research in the field, some explanation styles have been further subdivided into more specific types. These subdivisions arise because different explanation styles affect offline and online evaluations in distinct ways, drawing on different types of information to generate explanations. The identified classifications of the \textbf{Style} category include: \looseness=-1

\begin{itemize}
    \item \textbf{Personality}, that generates explanations based on psychological traits using techniques such as the Big Five \cite{takami_flanagan_dai_ogata_2023}. In the categorization of \cite{kouki2019personalized}, this can be classified as a user-based explanation. However, we propose this new explanation style because it represents an emerging sub-category of user-based explanations, where the user's and other users' personality traits are used to generate explanations that match their psychological aspects;\looseness=-1

    \item \textbf{Content}, previously defined in \cite{kouki2019personalized}, it uses metadata information from items to connect them to users and enrich \ac{RS}s with external information. In our literature, item metadata were used in two distinct ways: with key-value pairs, where the key represents an item and the value represents a list of the item's metadata, and with \ac{KG}. A \ac{KG} is defined as $KG = \{ (h,r,t) | h,t \in E, r \in R \}$ where $h$ and $t$ represent node entities, which are items and attributes and $r$ represents a relation link (also called as edge type) between these two entities \cite{10.1145/3269206.3271739}. In Figure \ref{fig:graph}, for instance, ``genre", ``actor" and ``costume designers" would be instances of $r$ in a KG since it connects two entities, an item node ``Saving Private Ryan" and an attribute node
    ``drama film". Similarly, ``Forest Gump" is another item node connected to the same ``drama" film' attribute node as ``Saving Private Ryan" item node. Entities can also represent users, creating a graph where nodes of users, items and metadata are all connected;\looseness=-1
    
    \item \textbf{Review}, a sub-category of the content explanation style, uses unstructured data from items in the form of user reviews. Review-style explanations use \ac{NLP} techniques to extract item features from text to enhance the recommendation accuracy and transparency. It was separated from it is an original category because review style explanations are prominent in the literature \cite{musto2019combining,AI2025129692,10.1145/3178876.3186070,10.1145/3178876.3186145,10.1145/3485447.3512029};\looseness=-1

    \item \textbf{User-based}, which generates user-based styles explanations as proposed in \cite{kouki2019personalized}. It uses information from other users as well as the active user to create explanations;\looseness=-1
    
    \item \textbf{Feature} explanations have gained prominence with the rise of model-agnostic feature importance methods such as \ac{SHAP} \cite{NIPS2017_8a20a862} and \ac{LIME} \cite{10.1145/2939672.2939778}. These methods are increasingly applied in the \ac{RS} domain, particularly in decision-making scenarios in which the input values differ from typical item and user embeddings. Examples include applications in manufacturing and agriculture \cite{10825771,10308154}. Such cases were not described in \cite{kouki2019personalized} since \ac{LIME} and \ac{SHAP} were recently proposed by then;\looseness=-1
    
    \item \textbf{\ac{LLM}} that leverages the power of generative models, more specifically, \ac{LLM}, to generate explanations for \ac{RS}s. Research in this area typically involves prompting LLMs to produce explanations for recommendations \cite{10658914}. Other studies have explored instruction-tuning prompts for LLMs to enhance both recommendation and explanation performance \cite{10.1145/3640457.3688137}. Because LLMs have the potential to utilize external and multimodal information \cite{wang2024nextgenerationllmbasedrecommendersystems}, they can be considered a subcategory of several other categories enumerated here. However, due to their versatility, they do not fit neatly into any existing category. Consequently, we introduce a dedicated category for them, and; \looseness=-1
    
    \item \textbf{Hybrid}, as proposed in \cite{kouki2019personalized}, encompasses explanation algorithms that use different explanation styles of information to generate explanations. \looseness=-1 

\end{itemize}

We also introduced three main methods (Method category) to generate explanations in \ac{RS}s: \textbf{Intrinsic}, \textbf{Agnostic}, and reordering. Because reordering methods change the ranking of a recommendation algorithm to prioritize those with more compelling explanations, in this rapid literature review these methods are categorized as agnostic approaches.\looseness=-1

In that regard, after explanations are generated, in Section \ref{sec:def_and_term} we also listed the two main ways of evaluation: with online experiments, by measuring and analyzing user responses when exposed to explanations and\slash or offline experiments, by evaluating mathematically explanations with metrics. \looseness=-1

If a paper performed offline evaluation, the categories \textbf{Offline Metric} and \textbf{\# of Users} detail how the offline evaluation was conducted. The first column describes the offline metric used and the second column indicates the number of users to whom the metric was applied. Specifically, metrics were either computed across the entire dataset (\textbf{All dataset}) or based on a sample of users (\textbf{Examples}). The following offline metrics were identified in the reviewed papers:\looseness=-1

\begin{itemize}
    \item \textbf{Precision\slash Recall} metrics are calculated based on the amount of relevant information generated in an explanation compared to ground-truth data. The use of these metrics varies depending on the context. For example, in the \ac{SHAP} and \ac{LIME} methods, error metrics are used to evaluate the surrogate model that outputs feature importance values for each input \cite{BRUNOT2022102021}. They are also applied in sentiment analysis \cite{li_liu_zhang_kou_liu_qu_2025}, to compare the co-occurrence of words in generated explanations with those in actual user reviews \cite{10884422}, and to assess the relevance of a user review using ground-truth annotated data \cite{YANG2021106687,10.1145/3178876.3186070};\looseness=-1
    
    \item \textbf{Path Metrics} explanations connect interacted and recommended items through attributes, forming a path between a user's historical items and recommended items in the same way as in Figure \ref{fig:graph}. Originally proposed in \cite{balloccu2022post}, path explanations measure two key elements: (a) interacted items and (b) item attributes (attributes), using three main metrics. For attributes, popularity and diversity across explanations were measured. For the interacted items, the measurement was based on recency. The main hypotheses for path metrics are that explanations should connect recently interacted items with recommended items, and attributes should be popular yet diverse across different explanations. Path explanations are common among content-type explanations that use graphs or \ac{KG}. For example, in \cite{ZANON2022109333}, a reordering approach was proposed based on the best explanation considering weighted paths between the items with which the user interacted and those recommended. The weights of the paths were measured according to a \ac{TF-IDF} of \ac{KG} attributes;\looseness=-1\looseness=-1
    
    \item \textbf{Anecdotal}, which are evidence of the functioning of an explanation algorithm based on examples that pass an ``face-validity" \cite{10.1145/3583558}. In \ac{RS}, a set of example explanations is generated for some users of the dataset.
    The studies by \cite{melchiorre2022protomf} and \cite{zhao2020leveraging} serve as examples of this offline metric. In \cite{melchiorre2022protomf}, a user-based explanation style was developed using a collaborative filtering algorithm that extends Matrix Factorization methods and creates prototypes, which are representative entities from users and items used in explanations. Conversely, \cite{zhao2020leveraging} employed an adversarial actor-critic reinforcement learning algorithm over a \ac{KG} to identify optimal paths based on user interactions, enhancing recommendations and explanations. In both studies, the proposals were evaluated based on their recommendation ability, with explanations being assessed through examples. This is also present in other intrinsic explainable recommendation engines \cite{10.1145/3269206.3271739,10.1145/3331184.3331203,balloccu2023reinforcement,tai2021user};\looseness=-1 
    
    \item \textbf{Counterfactual} measures the quality of explanations based on counterfactual metrics such as probability of sufficiency and probability of necessity. These metrics were used in  \cite{tan2021contefactual}, for instance, where a counterfactual explanation for the recommendation algorithm was produced as a black box with a soft optimization method sensitive to changes to the item’s aspects via solving a counterfactual optimization; \looseness=-1
    
    \item \textbf{\ac{BLEU}/\ac{ROUGE}} are metrics from \ac{NLP} based on the comparison of n-grams between a generated and a ground-truth text. \ac{BLEU} is a precision-focused metric, calculating the number of n-grams in the generated text that match a ground-truth text, divided by the total number of n-grams in the generated text. \ac{ROUGE}, on the other hand, is a recall-focused metric, computing the number of n-grams in the generated text that match a ground-truth text, divided by the total number of n-grams in the ground-truth text. These metrics are exclusive to explanation algorithms that use reviews as their source. An example of the use of such metrics is in \cite{hada2021rexplug}, which generated personalized reviews for users as explanations with an attention-based parallel network called cross-attention for selecting candidate users and item reviews for constructing the final sentence. The authors compared the ground-truth text with the generated review to evaluate the explanations;\looseness=-1
    
    \item \textbf{Explainable Items}, which measures the quantity of recommended items that can be explained by an explanation algorithm. As an example of this metric, authors in ~\cite{coba_confalonieri_zanker_2022} implemented a library with metrics such as Mean Explainability Precision, which measures the number of explainable items for a user, Model Fidelity, which evaluates recommendations with proxy predictions, and Explanation Score, which measures the number of interactions that support an explanation;\looseness=-1
    
    \item \textbf{Correlation}, which measures the strength and direction of the relationship between an explanation and another aspect. For instance, in \cite{10.1145/3357384.3357925}, explanations were evaluated using heatmaps to explore the relationship between the attention scores and features. In \cite{10.1145/3442381.3449788}, plots were generated to compare the embeddings from the proposed model with those from other methods. Similarly, in \cite{8594883} it was measured the correlation between ratings and explanation sentiments.\looseness=-1
\end{itemize}

Similarly, when the paper does online experiments, the \textbf{Online Metric} category divides such types of works into other two: A\slash B testing, where a comparison between algorithms is conducted in environments similar to real scenarios, as described in Section \ref{sec:def_and_term}; and (\textbf{User Trial}), where recruited participants evaluate explanations considering transparency, effectiveness, scrutability trust, persuasiveness, efficiency and satisfaction proposed in \cite{tintarev2015explaining} with a within-subjects or between-subjects experiment.\looseness=-1

\subsection{Analysis and Insights}
\label{sec:lite_review_analysis}

To answer \ac{RQ}1 on how explanations are evaluated offline in \ac{RS}, Figure \ref{fig:results_lit_review} illustrates the results from Table \ref{tab:lit_review}, focusing on the 81 papers that conducted offline evaluations of the generated explanations. Specifically, we examined three main aspects of the offline evaluation: the metric applied, the number of users in the dataset for whom explanations were generated, and whether the method was intrinsic or agnostic. We analyze these perspectives considering the explanation style to understand how the evaluation for each is conducted.\looseness=-1

\begin{figure}[!b]
    \centering
    \includegraphics[width=.8\textwidth]{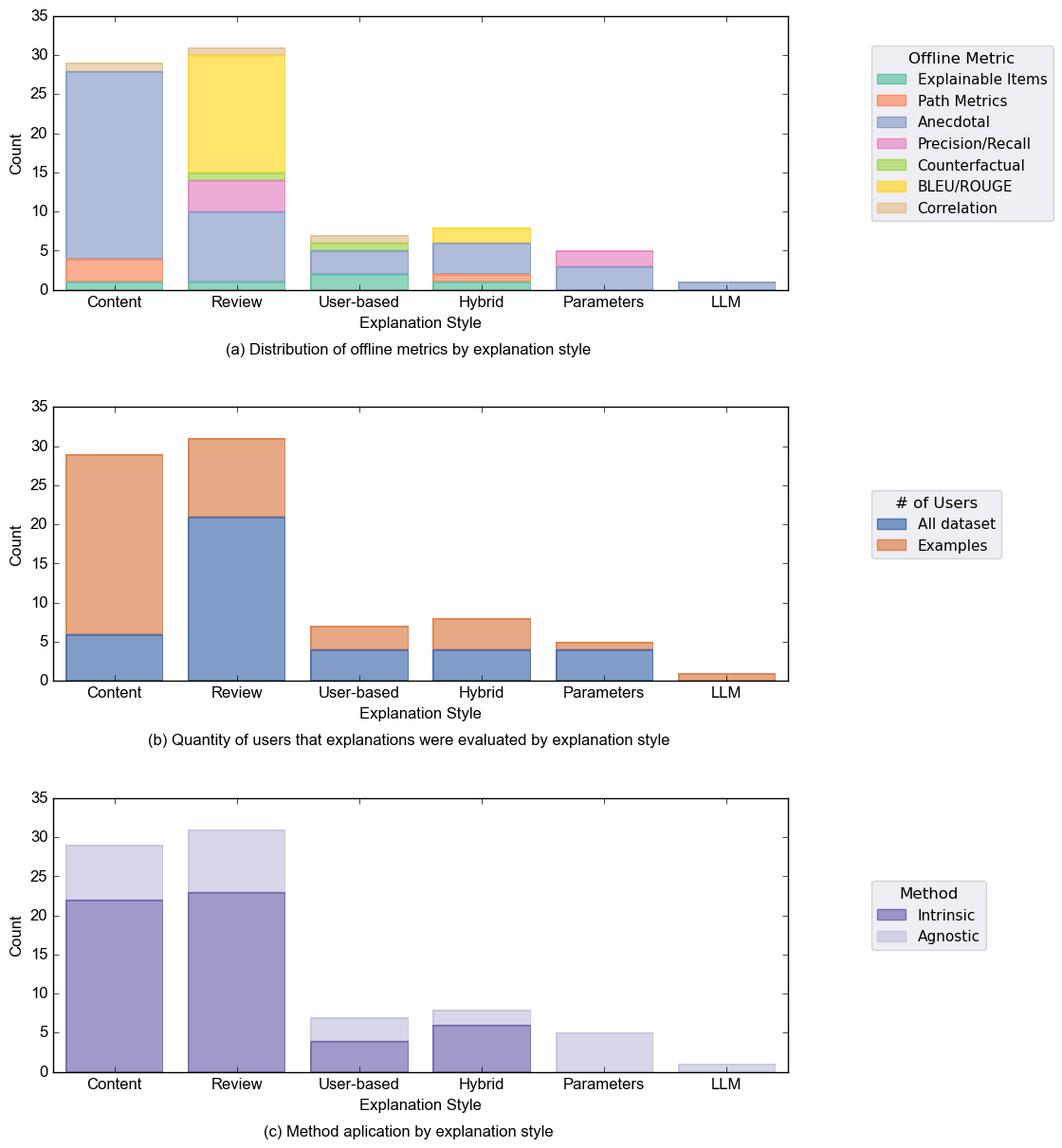}
    \caption{Distribution of papers that did offline evaluation of \ac{RS}s considering the metrics used (a), the number of users used to evaluate generated explanations (b) and chosen method (c) in relation to explanation styles.}
    \label{fig:results_lit_review}
\end{figure}

Analyzing the histogram in Figure \ref{fig:results_lit_review} (a), a stacked bar chart displays the proportion of offline metrics used to evaluate a specific explanation style. In this regard, there is no consensus on the evaluation metrics for each explanation style. Considering content-based explanations, most studies use anecdotal evidence from a small number of users to display the potential explainability effect of the proposed explanation algorithm. However, informally looking at a handful of examples can lead to sample bias.\looseness=-1

The combination of content-based explanations with anecdotal evidence evaluation occurred in 24 of the 29 manuscripts that performed offline evaluations. From the five remaining papers, 3 of them used path metrics, one used correlation, and one used the number of recommended items that could be explained (explainable items) as explanation metrics. A similar pattern can also be seen on User-based, Hybrid, Parameters and \ac{LLM} explanation styles, where the majority of works are evaluated on anecdotal evidence.\looseness=-1

These results align particularly with a literature review of the broader \ac{XAI} community in \ac{ML} \cite{10.1145/3583558}, where there is no consensus on metrics for explanations beyond anecdotal evidence, as illustrated by the example of explanations. Consequently, similar to the \ac{XAI} community in \ac{ML}, the \ac{RS} community has yet to agree on the explanation metrics for each explanation style. This problem also affects assessing state-of-the-art progress in the field, as explanations of the same type often use different metrics, making it difficult to demonstrate improvements over algorithms.\looseness=-1

Review-based explanations are the most common in the literature. Although anecdotal evidence is also largely used in this scenario, most studies evaluate their methods using \ac{BLEU} and\slash or \ac{ROUGE}, with a notable portion employing Precision and Recall. The main hypothesis for validating explanations under these four metrics is that users relate to explanations similar to their own or other user reviews. Nevertheless, recent evidence suggests that the correlation between the \ac{BLEU} and \ac{ROUGE} metrics is weak regarding user perception in the conversational \ac{RS} scenario \cite{10.1145/3627043.3659574}. Therefore, the validity of these explanations to user perception still needs further research.\looseness=-1

The observations in Figure \ref{fig:results_lit_review} (a) impact the analysis in Figure \ref{fig:results_lit_review} (b). Specifically, content, user-based, hybrid, and LLM explanation styles, which are mostly associated with anecdotal evaluations, primarily use sample users to evaluate explanations. Analogously, for the review explanation style, which mostly use \ac{BLEU} and \ac{ROUGE} \ac{NLP} metrics, evaluations are conducted on all users within a dataset.\looseness=-1 

Figure \ref{fig:results_lit_review} (c) reveals that most studies employ intrinsic methods to generate explanations. Of the 103 papers examined, 68 utilized this approach, with eight not conducting offline or online experiments of explanations. These studies emphasize contributions to recommendation ranking, accuracy, and beyond-accuracy metrics \cite{nunes2017systematic}. Although these works claim the recommendation engine is interpretable, they do not always specify an explicit algorithm for generating explanations.\looseness=-1

Of all the papers from our rapid literature review, only 14 of the 103 conducted online experiments. This highlights and confirms the discoveries in \cite{10.1145/3583558} and \cite{nunes2017systematic} that online experiments is not commonly used in the \ac{RS} \cite{nunes2017systematic} and \ac{XAI} communities \cite{10.1145/3583558} within \ac{ML}. This is surprising given that \ac{RS}s are closely tied to human decision-making, with explanation goals aimed at improving user perception of the \ac{RS} \cite{tintarev2015explaining,10.1145/2959100.2959186}.\looseness=-1

Furthermore, similar to the offline evaluation of item ranking in \ac{RS}, where improvements in offline metrics may not align with online metrics \cite{steck2021deep}, a similar issue arises when explaining recommendations \cite{10.1145/3627043.3659574}. Consequently, current offline explanation metrics fall short in addressing the improvement of explanation goals, as they lack validation through online user studies.\looseness=-1

In addition to the lack of user studies, research papers often evaluate explanations with users who do not adequately represent the target recommendation domain \cite{10.1145/3716394}. Consequently, theoretical frameworks for conducting online experiments, such as those in \cite{knijnenburg2015evaluating}, remain largely unutilized. This highlights the need for robust offline metrics that can correlate with and guide the development of effective recommendation explanations.\looseness=-1

Therefore, how explanations are evaluated offline and online are currently unrelated. One of the main reasons is due to the complexity of performing user evaluations since they require the development of a user application, recruitment of participants, and rigorous statistical analysis for evaluating and comparing explanations.\looseness=-1

Future directions of our literature review rely on the creation of a theoretical framework for the evaluation of explanations in \ac{RS}, considering the conduction of offline experiments, metrics for each source of information, and availability of data and generated explanations, which could help create explanation algorithms to develop a state-of-the-art evolution of algorithms. In addition works on hybrid and personality explanation styles, as well as counterfactual explanations are new emerging research topics that could be further explored.\looseness=-1

\subsection{New Offline Explanation Metrics}
\label{sec:relworks}

Recently, some studies have focused on proposing new methods for evaluating explanations through offline experiments in \ac{RS}. These metrics were not included in our rapid literature review because they were not used in papers introducing explanation algorithms. In this section, we discuss these works and highlight their main limitations.\looseness=-1

In \cite{Wen2022}, the authors introduced a new metric, ExpScore, designed to evaluate explanations in the absence of a ground truth. They used evaluation factors, such as recommendation relevance, explanation length, readability, word importance, repetition, subjectivity, polarity, grammatical correctness, and feature appearance, as inputs to a neural network trained on a large dataset collected by the authors. ExpScore outperformed \ac{BLEU} and \ac{ROUGE} in capturing user perceptions. However, the metrics do not relate on explanation goals as the dataset collected to fit the model was based on users evaluating a series of explanations on a Likert scale of 1 to 5 on quality, where 1 represented a user perception of ``low quality" of an explanation and 5 of ``high quality" of an explanation.\looseness=-1

Alternatively, \cite{balog2020measuring} measured the correlation between explanation goals by generating recommendations along with explanations designed by crowd workers to align with each specific explanation goal. The objective of this study was to determine whether optimizing explanations for a particular goal could affect users’ perceptions of another goal. Participants rated the explanations according to all explanation goals, resulting in moderate correlations across all metrics. However, the paper does not address how explanations should be constructed.\looseness=-1

With the same limitation, \cite{zhang2024large} proposed the use of \ac{LLM} to evaluate text explanations on \ac{RS}s and compared the score from the \ac{LLM} with offline metrics, such as \ac{BLEU} and \ac{ROUGE}, and online explanation goal metrics. In this regard, a medium correlation was found between the online explanation goal metrics and offline metrics, with the same effect occurring with the score provided by the \ac{LLM}s. Furthermore, the use of \ac{LLM}s to evaluate explanations in \ac{RS}s, despite promising, does not help researchers understand how explanations should be generated to captivate users.\looseness=-1 

The main limitation of current offline explanation metrics, such as \ac{BLEU} and \ac{ROUGE}, as well as the newly proposed scores, is that they do not consider the elements within explanations and treat them as sequences of words. In {\bf RQ2}, we investigate whether path metrics, which regard attributes and interacted items as elements with measurable properties—such as popularity and diversity for attributes and recency for interacted items—are related to explanation goals.\looseness=-1

\begin{tcolorbox}[colback=white!10!white,colframe=gray!75!black,title=Answer to \ac{RQ}1: How are explanations in \ac{RS}s evaluated with offline metrics in the literature?]

Based on our rapid literature review, we discovered three main results: 

\begin{enumerate}
    \item \emph{Explanations in \ac{RS} are typically evaluated according to their explanation style}. For instance, algorithms that generate review-based explanations are mostly assessed with \ac{BLEU} and \ac{ROUGE}. \emph{Yet, evidence shows that these \ac{NLP} metrics do not align with how users actually perceive explanations} \cite{10.1145/3627043.3659574};\looseness=-1 

    \item For other explanation styles, \emph{evaluation often relies on anecdotal evidence—sample explanations that appear convincing and illustrate the algorithm’s functioning. However, this approach lacks the rigor required for robust validation \cite{10.1145/3583558}}. Taken together, this means that the two most widely used offline evaluation strategies—applied in 61 of the 103 papers analyzed ($\approx$ 60\%)—do not reflect the user-centered goals of explanations proposed by \cite{tintarev2015explaining}. This limitation makes it difficult to track progress across explanation styles and to establish a clear timeline of advances in \ac{RS} explanation algorithms;\looseness=-1

    \item Since most algorithms that generate explanations for \ac{RS} rely on other contributions \cite{nunes2017systematic}, \emph{online evaluation is not common in the field}. Nevertheless, the main benefits of providing explanations, as described by the explanation goals of \cite{tintarev2015explaining}, come from improvements in user experience. Therefore, we highlight the importance of online user evaluation when validating explanations. \looseness=-1

\end{enumerate}

To address this, we recommend evaluating explanations in \ac{RS} through large-scale user studies that combine offline and online experiments. Offline metrics can serve as a first filter, indicating whether an algorithm is ready for validation in online settings.\looseness=-1

In summary, \emph{offline explanation evaluation practices in the \ac{RS} community share the same limitations observed in the broader \ac{XAI} field: the absence of consensus on offline metrics and the scarcity of online user studies \cite{10.1145/3583558}}.\looseness=-1

\end{tcolorbox}

\section{Methodology}
\label{sec:kg_metrics}

In Section \ref{sec:lite_review_analysis}, we verified that offline explanation metrics do not correlate with online experiment metrics. The most common metrics identified in our rapid literature review are \ac{BLEU} and \ac{ROUGE}, which measure the n-gram overlap between generated and reference texts. However, these metrics have been proven to be weakly correlated with user perception in the conversational domain \cite{10.1145/3627043.3659574}. Precision and Recall are similar, as they also assess the similarity between the generated and ground-truth explanations. Finally, another metric used is the number of recommendations that can be explained; this metric measures the robustness of an explanation algorithm in generating explanations for all the recommendations. Consequently, there is a gap in the literature regarding offline explanation metrics in \ac{RS} that correlate with user perception on online experiments under explanation goals.\looseness=-1  

\begin{figure}[!ht]
    \centering
    \includegraphics[width=.7\textwidth]{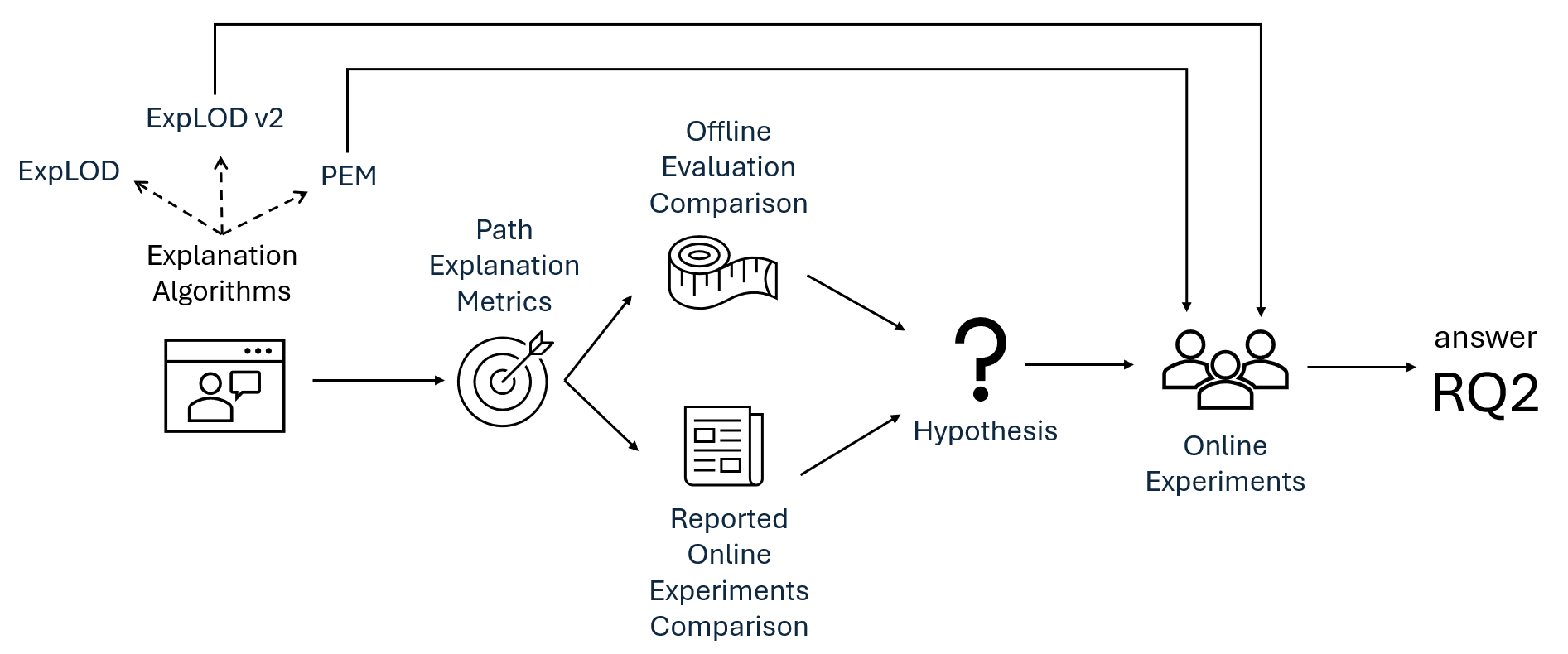}
    \caption{Methodology to validate the offline metrics}
    \label{fig:met}
\end{figure}

However, offline path metrics have not yet been thoroughly explored in the literature. Unlike \ac{BLEU} and \ac{ROUGE}, which treat explanations as a sequence of words, path metrics measure the different elements of an explanation. As described in Section \ref{sec:org_lit}, paths connect interacted and recommended thought-shared attributes, and have two main elements: (a) interacted items and (b) item attributes. For example, in Figure \ref{fig:graph}, if a user interacts with the movie ``Saving Private Ryan" and the \ac{RS} recommends ``Forest Gump,” there are three possible shared attributes: the genre ``drama,” actor ``Tom Hanks, " and costume designer ``Joana Johnston.” Additionally, path metrics could be applied to review explanations, as the popularity and diversity of attributes can be measured using a large corpus of text or ground truth.\looseness=-1 

Therefore, with \ac{RQ}2, we sought to answer whether the path elements that compose an explanation are related to the user’s perception of explanation goals. To evaluate the interacted items and attributes that connect interacted and recommended items, we used the metrics proposed by \cite{balloccu2022post} and measured three aspects of the two elements of path explanations: (1) the recency of interacted items; (2) the popularity of attributes; and (3) the diversity of attributes across different explanations. We also propose three new ones to evaluate aspects that are not covered by the metrics of \cite{balloccu2022post}.\looseness=-1

To address \ac{RQ}2, we applied the methodology shown in Figure \ref{fig:met}. We began by reproducing three explanation algorithms: ExpLOD \cite{musto2016explod}, ExpLOD v2 \cite{musto2019linked}, and \ac{PEM} \cite{du2022post}. All three algorithms are content-based, leverage \ac{KG}s, and are agnostic methods for explaining recommendations. They reflect the key advancements in the state-of-the-art for this explanation style and approach, identified exclusively through online experiments. In these prior user trials, one algorithm consistently outperformed the others in terms of explanation goal metrics. Specifically, ExpLOD v2 outperformed ExpLOD \cite{musto2019linked}, and under similar experimental conditions, \ac{PEM} outperformed ExpLOD v2 \cite{du2022post}.\looseness=-1  

We extended the results of these studies by conducting offline experiments in which explanations were generated for every user using the three state-of-the-art algorithms across two datasets. We then measured and compared the offline explanation path metrics. Finally, we compared these offline path metric results with the online experiment outcomes reported in \cite{musto2019linked} and \cite{du2022post} to formulate hypotheses regarding potential correlations between offline path metrics and explanation goals.\looseness=-1

To validate these hypotheses derived from our offline experiments and the online results reported in \cite{musto2016explod, musto2019linked, du2022post}, we conducted our own online experiments with the two most recent algorithms—ExpLOD v2 \cite{musto2019combining} and \ac{PEM} \cite{du2022post}—which also achieved the best results in our offline experiments, using a within-subjects user trial.\looseness=-1 

In Section \ref{sec:offline}, we present the offline experiments, including the explanation algorithms, knowledge graph extraction, and analysis. We then discuss each offline path metric and its potential influence on the user perception. The results of these experiments are reported in Section \ref{sec:online} along with the hypotheses that were subsequently tested in the online evaluation. In that section, we detail the online user study and analyze the relationship between offline path metrics and online explanation goals, providing evidence to address \ac{RQ}2.\looseness=-1 

\section{Offline Experiments}
\label{sec:offline} 

To understand the impact of explanation construction on user perception, we conducted offline experiments comparing three state-of-the-art model-agnostic \ac{KG} explanation algorithms. These algorithms were evaluated across six \ac{RS} algorithms for all users in two datasets. The generated explanations connect each recommendation to interacted items based on shared attributes in the \ac{KG}. Consequently, offline explanation metrics assess how each algorithm selects these interacted items and attributes (whereas recommendations are evaluated using ranking metrics such as \ac{NDCG}). This section describes the offline experimental procedure and compares the offline path metrics results across the three explanation algorithms.\looseness=-1 

\subsection{Offline Evaluation and Datasets}
\label{sec:off_eval_ds}

Figure \ref{fig:off_eval} illustrates the process of the offline evaluation, in which we assessed three model-agnostic algorithms that use a content-based \ac{KG} explanation style through offline path explanation metrics.\looseness=-1

\begin{figure}[!b]
    \centering
    \includegraphics[width=.85\textwidth]{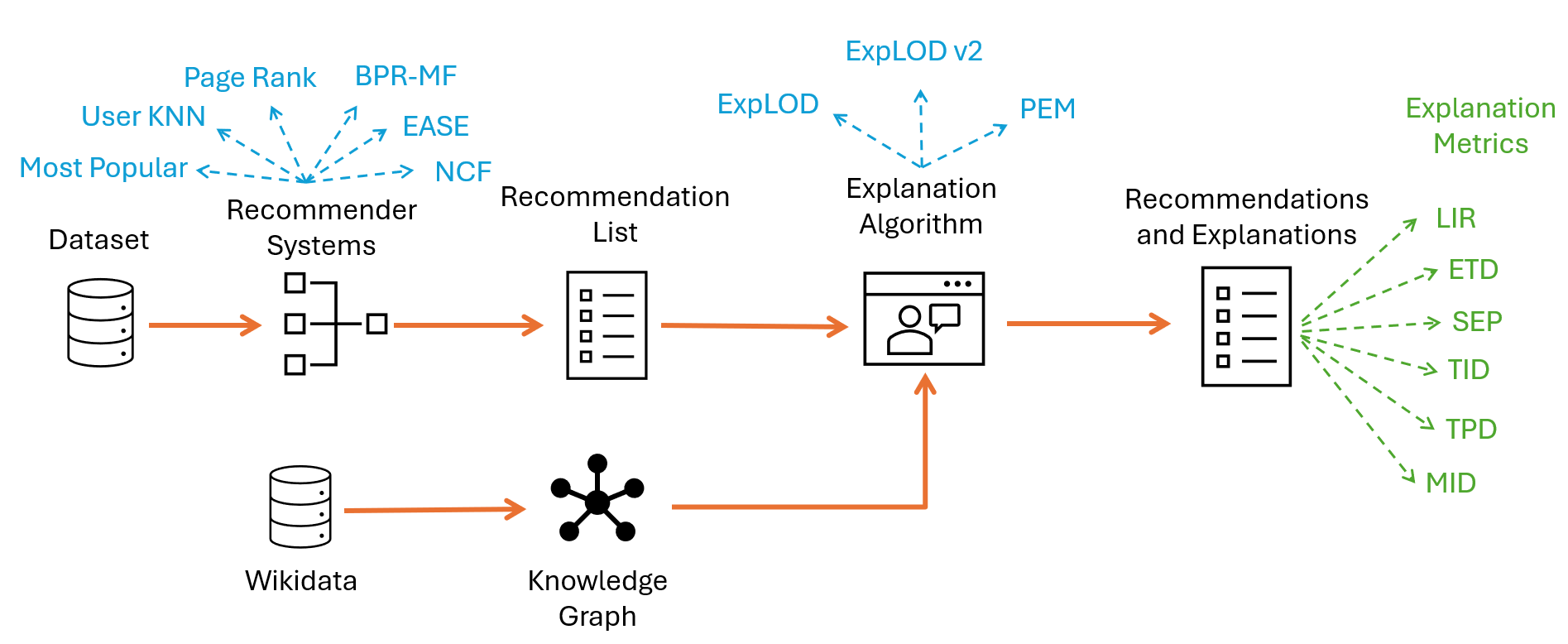}
    \caption{Offline evaluation experiment data flow}
    \label{fig:off_eval}
\end{figure}

Initially, we extracted a \ac{KG} for the movies and artists in the \ac{MovieLens} \cite{harper2015movielens} and \ac{LastFM} \cite{Cantador2011} datasets. We excluded interactions of items from the dataset that had no data in the \ac{KG} and binarized all interactions. We did not add a threshold for binarization because we considered that even if the user did not like the item, it still captured the user's attention.\looseness=-1

The dataset processed after excluding interactions of items without content on the \ac{KG} retained 99\% of the original interactions for the \ac{MovieLens} dataset and 89\% for the \ac{LastFM} dataset. Table \ref{tab:dataset} displays the differences between the original and processed datasets and Section \ref{sec:kg_extraction_analysis} details the KG data acquisition from Wikidata.\looseness=-1

Then, we executed six \ac{RS} using the reproducibility guidelines of \cite{ferrari2021troubling}. This guideline suggests that every recommendation algorithm should be evaluated by comparing it with different families of recommendation algorithms, following a rigorous evaluation process for statistically significant results. \looseness=-1  

This guideline was adopted because we used model-agnostic \ac{KG} algorithms to generate explanations. Consequently, an algorithm may perform well on offline explanation metrics for neural models but not for neighborhood-based recommenders. As a result, if one method is effective for a \ac{RS}, it does not necessarily mean that it will perform equally well across all others. This is because each of these methods relies on distinct mechanisms and structures to generate recommendations, which means that an explanation algorithm that is suited for one \ac{RS} may not align with the underlying principles or data representations of another. All implementations of our offline experiments and the processed datasets are available in a public source code repository\footnote{\url{https://github.com/andlzanon/lod-personalized-recommender}}.\looseness=-1 

\begin{table}[!t]
\caption{Statistics of Original and Processed Datasets.\looseness=-1}
\centering
\begin{tabular}{c|cc||cc|}
\cline{2-5}
                              & \multicolumn{2}{c||}{\ac{MovieLens}}                       & \multicolumn{2}{c|}{\ac{LastFM}}                               \\ \cline{2-5} 
\multicolumn{1}{l|}{}         & \multicolumn{1}{c|}{Original Dataset} & Processed Dataset & \multicolumn{1}{c|}{Original Dataset} & Processed Dataset \\ \hline
\multicolumn{1}{|c|}{users}   & \multicolumn{1}{c|}{610}              & 610               & \multicolumn{1}{c|}{1,892}            & 1,875             \\ \hline
\multicolumn{1}{|c|}{items}   & \multicolumn{1}{c|}{9,724}            & 9,517             & \multicolumn{1}{c|}{17,632}           & 11,641            \\ \hline
\multicolumn{1}{|c|}{ratings} & \multicolumn{1}{c|}{100,836}          & 100,521           & \multicolumn{1}{c|}{92,834}           & 83,017            \\ \hline
\end{tabular}
\label{tab:dataset}
\end{table}

\subsection{Recommender Systems Algorithms}
\label{sec:rec_algs}

The different families and recommendation algorithms applied are as follows.
\begin{itemize}
    \item \textbf{Most Popular} \citep{Cremonesi2010} offers non-personalized recommendations by suggesting the most popular items that a user has not yet interacted with.

    \item \textbf{Personalized PageRank} algorithm \citep{musto2016explod}, for graph-based recommendations. To generate recommendations, it leverages the Wikidata graph and uses random walks, allocating 80\% weight to items previously interacted with by the user and 20\% to all other nodes.\looseness=-1 

    \item \textbf{User-KNN} \citep{Resnick1994} provides neighborhood-based recommendations through cosine similarity, suggesting items interacted with by similar users. The parameter $K$ is set based on the square root of the total number of users.\looseness=-1 

    \item \textbf{Embarrassingly Shallow AutoEncoder (EASE)} \citep{Steck2019} is a non-neural algorithm and employs a linear auto-encoder approach. Parameter lambda was set to 500 according to the paper's original results.\looseness=-1  
    
    \item \textbf{Bayesian Personalized Ranking Matrix Factorization (BPR-MF)} \citep{Rendle2009} (also a representative of non-neural algorithms) is optimized for implicit feedback using matrix factorization using a pairwise ranking approach to rank items a user has interacted with over those they have not. The embedding size for the BPR-MF was set to 32.\looseness=-1  

    \item The \textbf{Neural Collaborative Filtering (NCF)} \citep{He2017} represent neural algorithms and integrates an Artificial Neural Network with Matrix Factorization, using specific configurations for embeddings, layers, epochs, and batch sizes. The testing phase followed a leave-one-out evaluation consistent with the original methodology. The parameters of the algorithm used were as follows: embedding sizes of users and items with size 32, four layers of 64, 32, 16, and 8 neurons, 10 epochs, and a batch size of 256. Negative sampling was also employed, where for each positive sample on the training set, four negative samples were added based on unseen items.\looseness=-1 
\end{itemize}

Most Popular, User-KNN, and BPR-MF were implemented using the library proposed by \cite{da2018case}, whereas the authors implemented EASE, NeuMF, and PageRank according to the corresponding papers, and they are available in our public repository. \ac{MovieLens} \cite{harper2015movielens} was executed using 10-fold cross-validation for the top-1 and top-5 recommendations of every algorithm on every fold. We also evaluated the \ac{LastFM} dataset \cite{Cantador2011} on the first fold to analyze if explanation algorithms' results varied on different domains.\looseness=-1 

\subsection{Knowledge Graph Extraction and Analysis}
\label{sec:kg_extraction_analysis}
 
Data used to generate explanations was extracted from the movie and artist domains on Wikidata\footnote{\url{https://www.wikidata.org}}, as it is more up-to-date and complete compared to DBPedia\footnote{\url{https://www.dbpedia.org/}} \cite{pillai2019comparing}. Descriptive information (e.g., box office data) and identification links (e.g., IMDb IDs) were removed during data retrieval from the \ac{LOD} because they are unique to specific items. The knowledge graph generated for \ac{MovieLens} includes 78,703 entities, 295,787 triples, and 23 edge types, while the graph for \ac{LastFM} comprises 34,297 entities, 134,197 triples, and 33 edge types.\looseness=-1

To extract information from the \ac{LOD} for the movie domain, we used the imdbId provided by the \ac{MovieLens} Latest dataset, which is also available on Wikidata as an identifier. Using this information, we constructed a SPARQL query on the Wikidata endpoint\footnote{\url{https://query.wikidata.org/sparql}} to create a movie domain-specific \ac{KG}. The edge types, representing attributes of items extracted from Wikidata, included: director, screenwriter, composer, genre, cast member, producer, award received, director of photography, country of origin, filming location, main subject, film editor, nominated for, title, creator, narrative location, costume designer, performer, production company, part of a series, voice actor, executive producer, and production designer.\looseness=-1 

In contrast, when constructing the artist domain-specific \ac{KG} for the \ac{LastFM} dataset, there was no direct connection between the dataset metadata and Wikidata. Therefore, we first constructed a SPARQL query to extract the \ac{LOD} URI from the artist based on the artist's name. In a second step, using another SPARQL query, we extracted all data associated with the artist from Wikidata using the URI obtained in the previous step.\looseness=-1

\begin{figure}[!t]
    \centering
    \includegraphics[width=.8\textwidth]{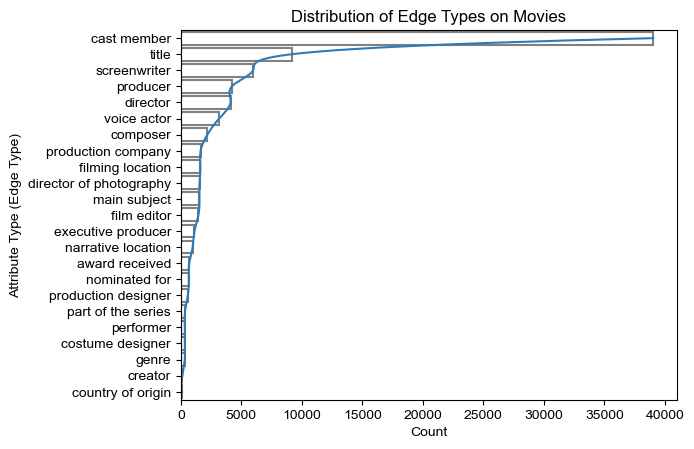}
    \caption{Distribution of Edge Types References from Item and Attribute Nodes on the Movie Wikidata KG}
    \label{fig:edge_type_movies}
\end{figure}

The edge types (or attribute types) of items extracted from the \ac{LOD} for the artist domain included: work period (start), has part, country of origin, record label, genre, inception, location of formation, country, languages spoken, written or signed, instrument, occupation, date of birth, voice type, member of, place of birth, sex or gender, educated at, country of citizenship, notable work, award received, field of work, residence, work location, religion, native language, participant in, influenced by, director\slash manager, nominated for, represented by, wears, sport, and participant.\looseness=-1 

\begin{figure}[!t]
    \centering
    \includegraphics[width=.7\textwidth]{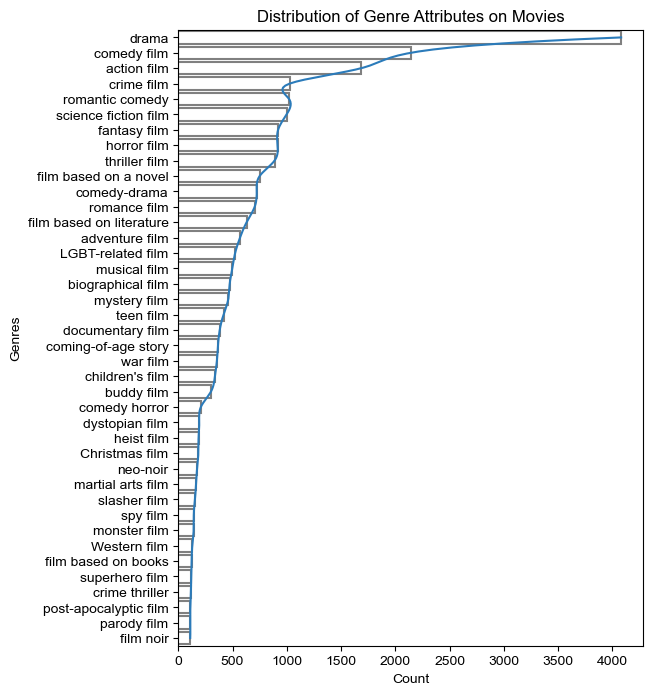}
    \caption{Distribution of Genre Attributes References from Item Nodes on the Movie Wikidata KG}
    \label{fig:genre_movies}
\end{figure}

One of the findings from Section \ref{sec:lite_review_analysis} is that evaluating explanation algorithms solely with anecdotal evidence or limited user trials is insufficient for a robust evaluation \cite{10.1145/3583558}. To explore this further from a data perspective, we analyzed the distribution of edge types and attributes within the KGs extracted for the movie and artist domains.\looseness=-1

Figure \ref{fig:edge_type_movies} represents the distribution of edge types (or relations) from interacted and recommended items to the attributes. The distribution is characterized by a long tail, where many items have some common edge types, whereas others are less referenced. Notably, most edges are from ``cast member", connecting movie item nodes with their respective actors and actresses, followed by ``title", ``screenwriter", and ``producer'. This pattern is understandable, as most items may have producers, directors, and cast members, but not all movies have features like ``awards received."\looseness=-1 

We also analyzed the distribution of attribute nodes connected to relations. For example, the ``filming location" edge type connects item nodes to places where movies were shot. As a result, locations like ``United States of America" appear more frequently than ``Brazil," reflecting a disparity in the number of movies filmed in each location.\looseness=-1 

In this context, Figure \ref{fig:genre_movies} shows the truncated distribution of the 40 most frequent genre attribute nodes connected to the 'genre' relation in the extracted movie \ac{KG} (less frequent attribute nodes are omitted due to space constraints). Similar to the ``filming location" edge type, the results also display a long-tail distribution, where the number of item nodes related to the ``drama" attribute with the ``genre" edge type is almost twice as high as the second most common attribute node, ``comedy''.\looseness=-1

Consequently, when explaining recommendations with attributes and edge types, the same long-tail distribution pattern observed in \ac{RS} interactions is present in the metadata. This bias can impact both the explanation algorithm and user perception, highlighting the importance of moving beyond anecdotal examples and robustly evaluating explanations.\looseness=-1

In Appendix \ref{apen:type_movies}, Figure \ref{fig:edge_type_artists} shows the distribution of relations in the artist domain. Additionally, in Appendix \ref{apen:genre_distribution}, Figure \ref{fig:genre_artists} presents the truncated distribution of the 90 most frequent genre attribute nodes connected to the 'genre' relation in the extracted artist \ac{KG}, displaying the same behavior as the edge type and attribute distributions in the movie \ac{KG} dataset discussed in this section.\looseness=-1

The code for the \ac{KG} extraction from Wikitada and the resulting \ac{KG}s for the \ac{MovieLens}\footnote{\url{https://github.com/andlzanon/lod-personalized-recommender/blob/main/generated\_files/wikidata/props\_wikidata\_movielens\_small.csv}} and \ac{LastFM}\footnote{\url{https://github.com/andlzanon/lod-personalized-recommender/blob/main/generated\_files/wikidata/last-fm/props\_artists\_id.csv}} datasets, along with the SPARQL queries\footnote{\url{https://github.com/andlzanon/lod-personalized-recommender/blob/main/preprocessing/wikidata\_utils.py}} used to obtain the \ac{LOD} data for explanations are available on the source code repository of the project. The analysis presented in this section is also available \footnote{\url{https://github.com/andlzanon/offline_metrics_explanation_rec_goals/blob/main/KnowledgeGraph/kg_exploration.ipynb}}\looseness=-1 

\subsection{Explanation Algorithms}
\label{sec:expl_algs}

With the \ac{KG} extracted, we applied the six recommendation algorithms described in Section \ref{sec:rec_algs}. For each, we generated explanations using three content-based, model-agnostic explanation algorithms for the top-1 and top-5 items, and then computed the offline path explanation metrics for every set of explanations.\looseness=-1

The three explanation algorithms implemented were: ExpLOD\cite{musto2016explod}, ExpLOD v2\cite{musto2019combining}, and \ac{PEM}\cite{du2022post}. All of them are content explanation styles algorithms using \ac{KG} and agnostic methods and generate sentences based on the ranking of common attributes on a \ac{KG} between historical and recommended items.\looseness=-1

Online experimental results from the literature show that state-of-the-art \ac{KG}-agnostic explanation algorithms have progressed in a well-defined manner in terms of their explanation goals. ExpLOD v2 extends ExpLOD \cite{musto2019linked} and validates its improvements through online experiments. Similarly, \ac{PEM}, the most recent algorithm, builds on ExpLOD v2 and has been validated through online experiments \cite{du2022post}. Results indicate that \ac{PEM} offers enhanced transparency, persuasiveness, engagement, trust, and effectiveness and, therefore, the current state-of-the-art.\looseness=-1

Consequently, the goal of the offline experiments was to evaluate these algorithms using offline path explanation metrics and to compare the results with those reported in \cite{musto2016explod}, \cite{musto2019linked}, and \cite{du2022post}. This comparison allowed us to formulate hypotheses regarding potential correlations between offline path explanation metrics and online explanation goal metrics. Based on these hypotheses, we then select one \ac{RS} and two explanation algorithms to test whether the hypotheses hold in an online user trial, as described in Section \ref{sec:online}.\looseness=-1

Explanations of these three \ac{KG}-agnostic algorithms were evaluated offline on user-level explanation metrics for all users in the \ac{LastFM} \cite{Cantador2011} and \ac{MovieLens} Latest \cite{harper2015movielens} datasets. Although both datasets are small in terms of the number of ratings, as discussed in our rapid literature review in Section \ref{sec:lite_review_analysis}, offline evaluation of content-based explanations is mostly performed using anecdotal examples. In our work, instead, we evaluate on a large quantity of users: 610 on \ac{MovieLens} and 1,875 on \ac{LastFM}.\looseness=-1

As described in Section \ref{sec:org_lit}, a \ac{KG} is defined as $KG = \{ (h,r,t) \mid h,t \in E, , r \in R \}$, where $h$ and $t$ represent entity nodes (items and attributes), and $r$ denotes a relation link (also referred to as an edge type) between these two entities \cite{10.1145/3269206.3271739}. For instance, in Figure \ref{fig:graph}, relations such as genre, actor, and costume designer are instances of $r$ in the \ac{KG}, as they connect an item node (e.g., Saving Private Ryan) to an attribute node (e.g., drama film). Similarly, ``Forrest Gump'' is another item node connected to the same attribute node (drama film) as ``Saving Private Ryan''. Based on this data structure, ExpLOD \cite{musto2016explod}, ExpLOD v2 \cite{musto2019linked}, and \ac{PEM} \cite{du2022post} rank attributes according to the following scoring functions: \looseness=-1 

\begin{itemize}
    \item \textbf{ExpLOD \cite{musto2016explod}:} The ExpLOD method ranks attributes of the \ac{KG} using Equation \ref{eq:explod_rank} ($score\_explod$), where $n_{p,I_u}$ and $n_{p,I_r}$ represent the number of links of attribute $p$ to the sets of interacted items ($I_{u}$) and recommended items ($I_{r}$), respectively. These are weighted by $\alpha$ and $\beta$ and multiplied by the inverse document frequency of $p$ ($IDF(p)$). Essentially, this approach adapts the \ac{TF-IDF} weighting scheme to graphs, where the first term captures the frequency of an attribute with respect to interacted and recommended items, and the IDF component accounts for the distribution of attributes across the entire set of items. Specifically, in the IDF equation, $IDF(p) = \log(\frac{N}{df_p})$, $N$ is the set of items (rather than documents) and $df_p$ is the number of items linked to attribute $p$. We set parameters $\alpha$ and $\beta$ to 0.5, as proposed originally in \cite{musto2016explod}. \looseness=-1 

    \begin{equation}
    \label{eq:explod_rank}
        score\_explod(p, I_{u}, I_{r}) = (\alpha \frac{{n_{p, I_{u}}}}{|I_{u}|}) + (\beta \frac{n_{p, I_{r}}}{{|I_{r}|}}) \times IDF(p)
    \end{equation}

    \item \textbf{ExpLOD v2 \cite{musto2019linked}:} The key difference in ExpLOD v2 is the inclusion of broader attributes from the \ac{KG} hierarchy. For example, the attribute ``Sci-Fi Comedy" is linked to two broader attributes: ``Science Fiction" and ``Comedy." This indicates that ``Sci-Fi Comedy" is an instance on the graph of both the ``Science Fiction" and ``Comedy" attribute genres. Consequently, more attributes can be considered potential explanation paths between interacted and recommended items. To achieve this, ExpLOD v2 extends Equation \ref{eq:explod_rank} by summing across ``instance of" (also called child) attributes of broader attributes, as shown in Equation \ref{eq:explod_v2_rank} where $b$ is a broader attribute, $P_{c}(b)$ is the set of attributes connected to $b$, and $p_{i}$ is the $i^{th}$ child attribute of $b$. Therefore, attributes that do not have child attributes are scored based on Equation \ref{eq:explod_rank}, while the broader attributes are scored based on Equation \ref{eq:explod_v2_rank}.\looseness=-1 

    \begin{equation}
    \label{eq:explod_v2_rank}
        score\_explod(b, I_{u}, I_{r}) = \sum_{i=1}^{|P_{c}(b)|} score\_explod(p_{i}, I_{u}, I_{r}) \times IDF(b)
    \end{equation}

    \item \textbf{\ac{PEM} \cite{du2022post}:} The recently proposed \ac{PEM} represents a syntactic baseline method that balances attribute popularity within the bipartite graph of interacted and recommended items. Unlike ExpLOD and ExpLOD v2, \ac{PEM} uses the number of interacted nodes connected to an attribute rather than the number of links connected to the attribute. It replaces the IDF penalization of the ExpLOD algorithms with a logarithmic function. Similar to ExpLOD v2, \ac{PEM} also considers broader properties for generating explanations. Equation \ref{eq:pem_rank} shows the \ac{PEM} calculation, where ($I_{u}$) and ($I_{r}$) represent the sets of interacted and recommended items, respectively. The term ($|\overline{I(p, I_{u})}|$) denotes the number of items to which a property is directly or indirectly connected ($I_{u}$), and ($|\overline{I(p, C)}|$) represents the number of items connected within the set of all items ($C$). The penalization term ($\log(|I(p, C)|)$) is applied to penalize a property if it is not frequently used in the item catalog.\looseness=-1 

    \begin{equation}
    \label{eq:pem_rank}
        score\_pem(p, I_{u}, I_{r}, C) = \frac{|\overline{I(p, I_{u})}|/|I_{u}|}{|\overline{I(p, C)}|/|C|} * \log(|I(p, C)|)
    \end{equation}
    
\end{itemize}

For all the algorithms, there are two main parameters for constructing explanation sentences that impact the metrics. The number of attributes shown in explanations and the interacted items connected to the attribute. For both, we set to three. This number was inspired by the work of \cite{kouki2019personalized}, who discovered that three to four explanations were sufficient for most people to not be overwelmed with information. Hence, three paths can be shown to users as explanations. Multiple attributes are only shown in the explanation when there is a tie in the scores between attributes. More than one interacted item is shown if there are multiple items connected to the top-ranked attribute by the scoring function of the explanation algorithms.\looseness=-1

\subsection{Offline Metrics}
\label{sec:metrics}

As previously explained, \ac{RS} generates recommendations for users based on their interacted items. One way to generate explanations is to find common attributes between users' interactions and recommended items. It typically consists of two components, namely, (a) a user's interacted item; and (b) the attribute that links the recommendation generated by the \ac{RS} to the user's interacted item.\looseness=-1

Considering the example from Figure \ref{fig:graph}, if the user interacted with the item ``Saving Private Ryan'' and the \ac{RS} suggested ``Forest Gump'', the explanation can be generated with or without the interacted items, therefore, in a sentence considering all elements ``Because you like drama films such as Saving Private Ryan, watch Forest Gump'', the connection to the interacted item can be ignored and the explanation become ``Because you like drama films, watch Forest Gump''. The latter is more related to review explanations because it does not consider intersections between different item metadata and instead uses user-provided information to generate explanations. With \ac{KG}s, instead, since paths on a graph can connect and find metadata shared across interacted and recommended items, the display of interacted items in explanations is more frequent.\looseness=-1 

In the rapid literature review in Section \ref{sec:lib}, we identified three offline path metrics proposed by \cite{balloccu2022post} that were also adopted in subsequent studies \cite{zanon2024model,balloccu_boratto_fenu_marras_2023}. To the best of our knowledge, these are the only metrics specifically designed for path-based explanations. They build on the idea that an explanation should (i) connect \emph{recently interacted items} with the recommended item through (ii) \emph{popular attributes}, and (iii) when multiple explanations are provided, highlight \emph{diverse attributes across them}. Accordingly, these metrics capture the recency of interacted items, the popularity of attributes along explanation paths, and the diversity of attributes across multiple explanation paths.\looseness=-1 

The main limitation of these metrics is that they are user-level, and as a result, when an explanation algorithm identifies a single local explanation that optimizes the three metrics and reproduces it consistently across all users in a dataset, they achieve high values. Nevertheless, this uniformity reduces the diversity of attributes and interacted items shown across all users. In this regard, we also propose three other metrics to address these limitations. The six offline path explanation metrics are shown in Equations 1 to 6.\looseness=-1

When recommending items, prior work has shown that user trust increases when suggestions are familiar \cite{kouki2019personalized}. Extending this assumption to explanations, trust can also be associated with popular attributes, while recently interacted items are more likely to be known and thus enhance trust in \ac{RS}s. In \cite{balloccu2022post}, Shared Entity Popularity ($SEP$) and Linking Interaction Recency ($LIR$) were proposed to measure the popularity of attributes and the recency of interacted items in an explanation, respectively. They are defined in Equations \ref{eq:sep} and \ref{eq:lir}.\looseness=-1

Both metrics rely on the mean of a min–max normalized exponentially weighted moving average: for $LIR$, applied to the timestamp ($t$) of the interacted item ($p$) included in the explanation, and for $SEP$, applied to the number of times ($v$) an attribute ($e$) is referenced in the graph. The term $i$ denotes the index of the item/attribute in the ordered array based on recency or popularity, and $\beta$ is a smoothing parameter set to $0.3$ as suggested in \cite{balloccu2022post}.\looseness=-1 

The values ranged from 0 to 1. Lower values indicate explanations based on less popular attributes and less recently interacted items, while higher values reflect explanations that connect recommended items to more popular attributes and more recent interactions.\looseness=-1

\begin{equation}
\label{eq:sep}
    SEP(e^{i}, v^{i}) = (1 - \beta) \times SEP(e^{i-1}, v^{i-1}) + \beta \times v^{i}
\end{equation}

\begin{equation}
\label{eq:lir}
    LIR(p^{i}, t^{i}) = (1 - \beta) \times LIR(p^{i-1}, t^{i-1}) + \beta \times t^{i} 
\end{equation}

The Explanation Type Diversity ($ETD$) metric in Equation \ref{eq:etd}, proposed in \cite{balloccu2022post}, accounts for the diversity of attributes across different explanations to prevent bias toward explaining all users' recommendations with the same attribute. It is calculated as the number of unique attributes in explanations ($\omega_{L_{u}}$) divided by the minimum between the size of the recommendation list $k$ and possible explanation attributes $\omega_{L}$. The main motivation for measuring diversity is that the engagement goal of explanation accounts for the discovery of new information \cite{tintarev2015explaining} and diversifying attributes shown across users can increase the chance of displaying attributes that the user finds relevant, but unfamiliar. Similar to the metrics $SEP$ and $ETD,$ values range from 0 to 1, with 1 being the optimal value where all attributes across different explanations are different.\looseness=-1  

\begin{equation}
\label{eq:etd}
    ETD(S) = \frac{|\omega_{L_{u}}|}{min(k, |\omega_{L}|)}
\end{equation}

In conclusion, the three metrics proposed in \cite{balloccu2022post} define that good explanations connect recently interacted items with the recommended item through popular shared attributes, which are not repetitive across different explanations.\looseness=-1 

A limitation of these metrics is that they do not consider the number of interacted items. When an explanation algorithm shows a low number of items that may be connected to many attributes, explanations can be repetitive towards a small set of interacted items and, consequently, less convincing to the users. To this end, we also propose Mean Item Diversity ($MID$), in Equation \ref{eq:mid}, as an equivalent of $ETD$ but for items shown in explanations. Term $E_{u}$ is the set of all explanations for all users, $e$ is an explanation in $E_{u}$ and $L_{i_{e}}$ is the set of profile items used for the explanations. Therefore, it is the mean quantity of interacted items shown in the set of explanations shown for each user. Higher values indicate that a larger number of historical items are shown across explanations.\looseness=-1

\begin{equation}
\label{eq:mid}
    MID(S) = mean(\forall_{e \in E_{u}} L_{i_{e}}) 
\end{equation}

All the metrics presented thus far evaluate attributes and interacted items at the individual user level without accounting for patterns across a set of users. This introduces a limitation: if an explanation algorithm identifies a single local explanation that maximizes the three metrics and replicates it for all users in the dataset, $ETD$, $LIR$, and $SEP$ will still yield high values, even though all users receive similar explanations. In other words, the metrics proposed by \cite{balloccu2022post} are intra-list and therefore capture only the diversity, popularity, and recency of explanations for a single user, without considering variations across the user base. To address this gap, we adapted catalog coverage, which complements these metrics by providing insights into the diversity of explanations at the global level.\looseness=-1 

First proposed by \cite{adomavicius2011improving}, the aggregate diversity of an \ac{RS} algorithm $S$ measures the number of distinct items exposed to all the users. When adapted to explanations, this metric differs from $ETD$, which captures the number of attributes within a single user's explanations. Instead, catalog-level metrics evaluate the number of distinct attributes and interacted items appearing across all generated explanations, providing insights into the extent to which explanation algorithms may bias toward a limited subset of attributes and items across the entire user base.\looseness=-1 

Similar to item ranking, if an explanation algorithm repeatedly selects the same set of attributes or interacted items to generate explanations, catalog coverage will be low. This indicates that the algorithm is overspecialized in a small subset of popular attributes and items across all users. Ideally, however, explanations should leverage user-specific and personalized attributes, thereby increasing the overall catalog size of attributes presented across different users. \looseness=-1 

Hence, adapting the catalog coverage metric, we propose Total Items Aggregate Diversity ($TID$) and Total Property Aggregate Diversity ($TPD$), defined by Equations \ref{eq:tid} and \ref{eq:tpd}, respectively, to measure the total amount of interacted items and total amount of attributes shown across all explanations for all users on a dataset.\looseness=-1  

\begin{equation}
\label{eq:tid}
    TID(S) = \left | \bigcup_{e \in E_{u}} L_{i_{e}} \right | 
\end{equation}

\begin{equation}
\label{eq:tpd}
    TPD(S) = \left | \bigcup_{e \in E_{u}} L_{p_{e}} \right |
\end{equation}

Again, term $E_{u}$ is the set of all explanations for all users, $e$ is an explanation in $E$, $L_{i_{e}}$ is the set of profile items used for the explanations, and $L_{p_{e}}$ is the set of attributes used for the explanations. Similar to the aggregate diversity of items, the idea behind $TPD$ and $TID$ is to verify the total number of attributes\slash items shown in explanations. Unlike the other metrics, $TID$ and $TPD$ do not have a fixed range. However, higher values indicate that a greater variety of attributes and interacted items are being presented to users, which in turn enhances personalization.\looseness=-1    

Table \ref{tab:metrics} summarizes the metrics and their corresponding objectives for analyzing the different aspects of the explanations. Metrics $ETD$, $TPD$ and $SEP$ evaluate attributes shown in explanations, which, in the case of Figure \ref{fig:graph} are ``drama'', ``Tom Hanks'' and ``Joana Johnston''. Metrics $LIR$, $MID$ and $TID$ evaluate the interacted item that composes and explanation (``Saving Private Ryan'' in Figure \ref{fig:graph}). The adopted explanation metrics do not cover recommended items since they are generated by the \ac{RS} and are best evaluated by ranking metrics such as \ac{NDCG}. \looseness=-1

\begin{table*}[h!]
\caption{Table of explanation path metrics.\looseness=-1}
\begin{adjustbox}{}
    \begin{tabular}{|c|c|}
        \hline
        \textbf{Objective} & \textbf{Equation} \\ \hline \hline
        Popularity of attributes   & \multicolumn{1}{c|}{$SEP(e^{i}, v^{i}) = (1 - \beta) \times SEP(e^{i-1}, v^{i-1}) + \beta \times v^{i}$}  \\ [3pt] \hline  
       Recency of Items   & \multicolumn{1}{c|}{$LIR(p^{i}, t^{i}) = (1 - \beta) \times LIR(p^{i-1}, t^{i-1}) + \beta \times t^{i}$}  \\ [3pt] \hline
        Diversity of attributes  & \multicolumn{1}{c|}{$ETD(S) = \frac{|\omega_{L_{u}}|}{min(k, |\omega_{L}|)}$}     \\ [3pt] \hline
         Diversity of interacted items      & $MID(S) = mean(\forall_{e \in E_{u}} L_{i_{e}})$  \\ [3pt] \hline
        Total number of items shown to all users     & $TID(S) = \left | \bigcup_{e \in E_{u}} L_{i_{e}} \right |$               \\ [3pt] \hline
         Total number of attributes shown to all users & $TPD(S) = \left | \bigcup_{e \in E_{u}} L_{p_{e}} \right |$               \\ [3pt] \hline
        \cline{1-2}
    \end{tabular}
\end{adjustbox}
\label{tab:metrics}
\end{table*}

The offline explanation path metrics presented in this section will be compared with the online user trials reported in \cite{musto2019linked} and \cite{du2022post} to generate hypotheses about how the presentation of interacted items and attributes by an explanation algorithm influences explanation goals. We then conducted our own user trial to validate these hypotheses and further address \textbf{RQ2}.\looseness=-1

\subsection{Offline Results}
\label{sec:offline_results}

To align the analysis of offline metrics with the online experiments in \cite{musto2019combining}, which compared ExpLOD to ExpLOD v2, and \cite{du2022post}, which compared ExpLOD v2 to \ac{PEM}, Table \ref{tab:offresults_n1} presents the values of offline explanation path metrics for the top-1 recommendation on the \ac{MovieLens} dataset. In both online studies by \cite{musto2019combining,du2022post}, the participants compared explanation algorithms based on a single recommendation. Each column in the table represents one of the metrics described in Section \ref{sec:metrics}, while each row corresponds to a recommendation algorithm with the three explanation methods.\looseness=-1  

Appendix \ref{apen:ranking} presents the results of the accuracy and beyond-accuracy ranking metrics for each recommendation algorithm. Additionally, Appendix \ref{apen:lastfm} includes Table \ref{tab:offresultslastfm_n1} and Table \ref{tab:offresultslastfm_n5}, which display the offline path metric results for the \ac{LastFM} dataset. These were omitted from the main text because the patterns and conclusions in \ac{LastFM} were consistent with those in \ac{MovieLens}, demonstrating the methods' robustness across different domains.\looseness=-1 

\begin{table*}[!b]
\caption{Offline results for the metrics for the top-1 recommendation for the \ac{MovieLens} dataset. Bold results with the symbol \greentriangleup are the best values considering the three explanation algorithms for the recommendation algorithm. Two values in column for a \ac{RS} with the symbol \yellowcircle represent Wilcoxon's p-value above 0.05 between them, meaning non-significant statistical differences.\looseness=-1}
    \label{tab:offresults_n1}
\begin{tabular}{cc|ccc||ccc|}
\cline{3-8}
\multicolumn{1}{l}{\textbf{}}                   & \multicolumn{1}{l|}{\textbf{}} & \multicolumn{3}{c||}{Item Metrics}                                                                        & \multicolumn{3}{c|}{Attribute Metrics}                                                 \\ \cline{3-8} 
\textbf{}                                       & \textbf{}                      & \multicolumn{1}{c|}{MID}             & \multicolumn{1}{c|}{TID}                  & LIR                   & \multicolumn{1}{c|}{ETD} & \multicolumn{1}{c|}{TPD}            & SEP                   \\ \hline
\multicolumn{1}{|c|}{\multirow{3}{*}{MostPOP}}  & ExpLOD                         & \multicolumn{1}{c|}{\textbf{2,9956} \greentriangleup} & \multicolumn{1}{c|}{678,2}                & 0,0827                & \multicolumn{1}{c|}{1}   & \multicolumn{1}{c|}{{ 35,2} \yellowcircle}     & { \textbf{0,6611} \greentriangleup \yellowcircle} \\ \cline{2-8} 
\multicolumn{1}{|c|}{}                          & ExpLOD v2                      & \multicolumn{1}{c|}{2,9787}          & \multicolumn{1}{c|}{\textbf{766,4} \greentriangleup}       & \textbf{0,0893} \greentriangleup       & \multicolumn{1}{c|}{1}   & \multicolumn{1}{c|}{{ 31,8} \yellowcircle}     & { 0,6488 \yellowcircle}          \\ \cline{2-8} 
\multicolumn{1}{|c|}{}                          & PEM                            & \multicolumn{1}{c|}{1,9472}          & \multicolumn{1}{c|}{436,8}                & 0,0299                & \multicolumn{1}{c|}{1}   & \multicolumn{1}{c|}{\textbf{120,8} \greentriangleup} & 0,1418                \\ \hline \hline
\multicolumn{1}{|c|}{\multirow{3}{*}{UserKNN}}  & ExpLOD                         & \multicolumn{1}{c|}{\textbf{2,9733} \greentriangleup} & \multicolumn{1}{c|}{{ 810,1 \yellowcircle}}          & \textbf{0,0984} \greentriangleup       & \multicolumn{1}{c|}{1}   & \multicolumn{1}{c|}{56,8}           & { \textbf{0,5212} \greentriangleup \yellowcircle} \\ \cline{2-8} 
\multicolumn{1}{|c|}{}                          & ExpLOD v2                      & \multicolumn{1}{c|}{2,9046}          & \multicolumn{1}{c|}{{ \textbf{812,4} \greentriangleup \yellowcircle}} & 0,0889                & \multicolumn{1}{c|}{1}   & \multicolumn{1}{c|}{67,0}           & { 0,5011 \yellowcircle}          \\ \cline{2-8} 
\multicolumn{1}{|c|}{}                          & PEM                            & \multicolumn{1}{c|}{2,0370}          & \multicolumn{1}{c|}{531,1}                & 0,0271                & \multicolumn{1}{c|}{1}   & \multicolumn{1}{c|}{\textbf{272,1} \greentriangleup} & 0,1171                \\ \hline \hline
\multicolumn{1}{|c|}{\multirow{3}{*}{PageRank}} & ExpLOD                         & \multicolumn{1}{c|}{\textbf{2,9756} \greentriangleup} & \multicolumn{1}{c|}{714,8}                & { 0,0815 \yellowcircle}          & \multicolumn{1}{c|}{1}   & \multicolumn{1}{c|}{{ 54,9 \yellowcircle}}     & { 0,6003 \yellowcircle}          \\ \cline{2-8} 
\multicolumn{1}{|c|}{}                          & ExpLOD v2                      & \multicolumn{1}{c|}{2,9525}          & \multicolumn{1}{c|}{\textbf{801,9} \greentriangleup}       & { \textbf{0,0855} \greentriangleup \yellowcircle} & \multicolumn{1}{c|}{1}   & \multicolumn{1}{c|}{{ 54,0 \yellowcircle}}     & { \textbf{0,6006} \greentriangleup \yellowcircle} \\ \cline{2-8} 
\multicolumn{1}{|c|}{}                          & PEM                            & \multicolumn{1}{c|}{2,0388}          & \multicolumn{1}{c|}{443,3}                & 0,0253                & \multicolumn{1}{c|}{1}   & \multicolumn{1}{c|}{\textbf{152,0} \greentriangleup} & 0,1274                \\ \hline \hline
\multicolumn{1}{|c|}{\multirow{3}{*}{BPRMF}}    & ExpLOD                         & \multicolumn{1}{c|}{\textbf{2,9789} \greentriangleup} & \multicolumn{1}{c|}{{ 835,7 \yellowcircle}}          & { 0,0990 \yellowcircle}          & \multicolumn{1}{c|}{1}   & \multicolumn{1}{c|}{59,4}           & \textbf{0,5771} \greentriangleup       \\ \cline{2-8} 
\multicolumn{1}{|c|}{}                          & ExpLOD v2                      & \multicolumn{1}{c|}{2,9197}          & \multicolumn{1}{c|}{{ \textbf{845,1} \greentriangleup \yellowcircle}} & { \textbf{0,0948} \greentriangleup \yellowcircle} & \multicolumn{1}{c|}{1}   & \multicolumn{1}{c|}{74,4}           & 0,5555                \\ \cline{2-8} 
\multicolumn{1}{|c|}{}                          & PEM                            & \multicolumn{1}{c|}{1,9623}          & \multicolumn{1}{c|}{587,1}                & 0,0316                & \multicolumn{1}{c|}{1}   & \multicolumn{1}{c|}{\textbf{317,8} \greentriangleup} & 0,1446                \\ \hline \hline
\multicolumn{1}{|c|}{\multirow{3}{*}{EASE}}     & ExpLOD                         & \multicolumn{1}{c|}{\textbf{2,9679} \greentriangleup} & \multicolumn{1}{c|}{{ 786,8 \yellowcircle}}          & { \textbf{0,0950} \greentriangleup \yellowcircle} & \multicolumn{1}{c|}{1}   & \multicolumn{1}{c|}{60,2}           & \textbf{0,5961} \greentriangleup       \\ \cline{2-8} 
\multicolumn{1}{|c|}{}                          & ExpLOD v2                      & \multicolumn{1}{c|}{2,8944}          & \multicolumn{1}{c|}{{ \textbf{805,7} \greentriangleup \yellowcircle}} & { 0,0891 \yellowcircle}          & \multicolumn{1}{c|}{1}   & \multicolumn{1}{c|}{73,3}           & 0,5590                \\ \cline{2-8} 
\multicolumn{1}{|c|}{}                          & PEM                            & \multicolumn{1}{c|}{2,0674}          & \multicolumn{1}{c|}{529,5}                & 0,0254                & \multicolumn{1}{c|}{1}   & \multicolumn{1}{c|}{\textbf{257,5} \greentriangleup} & 0,1264                \\ \hline \hline
\multicolumn{1}{|c|}{\multirow{3}{*}{NCF}}      & ExpLOD                         & \multicolumn{1}{c|}{\textbf{2,9592} \greentriangleup} & \multicolumn{1}{c|}{\textbf{967,6} \greentriangleup}       & \textbf{0,1125} \greentriangleup       & \multicolumn{1}{c|}{1}   & \multicolumn{1}{c|}{79,7}           & { 0,5217 \yellowcircle}          \\ \cline{2-8} 
\multicolumn{1}{|c|}{}                          & ExpLOD v2                      & \multicolumn{1}{c|}{2,8727}          & \multicolumn{1}{c|}{951,5}                & 0,0996                & \multicolumn{1}{c|}{1}   & \multicolumn{1}{c|}{105,4}          & { \textbf{0,5266} \greentriangleup \yellowcircle} \\ \cline{2-8} 
\multicolumn{1}{|c|}{}                          & PEM                            & \multicolumn{1}{c|}{1,9243}          & \multicolumn{1}{c|}{671,7}                & 0,0387                & \multicolumn{1}{c|}{1}   & \multicolumn{1}{c|}{\textbf{456,0} \greentriangleup} & 0,1558                \\ \hline
\end{tabular}
\end{table*}

Regarding the item metrics, ExpLOD outperformed its updated version, ExpLOD v2, on the MID metric, which then outperformed \ac{PEM} on both datasets. This suggests that the number of items included in an explanation may have a weaker influence than the attributes linking historical and recommended items. This is because the results of explanation goals in \cite{musto2019linked} and \cite{du2022post} were in the opposite direction, favoring ExPLOD v2 and \ac{PEM} over their baselines, ExpLOD, and ExPLOD v2, respectively. Similarly, the total number of items shown to users ($TID$) was the highest for the ExpLOD v2 algorithm. This suggests that the algorithms were not biased toward some items, which would create a long-tail distribution of the items displayed.\looseness=-1  

Because the user studies in \cite{musto2019linked} and \cite{du2022post} involved participants interacting with the system only once, the effect of interaction recency on explanation goals— although theoretically important \cite{balloccu2022post}—could not be directly captured in users’ evaluations. Consequently, assessing users’ perception of item recency remains a direction for future work. \looseness=-1

\begin{table*}[!t]
\caption{Offline results for the metrics for the top-5 recommendations for the \ac{MovieLens} dataset. Bold results with the symbol \greentriangleup are the best values considering the three explanation algorithms for a recommendation algorithm. Two values in column for a \ac{RS} with the symbol \yellowcircle represent Wilcoxon's p-value above 0.05 between them, meaning non-significant statistical differences.\looseness=-1}
\begin{tabular}{cc|ccc||ccc|}
\cline{3-8}
\multicolumn{1}{l}{\textbf{}}                   & \multicolumn{1}{l|}{\textbf{}} & \multicolumn{3}{c||}{Item Metrics}                                                                   & \multicolumn{3}{c|}{Attribute Metrics}                                                              \\ \cline{3-8} 
\textbf{}                                       & \textbf{}                      & \multicolumn{1}{c|}{MID}             & \multicolumn{1}{c|}{TID}                   & LIR             & \multicolumn{1}{c|}{ETD}             & \multicolumn{1}{c|}{TPD}             & SEP                   \\ \hline
\multicolumn{1}{|c|}{\multirow{3}{*}{MostPop}}  & ExpLOD                         & \multicolumn{1}{c|}{6.4674}          & \multicolumn{1}{c|}{1320.4}                & \textbf{0.0890} \greentriangleup & \multicolumn{1}{c|}{0.5724}          & \multicolumn{1}{c|}{66.8}            & \textbf{0.6504} \greentriangleup       \\ \cline{2-8} 
\multicolumn{1}{|c|}{}                          & ExpLOD v2                      & \multicolumn{1}{c|}{6.7446}          & \multicolumn{1}{c|}{\textbf{1360.8} \greentriangleup}       & 0.0797          & \multicolumn{1}{c|}{0.5778}          & \multicolumn{1}{c|}{40.6}            & 0.5822                \\ \cline{2-8} 
\multicolumn{1}{|c|}{}                          & PEM                            & \multicolumn{1}{c|}{\textbf{7.7393} \greentriangleup} & \multicolumn{1}{c|}{1040.1}                & 0.0313          & \multicolumn{1}{c|}{\textbf{0.9381} \greentriangleup} & \multicolumn{1}{c|}{\textbf{378.6} \greentriangleup}  & 0.1427                \\ \hline \hline
\multicolumn{1}{|c|}{\multirow{3}{*}{UserKNN}}  & ExpLOD                         & \multicolumn{1}{c|}{7.3926}          & \multicolumn{1}{c|}{1578.8}                & \textbf{0.0976} \greentriangleup & \multicolumn{1}{c|}{0.6533}          & \multicolumn{1}{c|}{113.0}           & \textbf{0.5427} \greentriangleup       \\ \cline{2-8} 
\multicolumn{1}{|c|}{}                          & ExpLOD v2                      & \multicolumn{1}{c|}{7.4649}          & \multicolumn{1}{c|}{\textbf{1630.1} \greentriangleup}       & 0.0917          & \multicolumn{1}{c|}{0.6351}          & \multicolumn{1}{c|}{104.8}           & 0.4233                \\ \cline{2-8} 
\multicolumn{1}{|c|}{}                          & PEM                            & \multicolumn{1}{c|}{\textbf{8.2190} \greentriangleup} & \multicolumn{1}{c|}{1285.0}                & 0.0318          & \multicolumn{1}{c|}{\textbf{0.9430} \greentriangleup} & \multicolumn{1}{c|}{\textbf{844.3} \greentriangleup}  & 0.1333                \\ \hline \hline
\multicolumn{1}{|c|}{\multirow{3}{*}{PageRank}} & ExpLOD                         & \multicolumn{1}{c|}{6.4330}          & \multicolumn{1}{c|}{1349.5}                & \textbf{0.0892} \greentriangleup & \multicolumn{1}{c|}{0.5611}          & \multicolumn{1}{c|}{112.7}           & { \textbf{0.6099} \greentriangleup \yellowcircle} \\ \cline{2-8} 
\multicolumn{1}{|c|}{}                          & ExpLOD v2                      & \multicolumn{1}{c|}{7.1271}          & \multicolumn{1}{c|}{\textbf{1485.5} \greentriangleup}       & 0.0819          & \multicolumn{1}{c|}{0.5968}          & \multicolumn{1}{c|}{95.0}            & { 0.5688 \yellowcircle}          \\ \cline{2-8} 
\multicolumn{1}{|c|}{}                          & PEM                            & \multicolumn{1}{c|}{\textbf{7.9168} \greentriangleup} & \multicolumn{1}{c|}{1061.6}                & 0.0305          & \multicolumn{1}{c|}{\textbf{0.9335} \greentriangleup} & \multicolumn{1}{c|}{\textbf{509.2} \greentriangleup}  & 0.1177                \\ \hline \hline
\multicolumn{1}{|c|}{\multirow{3}{*}{BPRMF}}    & ExpLOD                         & \multicolumn{1}{c|}{7.7752}          & \multicolumn{1}{c|}{1698.8}                & \textbf{0.0970} \greentriangleup & \multicolumn{1}{c|}{0.6908}          & \multicolumn{1}{c|}{{ 118.3 \yellowcircle}}     & \textbf{0.6007} \greentriangleup       \\ \cline{2-8} 
\multicolumn{1}{|c|}{}                          & ExpLOD v2                      & \multicolumn{1}{c|}{7.9518}          & \multicolumn{1}{c|}{\textbf{1774.5} \greentriangleup}       & 0.0890          & \multicolumn{1}{c|}{0.6788}          & \multicolumn{1}{c|}{{ 119.3 \yellowcircle}}     & 0.5353                \\ \cline{2-8} 
\multicolumn{1}{|c|}{}                          & PEM                            & \multicolumn{1}{c|}{\textbf{8.1582} \greentriangleup} & \multicolumn{1}{c|}{1415.5}                & 0.0328          & \multicolumn{1}{c|}{\textbf{0.9542} \greentriangleup} & \multicolumn{1}{c|}{\textbf{1033.1} \greentriangleup} & 0.1452                \\ \hline \hline
\multicolumn{1}{|c|}{\multirow{3}{*}{EASE}}     & ExpLOD                         & \multicolumn{1}{c|}{7.0517}          & \multicolumn{1}{c|}{1530.7}                & \textbf{0.0960} \greentriangleup & \multicolumn{1}{c|}{0.6176}          & \multicolumn{1}{c|}{{ 125.8 \yellowcircle}}     & \textbf{0.6009} \greentriangleup       \\ \cline{2-8} 
\multicolumn{1}{|c|}{}                          & ExpLOD v2                      & \multicolumn{1}{c|}{7.3943}          & \multicolumn{1}{c|}{\textbf{1630.0} \greentriangleup}       & 0.0891          & \multicolumn{1}{c|}{0.6278}          & \multicolumn{1}{c|}{{ 121.3 \yellowcircle}}     & 0.5272                \\ \cline{2-8} 
\multicolumn{1}{|c|}{}                          & PEM                            & \multicolumn{1}{c|}{\textbf{8.2949} \greentriangleup} & \multicolumn{1}{c|}{1296.1}                & 0.0309          & \multicolumn{1}{c|}{\textbf{0.9405} \greentriangleup} & \multicolumn{1}{c|}{\textbf{863.5} \greentriangleup}  & 0.1335                \\ \hline \hline
\multicolumn{1}{|c|}{\multirow{3}{*}{NCF}}      & ExpLOD                         & \multicolumn{1}{c|}{\textbf{9.4064} \greentriangleup} & \multicolumn{1}{c|}{{ 2062.9 \yellowcircle}}          & \textbf{0.1163} \greentriangleup & \multicolumn{1}{c|}{0.8395}          & \multicolumn{1}{c|}{185.6}           & { \textbf{0.5453} \greentriangleup \yellowcircle} \\ \cline{2-8} 
\multicolumn{1}{|c|}{}                          & ExpLOD v2                      & \multicolumn{1}{c|}{9.2497}          & \multicolumn{1}{c|}{{ \textbf{2077.2} \greentriangleup \yellowcircle}} & 0.1026          & \multicolumn{1}{c|}{0.8096}          & \multicolumn{1}{c|}{235.6}           & { 0.5145 \yellowcircle}          \\ \cline{2-8} 
\multicolumn{1}{|c|}{}                          & PEM                            & \multicolumn{1}{c|}{8.5089}          & \multicolumn{1}{c|}{1749.1}                & 0.0375          & \multicolumn{1}{c|}{\textbf{0.9873} \greentriangleup} & \multicolumn{1}{c|}{\textbf{1796.3} \greentriangleup} & 0.1605                \\ \hline
\end{tabular}
\label{tab:offresults_n5}
\end{table*}

The progression of $TPD$ aligned with findings from prior user studies on explanation goals, indicating that the attribute connecting historical and recommended items plays a significant role in shaping users’ perception of explanation quality. However, the $ETD$ metric, which measures the diversity of attributes in an explanation list, always yields a value of 1. This is because only a single explanation was provided to the user, thereby only one attribute per user was shown. This result aligns with the findings reported in \cite{10.1145/3705328.3748028}, where the authors emphasized that evaluating counterfactual explanations beyond the top-1 recommendation enhances the consistency of explanation assessment.\looseness=-1  

Therefore, the first conclusion from our offline experiment is as follows: across all three algorithms, sentence-based explanations rely on connections between interacted and recommended items through common attributes. \emph{Consequently, when users evaluate only a single explanation, they may overlook potential algorithmic bias. As shown by the metadata distribution in Section \ref{sec:kg_extraction_analysis}, algorithms tend to favor attributes and items that are already popular within a user's profile, which can cause the same popular attribute to appear in multiple explanations. Although evaluating multiple explanations requires more user effort, providing longer recommendation lists in online experiments enables users to better assess explanation quality, particularly in terms of attribute repetition.}\looseness=-1   

To analyze the effects of attribute diversification, which could not be captured by the ETD metric for the top recommendation, Table \ref{tab:offresults_n5} presents the results of offline experiments for the top-5 recommendations on the \ac{MovieLens} dataset.\looseness=-1 

According to the results from the three algorithms, \ac{PEM} achieved the highest user mean diversity of items and attributes ($MID$ and $ETD$) for all recommendation algorithms. Due to this high attribute diversification, it also exhibited the highest Total Property Diversity ($TPD$). The decrease in $SEP$ is a consequence of diversification, as the algorithm's tendency to show a wider variety of attributes necessarily includes less popular ones. \looseness=-1

Furthermore, \ac{PEM} achieved the lowest total item diversity ($TID$), suggesting that despite its attribute diversification, the items associated with these attributes are more common across users. This finding is particularly relevant when considering the impact of a single recommendation (Table \ref{tab:offresults_n1}) versus multiple recommendations (\ref{tab:offresults_n5}) on user perception. For instance, unlike scenarios with a single recommendation, \ac{PEM} performed better in intra-list item diversity ($MID$) when multiple recommendations were shown, suggesting that for multiple recommendations, both ExpLOD algorithms are biased toward items with popular attributes.\looseness=-1

Consequently, increasing the number of interacted items and displayed attributes can positively influence user perception of explanations. This finding is consistent with industry practices where explanations are often presented in a list format, and diversity is prioritized to help users discover new interests, thereby increasing their engagement and long-term fidelity with the platform \cite{alvino2015learning}. \looseness=-1

In \cite{musto2019linked}, the authors conducted online experiments comparing ExpLOD and ExpLOD v2. The results indicated that, relative to the first version, ExpLOD v2 produced a statistically non-significant decrease in persuasion, non-significant improvements in transparency and engagement, and a significant improvement in trust. These findings align with the metrics, as our experiments revealed improvements in ExpLOD v2 over ExpLOD in the MID and ETD metrics (except for the User-KNN and NCF algorithms). They also underscore the importance of selecting interacted items and attributes in explanation algorithms on user perception.\looseness=-1  

This leads to the second conclusion of our offline experiments: \emph{When analyzing the popularity of attributes ($SEP$), a trade-off with the diversity of attributes ($ETD$) was observed in Table \ref{tab:offresults_n1} and Table \ref{tab:offresults_n5}, where \ac{PEM} was outperformed by its baselines in attribute popularity but outperformed them in diversity.} \looseness=-1 

These results indicate that popularity directly shapes users’ perception of explanations, whereas diversity exerts an indirect influence through its effect on popularity. When explanations are overly similar, users appear to prefer exposure to different attributes, rather than repetitive information. The impact of popularity, either positive or negative, on user perception will therefore be further investigated in the experiments described in Section \ref{sec:online_results}. \looseness=-1 

In summary, the online study in \cite{du2022post} showed that the \ac{PEM} algorithm significantly enhanced persuasiveness, engagement, and trust, with a slight, but not significant, improvement in transparency over ExpLOD v2. Our offline results, as shown in Table \ref{tab:offresults_n5}, are consistent with this, revealing that \ac{PEM} included more attributes and interacted items in its explanations ($ETD$ and $MID$) despite its lower total item diversity ($TID$). This demonstrates the importance of attribute diversity in explanations, as improvements in $ETD$ and $MID$ were consistently linked to better user-perceived outcomes, both when moving from ExpLOD to ExpLOD v2 and from ExpLOD v2 to \ac{PEM}. \looseness=-1

Therefore, based on our offline experiments, we highlight two major hypotheses regarding the relationship between offline path explanation metrics and online user perception of explanation goals: \looseness=-1

\begin{enumerate}
    \item Offline path explanation metrics serve as indicators of user perception, offering a valuable tool to evaluate the progression of explanation algorithms. \looseness=-1

    \item Users' perception of explanations is influenced by the diversity and popularity of attributes that connect interacted items to recommended items. \looseness=-1
\end{enumerate}

Considering these two main insights from our offline evaluation and comparing them to the online studies of each algorithm in \cite{musto2016explod}, \cite{musto2019combining}, and \cite{du2022post}, we conducted an online study to further validate these hypotheses and address our \ac{RQ}2 regarding the applicability of such metrics in offline experiments.

\section{Online Experiments}
\label{sec:online} 

To validate whether there exists a relation between offline path explanation metrics and explanation goals, we conducted an online user experiment.\looseness=-1 

Initially, we decided to use the EASE recommendation algorithm to generate recommendations for three main reasons: (i) in our offline results reported in Table \ref{tab:offresults_n1} and Table \ref{tab:offresults_n5}, all explanation algorithms were consistent across different \ac{RS} algorithms; (ii) in the offline ranking metrics in Appendix \ref{apen:ranking}, EASE had the best ranking \ac{NDCG} results across the different \ac{RS} algorithms of the different families; and (iii) it has fast training and testing times to compute recommendations online for the participants.\looseness=-1

In Section \ref{sec:offline_results}, we concluded that showing explanations for only one recommendation can hide algorithmic bias towards showing popular paths of the \ac{KG}. Therefore, we chose to show five recommendations and their respective explanations for users to also properly assess item and attribute diversity.\looseness=-1

As verified in Section \ref{sec:offline_results}, there is a trade-off between the diversity and popularity of attributes. While PEM \cite{du2022post} displays high diversity in attribute selection, both versions of ExpLOD \cite{musto2016explod,musto2019linked} have higher attribute popularity. To validate our second hypothesis regarding this behavior, we chose two algorithms for our user experiments: PEM and ExpLOD v2 \cite{musto2019linked}. We selected PEM for its high diversity, and ExpLOD v2 because it is an improved version of ExpLOD that was used in the most recent prior online evaluation \cite{du2022post}. \looseness=-1

\subsection{Online Experiment Protocol}
\label{sec:on_protocol}

We designed the online study as a within-subjects experiment, in which all participants directly compared the explanations produced by ExpLOD v2 \cite{musto2019linked} and \ac{PEM} \cite{du2022post}. Participants evaluated these side-by-side explanations using a 5-point Likert scale with labels ranging: Much More A, More A, Equal, More B, Much More B. The assignment of the anonymized labels A and B to each algorithm was randomized for different participants to prevent any positional bias.\looseness=-1

As recruiting participants can be challenging, we opted for a within-subjects design, which requires fewer participants to achieve meaningful results \cite{knijnenburg2015evaluating}. We conducted an \textit{a priori} power analysis in G*Power \cite{faul_erdfelder_buchner_lang_2009} for a within-subjects experiment, considering a Wilcoxon signed-rank test (one-sample case). We assumed a medium effect size ($d = 0.5$), significance level of $\alpha = 0.05$ (5\% chance of a false positive), and statistical power of $1 - \beta = 0.9$ (90\% chance of detecting a real difference). This analysis indicated a required minimum of 47 participants. By contrast, if we had conducted a between-subjects A/B experiment, a Wilcoxon–Mann–Whitney test with the same parameters and an allocation ratio of 1 would have required a much larger sample of 180 participants to reach the same statistical power.\looseness=-1 

In accordance with \cite{10.1145/3716394}, we aimed to recruit participants with diverse profiles by inviting individuals from different backgrounds. We recruited participants by reaching out to students of the authors, as well as to online and offline communities not involved in the research and unfamiliar with the study topics (e.g., social media groups and local organizations). The project was approved by the Brazilian Ethics Committee.\looseness=-1 

In the first step, participants read the consent terms from the ethical committee and filled in personal information, such as nationality, level of education, age, gender, and whether they were familiar with \ac{RS}s. Participants were asked to add ten liked films to build their profiles, simulating an interacted item set. Following the findings of \cite{rashid2002getting}, the top 100 items were displayed in random order and ranked by $\forall i \in I \log_{10}(popularity * entropy)$, where $I$ is the set of items in the dataset. Appendix \ref{apen:profile} shows a screen where users choose items to represent their history in the experiment.\looseness=-1

\begin{figure}[!b]
    \centering
    \includegraphics[width=.9\textwidth]{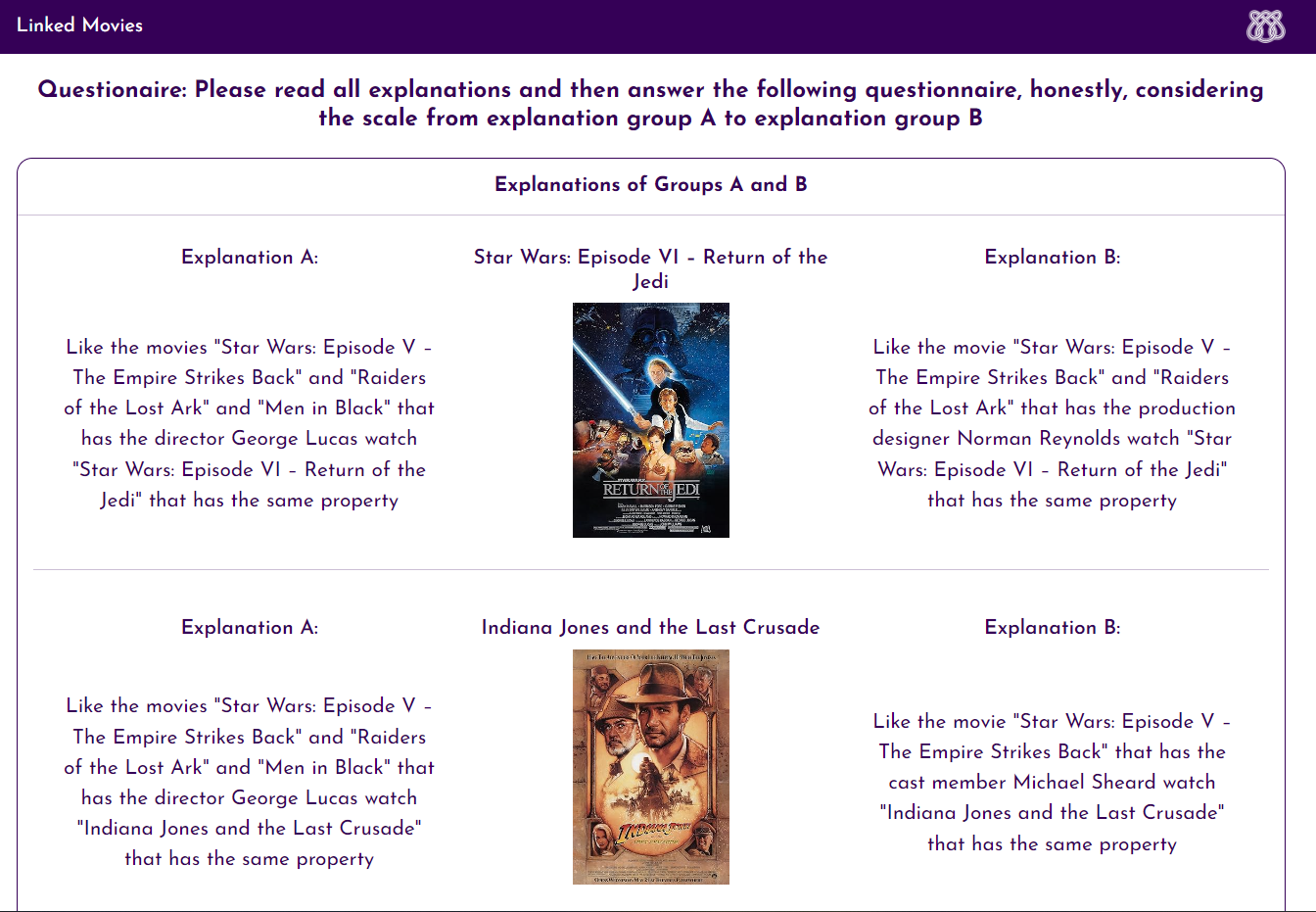}
    \caption{Example of a user's screen with recommendations and a set of questions to be answered}
    \label{fig:rec}
\end{figure}

The online study involved participants comparing and assessing five explanations from two algorithms: ExpLOD v2 \cite{musto2016explod} and \ac{PEM} \cite{du2022post}. The experiment utilized the \ac{MovieLens} dataset, in line with previous studies \cite{musto2016explod, du2022post}, with recommendations generated by the EASE algorithm \cite{Steck2019}. \looseness=-1

As described in Section \ref{sec:expl_algs}, both algorithms rank attributes based on a score function that considers the number of times an attribute appears in both interacted and recommended items, relative to its overall popularity within the entire dataset. To ensure a fair comparison of the explanations analyzed by the users, we created a single template for both algorithms. This template was based on the highest-ranked attribute from each algorithm's respective scoring function.\looseness=-1 

The template begins with ``Like the movies $<h>$," where $<h>$ is a list of films previously selected by the user. We then add the attribute edge type that connects profile items to recommended items, using the template ``that has the $<type>$ $<attribute>$." and concludes with ``watch $<recommendation>$, that has the same property," where $<recommendation>$ is the suggestion from the recommendation algorithm. Thus, the complete template becomes:\looseness=-1

\vspace{5mm}
\textit{``Like the movies $<h>$ that has the $<type>$ $<attribute>$ watch $<recommendation>$, that has the same property"}
\vspace{5mm}

Considering the example in Figure \ref{fig:graph} and using the attribute ``drama", the explanation becomes: Like the movie ``Saving Private Ryan which has the genre drama, watch ``Forest Gump", which has the same property.\looseness=-1 

Recommended movies by the EASE \cite{Steck2019} algorithm were placed in the center, with two columns of explanations: A on the left side of the screen and B on the right. Positions A and B for each algorithm (\ac{PEM} \cite{du2022post} and ExpLOD \cite{musto2019linked}) were randomly chosen at run-time to prevent positional bias. Each explanation was built as in the offline experiment to facilitate the user evaluation of items and attributes. Figure \ref{fig:rec} displays the screen with recommendations and explanations for groups A and B.\looseness=-1 

Finally, after analyzing all recommendations and explanations, participants were asked to answer six questions on a Likert scale with the options: Much More A, More A, Equal, More B, or Much More B. The questions were drawn from previous user studies of \ac{PEM} in \cite{du2022post} and ExpLOD v2 in \cite{musto2019linked} to maintain consistent evaluation criteria. Table \ref{tab:def_goals} lists the questions and their respective objectives: \looseness=-1 

\begin{table*}[!t]
\caption{Table of explanation questions and related goals}
\begin{adjustbox}{}
    \begin{tabular}{l|c}  \hline
         \textbf{Goal} & \textbf{Question}  \\ \hline \hline
         (1) Diversity & \makecell{Which explanation group (A or B) \\ has more diverse explanations?} \\ \hline
         (2) Popularity &  \makecell{Which explanation group (A or B) \\ has more familiar explanations?} \\ \hline
         (3) Persuasiveness   &   \makecell{Which explanation group (A or B) \\ is more convincing?}       \\ \hline
         (4) Transparency       &  \makecell{Which explanation group (A or B) made you understand \\  better why the recommendation was suggested to you?}  \\ \hline
         (5) Engagement  &  \makecell{Which explanation group (A or B) \\ made you discover new information about the movie?} \\ \hline
         (6) Trust &   \makecell{ Which explanation group (A or B) made you trust \\ more in the recommendation system? }  \\ \hline
    \end{tabular}
\end{adjustbox}
\label{tab:def_goals}
\end{table*}

Questions (1) and (2) were the only ones not included in \cite{du2022post} and \cite{musto2019linked}. These questions assess the perceived diversity and popularity of attributes and should directly reflect offline path metrics for the diversity and popularity of attributes. They aim to validate whether these offline metrics accurately reflect user perception. Questions (3) to (6) evaluate the goals of Persuasiveness, Transparency, Engagement, and Trust, as outlined in \cite{tintarev2015explaining}, and are aligned with the online experiments in \cite{du2022post} and \cite{musto2019linked}. Appendix \ref{apen:expl} shows a screen with explanations from groups A and B, along with some of the questions posed to users.\looseness=-1

\subsection{Online Results}
\label{sec:online_results}

Our online within-subject experiments were conducted with 55 participants, exceeding the 47 required according to our G*Power \textit{a priori} estimation in Section \ref{sec:on_protocol}. The majority of participants were between 25 and 50 years old (58\%), with one below 17, sixteen between 18 and 24, four between 50 and 60, and two over 60. Regarding gender, most were male (39, or 71\%), while 16 were female. Nearly all participants (96\%) had previously interacted with \ac{RS}. In terms of education, 21 held a bachelor's degree (38\%), seven were in high school, ten had a master's degree, and fifteen had a PhD; two participants did not fit into any of these categories. Except for one Portuguese participant, all were Brazilian. The participants' explanations for both algorithms, their responses and analysis presented here are available open source\footnote{\url{https://github.com/andlzanon/offline_metrics_explanation_rec_goals/tree/main/Online\%20User\%20Study}}.\looseness=-1 

Figure \ref{fig:onresults1} shows a divergent bar chart with the participants' overall choices regarding the four explanation goals, in addition to the perceived popularity and diversity of attributes in explanations. Each row represents the distribution of user choices for a specific question, with the row name corresponding to an explanation goal as defined in Table \ref{tab:def_goals}.\looseness=-1  

As described in Section \ref{sec:online}, users evaluated explanations with the algorithm names \ac{PEM} and ExpLOD v2 hidden, while explanations were randomly positioned on sides A (left) and B (right) with the recommendation in the center. For 26 participants, the \ac{PEM} \cite{du2022post} explanation algorithm was placed on side A and ExpLOD v2 \cite{musto2019linked} on side B. For the other 29 participants, the inverse occurred, with ExpLOD v2 on side A and \ac{PEM} on side B. The results presented in this section are based on the Likert scale, with `A' and `B' placeholders replaced according to each participant's algorithm positioning. \looseness=-1  

To validate whether the offline results were aligned with user perceptions of explanations, we used two specific questions. As noted in the previous section, the \ac{PEM} algorithm \cite{du2022post} outperformed ExpLOD v2 \cite{musto2019linked} in terms of attribute diversity in offline metrics, whereas ExpLOD v2 showed higher performance in terms of attribute popularity. A similar pattern was observed in user responses: a slightly higher number of participants favored \ac{PEM} for the "Diversity" question (25 out of 55), while a comparable number favored ExpLOD v2 for "Popularity" (24 out of 55). These results are shown in the last two rows on Figure \ref{fig:onresults1}.\looseness=-1 

\begin{figure}[!t]
    \centering
    \includegraphics[width=.9\textwidth]{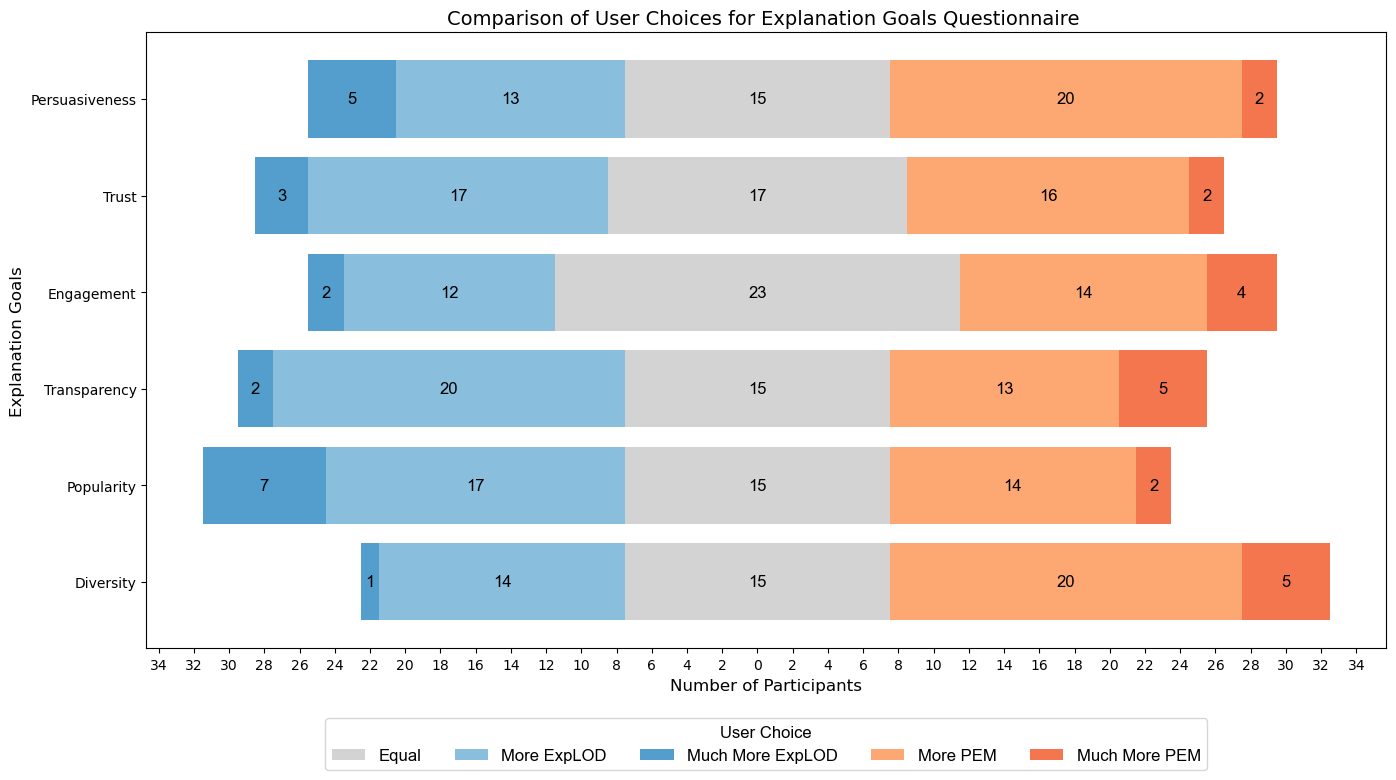}
    \caption{User response distribution on defined Likert scale.}
    \label{fig:onresults1}
\end{figure}

\begin{figure}[!b]
    \centering
    \includegraphics[width=.7\textwidth]{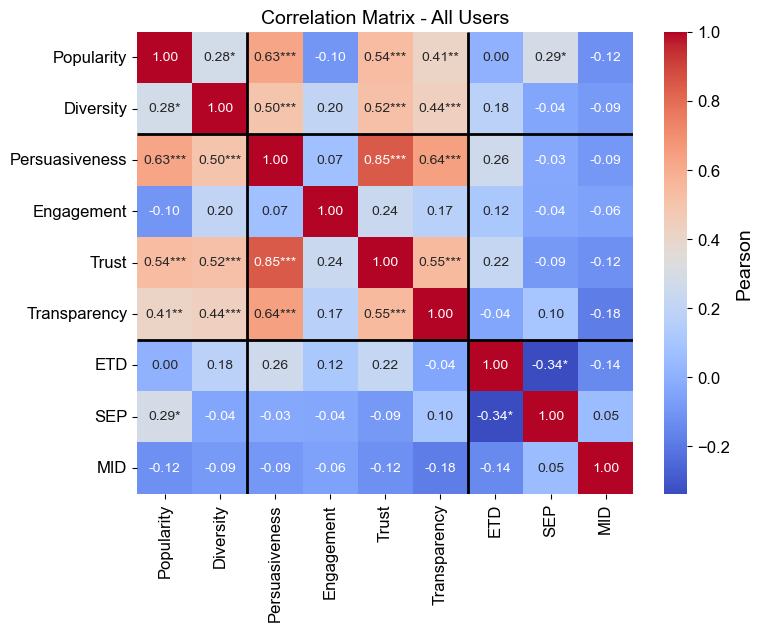}
    \caption{Correlation analysis between differences in user ratings for \ac{PEM} and ExpLOD v2 algorithms and the online explanation goals. The ETD, SEP, and MID columns denote the correlations for the differences between the corresponding \ac{PEM} and ExpLOD v2 offline metrics. Correlations for the individual offline metrics of each algorithm are also included (see lines and columns 7–12). The symbol * indicates a correlation with $p < 0.05$; ** indicates $p < 0.01$; and *** indicates $p < 0.001$.\looseness=-1}
    \label{fig:corr_all}
\end{figure}

These questions were also where users felt most confident, with ``Much More \ac{PEM}'' and ``Much More ExpLOD v2'' selected in 12 of the 55 sessions. Seven of these responses favored ``Much More ExpLOD v2'' for the ``Popularity'' question, and five favored ``Much More \ac{PEM}'' for ``Diversity''. To further validate these findings, we measured the diversity of attributes in the five explanations generated by \ac{PEM} and ExpLOD v2 for all participants. \ac{PEM} showed an average of 4.8 attributes per user, compared to 3.09 for ExpLOD v2. These results indicate our first major finding of the online experiments: \emph{users' perception of diversity and popularity of attributes aligned with their respective offline metrics results.}\looseness=-1 

Considering the four explanation goals, Figure \ref{fig:onresults1} shows that \ac{PEM} outperformed ExpLOD v2 in user perception of persuasiveness and engagement, aligning with the online experiment results reported in \cite{du2022post}. However, for trust and transparency, user perception was the opposite, with ExpLOD v2 outperforming \ac{PEM}. This result differs from those reported by \cite{du2022post}. We believe that this discrepancy arises because, in \cite{du2022post}, participants evaluated only the explanation for the top item and not a list of items. We argue that evaluating multiple explanations is important for users to assess the attribute bias of the explanation algorithm \cite{10.1145/3705328.3748028}. \looseness=-1

We performed a Wilcoxon test \cite{Wilcoxon1992} on the results for all questions to assess whether user preferences for one system were statistically significant. Likert-scale responses were mapped numerically as follows: ``Much More \ac{PEM}'' with 2; ``More \ac{PEM}'' with 1; ``Equal'' with 0; ``More ExpLOD'' with -1; and ``Much More ExpLOD'' with -2. The results showed no statistically significant preference for either system across any explanation goal or perceived popularity and diversity ($p \geq 0.05$), and thus should be interpreted with caution.\looseness=-1

In addition, we performed Pearson’s correlation analysis on users’ responses at the individual level, converting Likert-scale responses using the same numerical encoding applied in the Wilcoxon test \cite{Wilcoxon1992}. This analysis allowed us to examine the relationship between responses to explanation goals and users’ perceived diversity and popularity.\looseness=-1 

In addition, we computed the offline explanation metrics for participants’ explanations in the online experiment to directly examine the correlations between the offline and online metrics. Six path-based offline explanation metrics were defined in Table~\ref{tab:metrics}. However, recency of interacted items (LIR) could not be calculated because participants interacted with the system only once and provided a representation of their historical items, rather than interacting with items over time, which is necessary for computing the metric. Furthermore, TID and TPD quantify the total number of interacted items and attributes presented across all users, making them algorithm-level metrics. Consequently, they were calculated, but could not be meaningfully compared to user-level online explanation goal metrics. Therefore, our correlation analysis focuses on the offline metrics capturing attribute popularity (SEP), attribute diversity (ETD), and the diversity of interacted items (MID). \looseness=-1

Table~\ref{tab:off_metrics_online} reports the offline explanation metrics for the explanation algorithms in the user study. The results are consistent with those presented in Section~\ref{sec:offline_results} for the EASE \cite{Steck2019} algorithm: PEM exhibits higher ETD (diversity of attributes) and TPD (total number of attributes across all participants’ explanations), whereas ExpLOD v2 achieves higher SEP (attribute popularity). The only deviations are observed in the TID (total number of interacted items) and MID (diversity of interacted items) metrics. These differences arise from the considerably smaller number of interacted items provided by users in our online experiment compared to those in the \ac{MovieLens} dataset, as participants were instructed to select at least 10 items to represent their interaction history (see Appendix~\ref{apen:profile}). \looseness=-1

Figure~\ref{fig:corr_all} presents the correlation values between the offline explanation metrics and the online explanation goals. Following the same encoding applied to the Likert-scale responses, where values reflected comparisons between the \ac{PEM} and ExpLOD v2 algorithms, we computed for each user, the correlation based on the difference $PEM_{m} - ExpLODv2_{m}$ for each metric offline $m$. In the correlation matrix, the ETD, SEP, and MID columns correspond to these user-level metric differences. More clearly defined lines in the correlation matrix separate the three aspects we analyze: the perception of popularity and diversity in explanations, the online explanation goal metrics, and the offline explanation metrics. The complete correlation matrix is provided in Appendix~\ref{apen:correlations}, which also reports the correlations of each metric–algorithm pair with the online metrics and with users’ perceived diversity and popularity. Statistical significance is indicated as follows: * $p < 0.05$; ** $p < 0.01$; and *** $p < 0.001$.

\begin{table*}[t!]
\caption{Mean offline explanation metrics for \ac{PEM} and ExpLOD v2 for participants in the online experiment.\looseness=-1}
\begin{adjustbox}{}
\begin{tabular}{c|cc||ccc|}
\cline{2-6}
                                            & \multicolumn{2}{c||}{Item Metrics} & \multicolumn{3}{c|}{Attribute Metrics}                        \\ \cline{2-6} 
                                            & \multicolumn{1}{c|}{MID}    & TID & \multicolumn{1}{l|}{TPD} & \multicolumn{1}{c|}{ETD}   & SEP   \\ \hline
\multicolumn{1}{|c|}{\ac{PEM}} & \multicolumn{1}{c|}{2.218}  & 140 & \multicolumn{1}{c|}{144} & \multicolumn{1}{c|}{0.959} & 0.220 \\ \hline
\multicolumn{1}{|c|}{ExpLOD}                & \multicolumn{1}{c|}{2.592}  & 127 & \multicolumn{1}{c|}{49}  & \multicolumn{1}{c|}{0.618} & 0.472 \\ \hline
\end{tabular}
\end{adjustbox}
\label{tab:off_metrics_online}
\end{table*}

In this scenario, it is important to emphasize that we measure the correlation between the differences in Likert scale ratings assigned to the \ac{PEM} and ExpLOD v2 algorithms and the online explanation goals. In other words, our analysis focuses on relative user perception; that is, how users perceive the difference between the two explanation algorithms with respect to the diversity and popularity of attributes for each explanation goal. For example, a user may have considered both explanations poor overall, yet still rated one algorithm as more transparent than the other.\looseness=-1

Although the quality of the explanations generated by these algorithms has been demonstrated in their respective papers \cite{du2022post, musto2019linked}, this introduces a limitation for the correlation analysis, as our results rely solely on subjective user perceptions rather than on a direct numerical evaluation of the explanations, in which users would have individually assessed both algorithms. Furthermore, as noted earlier, these correlations were computed for algorithms that did not exhibit a statistically significant difference in user preference.\looseness=-1

To further support our findings, we compared the resulting correlations with those reported by \cite{balog2020measuring}, who analyzed relationships between the explanation goals proposed by \cite{tintarev2015explaining} using human-generated explanations. In addition to this comparison, we extend the analysis to include correlations with the perceived popularity and diversity of attributes in \ac{KG} explanations. The interpretation of correlation coefficients follows the guidelines of \cite{hinkle2003applied}, where 0.30–0.50 indicates a low correlation, 0.50–0.70 a moderate correlation, 0.70–0.90 a high correlation, and 0.90–1.0 a very high correlation.\looseness=-1

In \cite{balog2020measuring}, the authors found that most explanation goals defined in \cite{tintarev2015explaining} are moderately correlated, with some goals related to several others and others remaining relatively distinct. Our results are consistent with this pattern. Specifically, persuasiveness, trust, and transparency each showed at least moderate correlations ($r \geq 0.55$) with one another, all statistically significant.\looseness=-1

Furthermore, we observed the same strong correlation between the goals of persuasiveness and trust, with a statistically significant value of 0.85 in our online experiments, supporting the findings of \cite{balog2020measuring}. In contrast, the goal of engagement—introduced in ExpLOD \cite{musto2016explod, musto2019linked} and PEM \cite{du2022post}—was relatively distinct, showing low correlations with the other three explanation goals proposed in \cite{tintarev2015explaining}.\looseness=-1

\begin{figure}[!b]
    \centering
    \includegraphics[width=.9\textwidth]{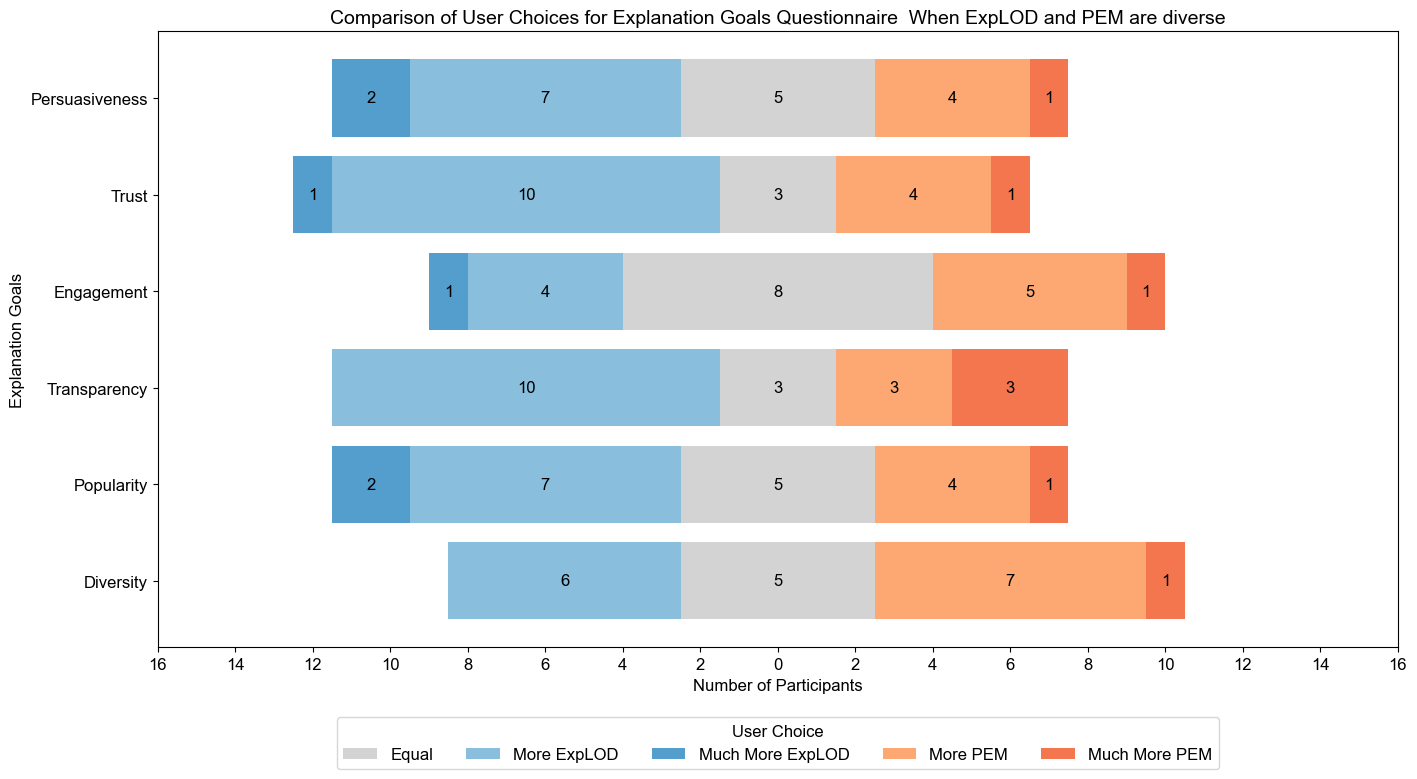}
    \caption{User response distribution on defined Likert scale when both ExpLOD v2 and \ac{PEM} display diverse explanation attributes.}
    \label{fig:onresults2}
\end{figure}

When analyzing the correlation between perceived diversity and popularity and the explanation goals, both factors were moderately correlated with persuasiveness and trust, with statistically significant values. In contrast, they showed only weak correlations with transparency, with Pearson coefficients of 0.41 for perceived popularity and 0.44 for perceived diversity, both statistically significant. Regarding engagement, we observed a correlation of 0.20 with perceived diversity, which helps explain why users preferred \ac{PEM} over ExpLOD in Figure~\ref{fig:onresults1}, as \ac{PEM} presented greater attribute diversity in its explanations. Overall, perceived diversity and perceived popularity were associated with all explanation goals, with the exception of engagement.\looseness=-1

Considering the offline metrics, the trade-off between SEP and ETD was reflected in a low negative correlation of -0.34, which was statistically significant. Interestingly, when examining the correlation between the offline metrics of popularity (SEP) and diversity (ETD) and their corresponding questionnaire items on perceived popularity and perceived diversity, SEP showed a statistically significant correlation of 0.29 with its respective perception measure, whereas ETD exhibited a non-significant value of 0.18. This suggests that users do not clearly understand what constitutes diversity and popularity in explanations. Such an interpretation is further supported by the low correlation between SEP and the explanation goals, which contrasts with the moderate correlations observed between perceived popularity and the explanation goals.

This is a key consideration for future research in the field, as the definition of offline metrics must be developed in ways that can be reliably interpreted and assessed by users. One limitation of our online experiment that may have contributed to the low correlations is the formulation of the questionnaire. However, we intentionally kept these questions open-ended to avoid introducing bias that could arise from more narrowly defined questions and to better observe how users naturally reason about the explanations.\looseness=-1

Finally, ETD was the only metric that showed any meaningful correlation with users’ perceptions of the explanation goals related to persuasiveness and trust. This may help explain why prior online experiments \cite{musto2019linked, du2022post} found that increasing explanation diversity also improved users’ overall perceptions of the explanations. Moreover, these findings suggest that current offline metrics do not accurately capture online explanation goals, and that developing metrics that more reliably reflect users’ perceptions of popularity and diversity remains an open research challenge.\looseness=-1

In our online experiments, \ac{PEM} consistently produced more diverse explanations than ExpLOD. To better understand the impact of diversity, we analyzed a subset of users for whom ExpLOD v2 also generated diverse explanations. This approach allowed us to isolate diversity as a variable, enabling us to examine its influence on explanation goals. Figure \ref{fig:onresults2} displays a divergent bar chart for this user subset, where the mean attribute diversity of ExpLOD was four or above. In this scenario, the primary difference between the two algorithms was attribute popularity, with ExpLOD outperforming \ac{PEM}.\looseness=-1

\begin{table*}[!t]
\caption{Mean offline explanation metrics for \ac{PEM} and ExpLOD v2 for participants in the online experiment, restricted to users for whom both algorithms produced diverse explanation attributes.\looseness=-1}
\begin{adjustbox}{}
\begin{tabular}{c|cc||ccc|}
\cline{2-6}
                                            & \multicolumn{2}{c||}{Item Metrics} & \multicolumn{3}{c|}{Attribute Metrics}                        \\ \cline{2-6} 
                                            & \multicolumn{1}{c|}{MID}    & TID & \multicolumn{1}{l|}{TPD} & \multicolumn{1}{c|}{ETD}   & SEP   \\ \hline
\multicolumn{1}{|c|}{\ac{PEM}} & \multicolumn{1}{c|}{2.063}  & 79  & \multicolumn{1}{c|}{69}  & \multicolumn{1}{c|}{0.978} & 0.151 \\ \hline
\multicolumn{1}{|c|}{ExpLOD}                & \multicolumn{1}{c|}{2.442}  & 77  & \multicolumn{1}{c|}{42}  & \multicolumn{1}{c|}{0.831} & 0.351 \\ \hline
\end{tabular}
\end{adjustbox}
\label{tab:off_f_metrics_online}
\end{table*}

\begin{figure}[!b]
    \centering
    \includegraphics[width=.7\textwidth]{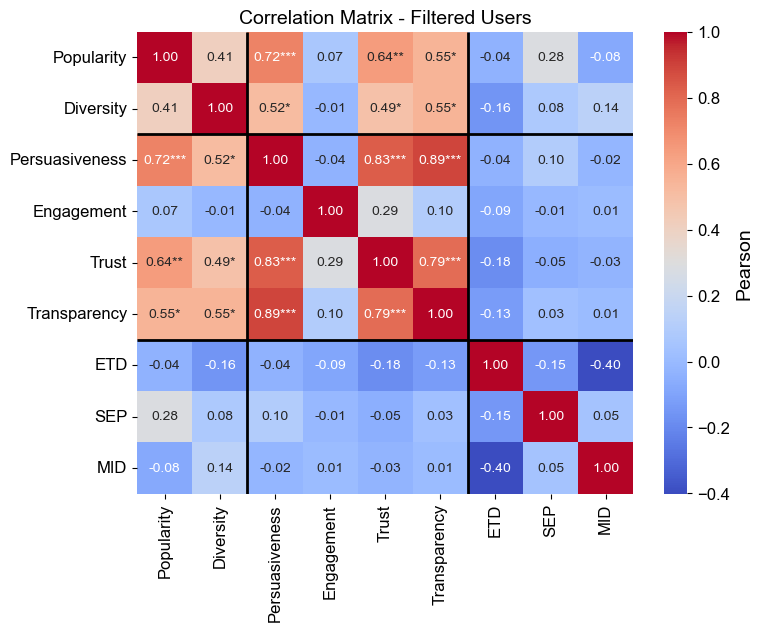}
    \caption{Correlation analysis between differences in user ratings for \ac{PEM} and ExpLOD v2 algorithms and the online explanation goals when they display diverse explanation attributes. The ETD, SEP, and MID columns denote the correlations for the differences between the corresponding \ac{PEM} and ExpLOD v2 offline metrics. Correlations for the individual offline metrics of each algorithm are also included (see lines and columns 7–12). The symbol * indicates a correlation with $p < 0.05$; ** indicates $p < 0.01$; and *** indicates $p < 0.001$.\looseness=-1}
    \label{fig:corr_f}
\end{figure}

We identified 19 users under this condition, where both \ac{PEM} and ExpLOD displayed at least four different attributes in the five recommendations. This means that, for these users, there was a maximum of one attribute repetition among the five explanations. Within this subset, the mean attribute diversity was 4.16 for ExpLOD v2 and 4.89 for \ac{PEM}.\looseness=-1

Based on these results, when both algorithms increase the number of attributes shown, users still recognize that \ac{PEM} presents more diverse and less popular attributes than ExpLOD v2. This supports our earlier finding that user perceptions of popularity and diversity align with the overall scores from the offline $SEP$ and $ETD$ metrics. Nevertheless, a Wilcoxon signed-rank test \cite{Wilcoxon1992} on all questions for this subset of users did not reveal a statistically significant preference for one system over the other ($p \geq 0.05$).\looseness=-1  

There was a noticeable shift in the explanation goal metrics for persuasiveness and engagement. Previously, users preferred \ac{PEM}, and these goals were associated with its higher attribute diversity. However, this effect disappeared once ExpLOD’s diversity matched that of \ac{PEM}. This was most evident for persuasiveness, where ExpLOD v2 outperformed \ac{PEM} in Figure~\ref{fig:onresults2}, in contrast to the results in Figure~\ref{fig:onresults1}. For engagement, although the effect was less pronounced, the gap between \ac{PEM} and ExpLOD v2 was slightly reduced. The influence of perceived diversity on engagement and persuasiveness is further emphasized by the stability of the transparency and trust results, which remained unchanged between Figures~\ref{fig:onresults1} and \ref{fig:onresults2} and thus appear to be linked to the perceived popularity of ExpLOD’s attributes. \looseness=-1  

Figure~\ref{fig:corr_f} presents the correlation analysis for these 19 users, while Table~\ref{tab:off_f_metrics_online} reports the overall path-based offline metric results for both algorithms on this subset. The results in Table~\ref{tab:off_f_metrics_online} follow the same trends observed in Table~\ref{tab:off_metrics_online}. In the correlation matrix, by controlling for the influence of diversity, and ensuring that both algorithms produced similarly high attribute diversity in their \ac{KG} explanations, the effect of perceived popularity became more pronounced. Notably, the correlation between perceived popularity and persuasiveness increased substantially, reaching a value of 0.72.\looseness=-1

Regarding perceived diversity, its influence on engagement decreased, with the correlation dropping from 0.20 to –0.01. This occurred because, for this subset of users, diversity no longer differentiated \ac{PEM} from ExpLOD, resulting in no observable relationship. These findings indicate that although engagement does not exhibit a strong correlation with diversity overall, it is still influenced by it to some extent.\looseness=-1

When both algorithms produce similarly diverse explanations, the role of the offline metrics in shaping explanation goals becomes more evident. This is reflected in the changes in the correlation values, particularly for ETD, where the correlation shifts from a positive value of around 0.20 in the full participant set to a small negative value in the filtered subset. Likewise, the correlation between ETD and perceived diversity changes from 0.16 to –0.16. Although these values are not statistically significant, they suggest that participants’ perceptions shift as ETD varies. In contrast, SEP remains consistent across both correlation matrices, indicating that users’ perceptions of the explanation goals with respect to the offline popularity metric were not influenced by the change in diversity.\looseness=-1

Another change visible when comparing Figure~\ref{fig:corr_f} with Figure~\ref{fig:corr_all} is the increase in the negative correlation between the diversity of items (MID) and the diversity of attributes (ETD). This can be explained by the fact that as explanations present more and varied attributes, those attributes tend to be less popular and therefore connected to fewer interacted items on the \ac{KG}. However, based on our online experiment and both correlation analyses, we find no clear evidence regarding the effect of interacted items in explanations on explanation goals, and we leave this aspect for future work.\looseness=-1

Consequently, our second major finding from the online experiment is that \emph{engagement is sensitive to the perceived diversity of explanations, whereas transparency, trust, and persuasiveness are influenced by both perceived popularity and perceived diversity. However, developing offline metrics that reliably correlate with explanation goals remains an open research challenge.}\looseness=-1

In Section \ref{sec:offline_results}, two hypotheses were formulated: first, that offline path explanation metrics are associated with user perception of explanation goals; and second, that users can perceive the diversity and popularity of attributes in explanations. Our online results partially confirmed the first hypothesis, showing that perceived differences in popularity and diversity between the explanation algorithms were moderately correlated with online explanation goals, nevertheless, the offline explanation metrics failed to capture correlation with online explanation goals. The second hypothesis was fully supported, as offline metrics for attribute popularity and diversity directly aligned with users’ responses and perceptions in the online experiments.\looseness=-1  

\begin{tcolorbox}[colback=white!10!white,colframe=gray!75!black,title=Answer to \ac{RQ}2: In explanations that connect interacted and recommended items based on shared attributes\text{,} how does the selection of item attributes and interacted items affect explanation goals?]

Based on the results of our offline and online experiments on path-based explanations we observed the findings:\looseness=-1

\begin{enumerate}
    \item \emph{Explanations should be evaluated in both online and offline settings using multiple explanations per user.} Because attribute occurrence across items follows a long-tail distribution, as shown in Section \ref{sec:kg_extraction_analysis}, relying on a single explanation can lead to a biased evaluation. In Table \ref{tab:offresults_n1}, for example, we could not evaluate attribute diversity because only a single recommendation was explained.\looseness=-1
    
    \item \emph{A trade-off exists between the offline explanation metrics of attribute diversity and popularity.} This was highlighted in Table \ref{tab:offresults_n5}, where \ac{PEM} \cite{du2022post}, which achieved higher diversity, exhibited lower attribute popularity compared to its baseline ExpLOD \cite{musto2016explod}. The same pattern appears in Figure~\ref{fig:corr_all}, where the ETD and SEP offline metrics that measure diversity and popularity, respectively, display a negative correlation;\looseness=-1
    
    \item \emph{Users are able to perceive the popularity and diversity of attributes within a set of content explanations.} In our online experiments, participants’ responses regarding perceived popularity and diversity aligned with the offline metrics, as shown in Figure \ref{fig:onresults1} and Figure \ref{fig:onresults2};\looseness=-1
    
    \item We measured correlations between users’ perceived differences in trust, transparency, persuasiveness, and engagement and their perceptions of diversity and popularity in explanations. As shown in Figure~\ref{fig:corr_all}, perceived popularity exhibited moderate correlations with persuasiveness, trust, and engagement across all users when comparing the results of \ac{PEM} and ExpLOD v2. In contrast, diversity showed slightly weaker correlations. However, when considering only users who received diverse explanations from both algorithms (Figure~\ref{fig:corr_f}), the influence of popularity on explanation goals increased, whereas the effect of diversity on engagement diminished, as its effect disappears (correlation~$\approx 0$) when both algorithms provide equally diverse explanations. Consequently, our results suggest that \emph{while engagement is sensitive to the perceived diversity in path explanations, the goals of transparency, trust, and persuasiveness are influenced by users’ perceptions of both diversity and popularity;}\looseness=-1

    \item Regarding correlations among explanation goals, our results align with those reported by \cite{balog2020measuring}, which found that \emph{most goals are moderately correlated, with some pairs showing particularly strong relationships. In particular, we observed the same high correlation between the persuasiveness and trust goals};\looseness=-1

    \item Finally, we also examined the correlations between the offline explanation metrics and the explanation goals. We found that diversity was the only metric that showed any notable correlation with the explanation goals. In addition, the correlations between the offline metrics of diversity and popularity and their corresponding perception-based questions were also low. These results indicate that \emph{current offline metrics do not adequately capture explanation goals or the user-perceived aspects such as familiarity, popularity, and diversity, and that developing metrics capable of doing so remains an open research challenge.} \looseness=-1
\end{enumerate}

\end{tcolorbox}

\section{Limitations and Future Directions}
\label{sec:limits}

There are many open directions for the offline evaluation of explanations in \ac{RS}s. Although differences in explanation path metrics affected user perception of explanation goals, these findings should be interpreted with caution, as the online user study comparing preferences between \ac{PEM} and ExpLOD did not yield statistically significant results. Consequently, an important research topic for future work is the development of new offline metrics that better align with explanation goals.\looseness=-1

In addition, as discussed in Section~\ref{sec:online_results}, the correlation analysis focused on users’ preferences between the two explanation algorithms. Accordingly, Figure~\ref{fig:corr_all} and Figure~\ref{fig:corr_f} present the correlations that illustrate how the offline metrics, together with perceived diversity and perceived popularity, relate to the explanation goals based on users’ comparative Likert-scale judgments of \ac{PEM} and ExpLOD v2. Consequently, although we concluded that users were able to perceive differences captured by the offline metrics—and that our findings align with those of \cite{balog2020measuring}—this conclusion assumes that the quality of the explanations produced by both algorithms was validated in their respective prior studies \cite{musto2016explod, musto2019combining, du2022post}. Furthermore, the phrasing of the questionnaire may also have contributed to the low correlations observed between the offline metrics and the online explanation goals. We intentionally opted for open questions to avoid biasing participants toward a specific interpretation.\looseness=-1

Another limitation of this study is that it relies on only one dataset to compare offline and online metrics from a user perspective. We reported offline results in Appendix \ref{apen:lastfm}; however, we leave the user evaluation of explanations for this domain for future research. Our rapid literature review revealed that the progression of algorithms from \cite{musto2016explod} to \cite{musto2019linked} to \cite{du2022post} is the only one that includes user studies on explanation goals, all of which rely on a dataset from the movie domain. Evaluation in other domains is left for future research. \looseness=-1

In addition, other types of explanation algorithms, such as review explanations, were not validated in this study because of the lack of an algorithm evolution timeline on explanation goals in the literature. However, we argue that path offline explanation metrics can be adapted to this domain by replacing the attributes of the KG with aspects extracted from reviews. We also leave the validation of these metrics to other families of explanation algorithms for future work. \looseness=-1

Offline explanation path metrics suggest that users prefer explanations that balance diversity with the popularity of an item’s attributes. Since current systems often provide explanations through categories, as rows of content sharing an attribute, an interesting research topic would be exploring the relationship between user engagement and changes in explanations over multiple visits. Additionally, the interacted item recency ($LIR$) metric should be evaluated in an online experiment where users assess explanations in a time-dependent scenario with different visits and interactions with the system. Finally, additional research is necessary to clarify the role of interacted items in path-based explanations—assessed here using the MID and TID metrics—and their connection to explanation goals. \looseness=-1

\section{Conclusions}
\label{sec:conclusions}

This study introduces the relationship between explanation path metrics and explanation goals. The explanations of three agnostic \ac{KG} content explanation style algorithms were assessed in offline and online experiments, and the results were compared, considering the diversification and popularity of attributes shown to users in paths that connect interacted items and recommended items. According to the results, users’ perceptions of the explanation goals are influenced by the attributes presented within the set of explanations they receive. \looseness=-1

We also conducted a survey on offline explanation metrics used in \ac{RS}, analyzing more than 100 papers on the subject. Our results showed that, similar to the field of \ac{XAI} in \ac{ML}, explanations in \ac{RS} are often evaluated with anecdotal evidence that passes ``face validity'' \cite{10.1145/3583558}. Additionally, popular offline explanation metrics such as \ac{BLEU} and \ac{ROUGE} do not correlate with user perception of explanations \cite{10.1145/3627043.3659574}. We also identified that research on hybrid and counterfactual explanation styles in \ac{RS}, and new offline metrics that correlate with explanation goals are emerging topics that could be further explored by researchers.\looseness=-1

The main objective of this study is to emphasize the importance of evaluating explanations to identify and address potential biases in explanation algorithms for \ac{RS}. This is crucial because explanations in \ac{RS} are primarily intended for the users. Most work on explanation algorithms in \ac{RS}s focuses on improvements in ranking metrics \cite{nunes2017systematic}, lacking evidence of the explanations' usefulness regarding explanation goals. In this context, similar to accuracy and beyond-accuracy offline metrics for ranking in \ac{RS}s, explanation path metrics have the potential to guide researchers in determining whether an explanation algorithm is ready for online A\slash B testing.

Moreover, this study underscores the need for a thorough evaluation of the explanations in \ac{RS}. When comparing algorithms, it is essential to analyze metrics across large databases and different ranking sizes to ensure the consistency and robustness of the results.\looseness=-1

Finally, the offline path metrics presented in this study assess the distribution of items and attributes in an explanation algorithm. They do not replace or suppress the necessity of a user study; however, they evaluate the algorithmic bias in generating explanations, providing researchers with a tool to analyze the quality of the generated explanations and how the algorithm performs compared to others. They also contribute to develop a state-of-the-art algorithm timeline. As a result, we hope this study can help raise the problem of analyzing, evaluating and creating new offline explanation metrics in \ac{RS}s.\looseness=-1

\section*{Acknowledgments}
This publication has emanated from research conducted with the financial support of Research Ireland under Grant number 12-RC-2289-P2, which is co-funded under the European Regional Development Fund. For the purpose of Open Access, the author has applied a CC BY public copyright license to any Author Accepted Manuscript version arising from this submission. The authors also acknowledge CAPES, CNPq, Fapesp, Fapemig, INCT-TILD-IAR for their funding and support of this research.\looseness=-1

\bibliographystyle{ACM-Reference-Format}
\bibliography{refs}

@article{takami_flanagan_dai_ogata_2023, title={Personality-based tailored explainable recommendation for trustworthy smart learning system in the age of artificial intelligence}, volume={10}, ISSN={2196-7091}, DOI={10.1186/s40561-023-00282-6}, number={1}, journal={Smart Learning Environments}, publisher={Smart Learning Environments}, author={Takami, Kyosuke and Flanagan, Brendan and Dai, Yiling and Ogata, Hiroaki}, year={2023} }

@article{lin_zhang_lin_zeng_zhou_wu_2024, title={Knowledge-aware reasoning with self-supervised reinforcement learning for explainable recommendation in MOOCs}, volume={36}, ISSN={0941-0643}, DOI={10.1007/s00521-023-09257-7}, number={8}, journal={Neural Computing and Applications}, publisher={Neural Computing and Applications}, author={Lin, Yuanguo and Zhang, Wei and Lin, Fan and Zeng, Wenhua and Zhou, Xiuze and Wu, Pengcheng}, year={2024}, pages={4115–4132} }

@inbook{zanon_da_rocha_manzato_2024, title={Model-Agnostic Knowledge Graph Embedding Explanations for Recommender Systems}, ISSN={1865-0929}, DOI={10.1007/978-3-031-63797-1_1}, booktitle={Communications in Computer and Information Science}, publisher={Communications in Computer and Information Science}, author={Zanon, André Levi and Da Rocha, Leonardo Chaves Dutra and Manzato, Marcelo Garcia}, year={2024}, pages={3–27} }

@inbook{samih_ghadi_fennan_2023, title={Knowledge Embeddings for Explainable Recommendation}, ISSN={2367-3370}, DOI={10.1007/978-3-031-28387-1_11}, booktitle={Lecture Notes in Networks and Systems}, publisher={Lecture Notes in Networks and Systems}, author={Samih, Amina and Ghadi, Abderrahim and Fennan, Abdelhadi}, year={2023}, pages={116–126} }

@inbook{li_liu_zhang_kou_liu_qu_2025, title={Integrating User Sentiment and Behavior for Explainable Recommendation}, ISSN={1865-0929}, DOI={10.1007/978-981-96-0055-7_12}, booktitle={Communications in Computer and Information Science}, publisher={Communications in Computer and Information Science}, author={Li, Dong and Liu, Zhicong and Zhang, Qingyu and Kou, Yue and Liu, Tingting and Qu, Haoran}, year={2025}, pages={135–148} }

@article{wen_liu_jing_yu_2024, title={Learning-based counterfactual explanations for recommendation}, volume={67}, ISSN={1674-733X}, DOI={10.1007/s11432-023-3974-2}, number={8}, journal={Science China Information Sciences}, publisher={Science China Information Sciences}, author={Wen, Jingxuan and Liu, Huafeng and Jing, Liping and Yu, Jian}, year={2024} }

@inbook{long_jin_2024, title={Prompt Tuning Models on Sentiment-Aware for Explainable Recommendation}, ISSN={0302-9743}, DOI={10.1007/978-3-031-51671-9_9}, booktitle={Lecture Notes in Computer Science}, publisher={Lecture Notes in Computer Science}, author={Long, Xiuhua and Jin, Ting}, year={2024}, pages={116–132} }

@inbook{jendal_le_lauw_lissandrini_dolog_hose_2024, title={Hypergraphs with Attention on Reviews for Explainable Recommendation}, ISSN={0302-9743}, DOI={10.1007/978-3-031-56027-9_14}, booktitle={Lecture Notes in Computer Science}, publisher={Lecture Notes in Computer Science}, author={Jendal, Theis E. and Le, Trung-Hoang and Lauw, Hady W. and Lissandrini, Matteo and Dolog, Peter and Hose, Katja}, year={2024}, pages={230–246} }

@inbook{zhong_negre_2022, title={Context-Aware Explanations in Recommender Systems}, ISSN={2367-3370}, DOI={10.1007/978-3-030-98531-8_8}, booktitle={Lecture Notes in Networks and Systems}, publisher={Lecture Notes in Networks and Systems}, author={Zhong, Jinfeng and Negre, Elsa}, year={2022}, pages={76–85} }

@article{alizadeh_noughabi_behkamal_zarrinkalam_kahani_2024, title={Persuasive explanations for path reasoning recommendations}, ISSN={0925-9902}, DOI={10.1007/s10844-024-00896-3}, journal={Journal of Intelligent Information Systems}, publisher={Journal of Intelligent Information Systems}, author={Alizadeh Noughabi, Havva and Behkamal, Behshid and Zarrinkalam, Fattane and Kahani, Mohsen}, year={2024} }

@inbook{zhang_zhu_wang_2023, title={Neighborhood Constraints Based Bayesian Personalized Ranking for Explainable Recommendation}, ISSN={0302-9743}, DOI={10.1007/978-3-031-25201-3_12}, booktitle={Lecture Notes in Computer Science}, publisher={Lecture Notes in Computer Science}, author={Zhang, Tingxuan and Zhu, Li and Wang, Jie}, year={2023}, pages={166–173} }

@article{zheng_chen_cao_peng_huang_2024, title={Explainable recommendation based on fusion representation of multi-type feature embedding}, volume={80}, ISSN={0920-8542}, DOI={10.1007/s11227-023-05831-x}, number={8}, journal={The Journal of Supercomputing}, publisher={The Journal of Supercomputing}, author={Zheng, Jianxing and Chen, Sen and Cao, Feng and Peng, Furong and Huang, Mingqing}, year={2024}, pages={10370–10393} }

@article{wang_xie_ding_chen_xiang_2025, title={Reinforced logical reasoning over KGs for interpretable recommendation system}, volume={114}, ISSN={0885-6125}, DOI={10.1007/s10994-024-06646-4}, number={4}, journal={Machine Learning}, publisher={Machine Learning}, author={Wang, Shirui and Xie, Bohan and Ding, Ling and Chen, Jianting and Xiang, Yang}, year={2025} }

@article{sang_yang_zhang_liao_2025, title={A user preference knowledge graph incorporating spatio-temporal transfer features for next POI recommendation}, volume={55}, ISSN={0924-669X}, DOI={10.1007/s10489-025-06290-y}, number={6}, journal={Applied Intelligence}, publisher={Applied Intelligence}, author={Sang, Chun-Yan and Yang, Yang and Zhang, Yi-Bo and Liao, Shi-Gen}, year={2025} }

@INPROCEEDINGS{9811151,
  author={Zarzour, Hafed and Alsmirat, Mohammad and Jararweh, Yaser},
  booktitle={2022 13th International Conference on Information and Communication Systems (ICICS)}, 
  title={Using Deep Learning for Positive Reviews Prediction in Explainable Recommendation Systems}, 
  year={2022},
  volume={},
  number={},
  pages={358-362},
  doi={10.1109/ICICS55353.2022.9811151},
  ISSN={2573-3346},
  month={June},}

@INPROCEEDINGS{8622439,
  author={Suzuki, Takafumi and Oyama, Satoshi and Kurihara, Masahito},
  booktitle={2018 IEEE International Conference on Big Data (Big Data)}, 
  title={Toward Explainable Recommendations: Generating Review Text from Multicriteria Evaluation Data}, 
  year={2018},
  volume={},
  number={},
  pages={3549-3551},
  doi={10.1109/BigData.2018.8622439},
  ISSN={},
  month={Dec},}

@INPROCEEDINGS{10741116,
  author={Tohidi, Nasim and Beheshti, Maedeh},
  booktitle={2024 IEEE International Symposium on Systems Engineering (ISSE)}, 
  title={Enhanced Explanations in Recommendation Systems}, 
  year={2024},
  volume={},
  number={},
  pages={1-5},
  doi={10.1109/ISSE63315.2024.10741116},
  ISSN={2687-8828},
  month={Oct},}

@INPROCEEDINGS{9607106,
  author={Vultureanu-Albişi, Alexandra and Bădică, Costin},
  booktitle={2021 25th International Conference on System Theory, Control and Computing (ICSTCC)}, 
  title={Explainable Collaborative Filtering Recommendations Enriched with Contextual Information}, 
  year={2021},
  volume={},
  number={},
  pages={701-706},
  doi={10.1109/ICSTCC52150.2021.9607106},
  ISSN={2372-1618},
  month={Oct},}

@INPROCEEDINGS{9260076,
  author={Lonjarret, Corentin and Robardet, Céline and Plantevit, Marc and Auburtin, Roch and Atzmueller, Martin},
  booktitle={2020 IEEE 7th International Conference on Data Science and Advanced Analytics (DSAA)}, 
  title={Why Should I Trust This Item? Explaining the Recommendations of any Model}, 
  year={2020},
  volume={},
  number={},
  pages={526-535},
  doi={10.1109/DSAA49011.2020.00067},
  ISSN={},
  month={Oct},}

@INPROCEEDINGS{10776491,
  author={Praseptiawan, Mugi and Muchtarom, M. Fikri Damar and Putri, Nabila Muthia and Pee, Ahmad Naim Che and Zakaria, Mohd Hafiz and Untoro, Meida Cahyo},
  booktitle={2024 11th International Conference on Electrical Engineering, Computer Science and Informatics (EECSI)}, 
  title={Mooc Course Recommendation System Model with Explainable AI (XAI) Using Content Based Filtering Method}, 
  year={2024},
  volume={},
  number={},
  pages={144-147},
  doi={10.1109/EECSI63442.2024.10776491},
  ISSN={},
  month={Sep.},}

@INPROCEEDINGS{9005590,
  author={Suzuki, Takafumi and Oyama, Satoshi and Kurihara, Masahito},
  booktitle={2019 IEEE International Conference on Big Data (Big Data)}, 
  title={Explainable Recommendation Using Review Text and a Knowledge Graph}, 
  year={2019},
  volume={},
  number={},
  pages={4638-4643},
  doi={10.1109/BigData47090.2019.9005590},
  ISSN={},
  month={Dec},}

@INPROCEEDINGS{10260804,
  author={Rani, Neha and Qian, Yadi and Chu, Sharon Lynn},
  booktitle={2023 IEEE International Conference on Advanced Learning Technologies (ICALT)}, 
  title={Explanation for User Trust in Context-Aware Recommender Systems for Search-As-Learning}, 
  year={2023},
  volume={},
  number={},
  pages={47-49},
  doi={10.1109/ICALT58122.2023.00019},
  ISSN={2161-377X},
  month={July},}

@INPROCEEDINGS{10683822,
  author={Vultureanu-Albişi, Alexandra and Murareţu, Ionuţ and Bădică, Costin},
  booktitle={2024 International Conference on INnovations in Intelligent SysTems and Applications (INISTA)}, 
  title={A Trustworthy and Explainable AI Recommender System: Job Domain Case Study}, 
  year={2024},
  volume={},
  number={},
  pages={1-7},
  doi={10.1109/INISTA62901.2024.10683822},
  ISSN={2768-7295},
  month={Sep.},}

@INPROCEEDINGS{9079084,
  author={Zarzour, Hafed and Jararweh, Yaser and Hammad, Mahmoud M. and Al-Smadi, Mohammed},
  booktitle={2020 11th International Conference on Information and Communication Systems (ICICS)}, 
  title={A long short-term memory deep learning framework for explainable recommendation}, 
  year={2020},
  volume={},
  number={},
  pages={233-237},
  doi={10.1109/ICICS49469.2020.239553},
  ISSN={2573-3346},
  month={April},}

@INPROCEEDINGS{9836983,
  author={Walek, Bogdan and Fajmon, Petr},
  booktitle={2022 3rd International Conference on Artificial Intelligence, Robotics and Control (AIRC)}, 
  title={A Recommender System for Recommending Suitable Products in E-shop Using Explanations}, 
  year={2022},
  volume={},
  number={},
  pages={16-20},
  doi={10.1109/AIRC56195.2022.9836983},
  ISSN={},
  month={May},}

@INPROCEEDINGS{10873804,
  author={Xiaolong, Zhou and Shijiao, Han and Zhenze, Li},
  booktitle={2024 21st International Computer Conference on Wavelet Active Media Technology and Information Processing (ICCWAMTIP)}, 
  title={Explainable Recommendation System Based on Aspect-Based Sentiment Analysis}, 
  year={2024},
  volume={},
  number={},
  pages={1-4},
  doi={10.1109/ICCWAMTIP64812.2024.10873804},
  ISSN={2576-8964},
  month={Dec},}

@INPROCEEDINGS{8594883,
  author={Wang, Xiting and Chen, Yiru and Yang, Jie and Wu, Le and Wu, Zhengtao and Xie, Xing},
  booktitle={2018 IEEE International Conference on Data Mining (ICDM)}, 
  title={A Reinforcement Learning Framework for Explainable Recommendation}, 
  year={2018},
  volume={},
  number={},
  pages={587-596},
  doi={10.1109/ICDM.2018.00074},
  ISSN={2374-8486},
  month={Nov},}

@ARTICLE{10048787,
  author={Yang, Zhe-Rui and He, Zhen-Yu and Wang, Chang-Dong and Lai, Jian-Huang and Tian, Zhihong},
  journal={IEEE Transactions on Computational Social Systems}, 
  title={Collaborative Meta-Path Modeling for Explainable Recommendation}, 
  year={2024},
  volume={11},
  number={2},
  pages={1805-1815},
  doi={10.1109/TCSS.2023.3243939},
  ISSN={2329-924X},
  month={April},}

@INPROCEEDINGS{10446052,
  author={Zhang, Jingsen and Bo, Xiaohe and Wang, Chenxi and Dai, Quanyu and Dong, Zhenhua and Tang, Ruiming and Chen, Xu},
  booktitle={ICASSP 2024 - 2024 IEEE International Conference on Acoustics, Speech and Signal Processing (ICASSP)}, 
  title={Active Explainable Recommendation with Limited Labeling Budgets}, 
  year={2024},
  volume={},
  number={},
  pages={5375-5379},
  doi={10.1109/ICASSP48485.2024.10446052},
  ISSN={2379-190X},
  month={April},}

@INPROCEEDINGS{10827288,
  author={Bhatti, Uzair Aslam and Yu, Yang Ke and Mamyrbayev, O.Zh. and Aitkazina, A.A. and Hao, Tang and Zhumazhan, N.O.},
  booktitle={2024 7th International Conference on Pattern Recognition and Artificial Intelligence (PRAI)}, 
  title={Recommendations for Healthcare: An Interpretable Approach Using Deep Learning}, 
  year={2024},
  volume={},
  number={},
  pages={529-535},
  doi={10.1109/PRAI62207.2024.10827288},
  ISSN={},
  month={Aug},}

@INPROCEEDINGS{9482221,
  author={Song, Wei and Wang, Chenglong and Ning, Keqing},
  booktitle={2021 IEEE 4th Advanced Information Management, Communicates, Electronic and Automation Control Conference (IMCEC)}, 
  title={Generate Personalized Explanations for Recommendation based on Keywords}, 
  year={2021},
  volume={4},
  number={},
  pages={51-57},
  doi={10.1109/IMCEC51613.2021.9482221},
  ISSN={2693-2776},
  month={June},}

@INPROCEEDINGS{10658914,
  author={Chun, Hong Wei and Ong, Rongqing Kenneth and Khong, Andy W. H.},
  booktitle={2024 IEEE 67th International Midwest Symposium on Circuits and Systems (MWSCAS)}, 
  title={Reasonable Sense of Direction: Making Course Recommendations Understandable with LLMs}, 
  year={2024},
  volume={},
  number={},
  pages={1408-1412},
  doi={10.1109/MWSCAS60917.2024.10658914},
  ISSN={1558-3899},
  month={Aug},}

@INPROCEEDINGS{10334552,
  author={Chen, Zhanghui and Ai, Xinbo and Guo, Yanjun and Huang, Yitian and Yang, Jing},
  booktitle={2023 IEEE 11th International Conference on Computer Science and Network Technology (ICCSNT)}, 
  title={Explainable Recommendation for Hazard Inspection Reasoning Through Knowledge Graph}, 
  year={2023},
  volume={},
  number={},
  pages={37-42},
  doi={10.1109/ICCSNT58790.2023.10334552},
  ISSN={2690-5892},
  month={Oct},}

@INPROCEEDINGS{10825771,
  author={Turgut, Özlem and Kök, İbrahim and Özdemir, Suat},
  booktitle={2024 IEEE International Conference on Big Data (BigData)}, 
  title={AgroXAI: Explainable AI-Driven Crop Recommendation System for Agriculture 4.0}, 
  year={2024},
  volume={},
  number={},
  pages={7208-7217},
  doi={10.1109/BigData62323.2024.10825771},
  ISSN={2573-2978},
  month={Dec},}

@INPROCEEDINGS{10308154,
  author={Das, Samiran and Chatterjee, Sujoy},
  booktitle={2023 14th International Conference on Computing Communication and Networking Technologies (ICCCNT)}, 
  title={Explainable Machine Learning for Crop Recommendation from Agriculture Sensor Data- a New Paradigm}, 
  year={2023},
  volume={},
  number={},
  pages={1-7},
  doi={10.1109/ICCCNT56998.2023.10308154},
  ISSN={2473-7674},
  month={July},}

@ARTICLE{10742303,
  author={Cao, Yang and Shang, Shuo and Wang, Jun and Zhang, Wei},
  journal={IEEE Transactions on Knowledge and Data Engineering}, 
  title={Explainable Session-Based Recommendation via Path Reasoning}, 
  year={2025},
  volume={37},
  number={1},
  pages={278-290},
  doi={10.1109/TKDE.2024.3486326},
  ISSN={1558-2191},
  month={Jan},}

@ARTICLE{10623784,
  author={Guo, Feipeng and Wang, Zifan},
  journal={IEEE Internet of Things Journal}, 
  title={KEMB-Rec: Knowledge-Enhanced Explainable Multibehavior Recommendation With Graph Contrastive Learning}, 
  year={2025},
  volume={12},
  number={4},
  pages={3563-3576},
  doi={10.1109/JIOT.2024.3439527},
  ISSN={2327-4662},
  month={Feb},}

@INPROCEEDINGS{10884422,
  author={Batmani, Sahar and Moradi, Parham and Heidari, Narges and Jalili, Mahdi},
  booktitle={2024 IEEE International Conference on Data Mining (ICDM)}, 
  title={An Explainable Recommender System by Integrating Graph Neural Networks and User Reviews}, 
  year={2024},
  volume={},
  number={},
  pages={669-674},
  doi={10.1109/ICDM59182.2024.00074},
  ISSN={2374-8486},
  month={Dec},}

@article{SHIMIZU2022107970,
title = {An explainable recommendation framework based on an improved knowledge graph attention network with massive volumes of side information},
journal = {Knowledge-Based Systems},
volume = {239},
pages = {107970},
year = {2022},
issn = {0950-7051},
doi = {https://doi.org/10.1016/j.knosys.2021.107970},
url = {https://www.sciencedirect.com/science/article/pii/S0950705121010959},
author = {Ryotaro Shimizu and Megumi Matsutani and Masayuki Goto}
}

@article{BRUNOT2022102021,
title = {Preference-based and local post-hoc explanations for recommender systems},
journal = {Information Systems},
volume = {108},
pages = {102021},
year = {2022},
issn = {0306-4379},
doi = {https://doi.org/10.1016/j.is.2022.102021},
url = {https://www.sciencedirect.com/science/article/pii/S0306437922000254},
author = {Léo Brunot and Nicolas Canovas and Alexandre Chanson and Nicolas Labroche and Willème Verdeaux}
}

@article{TAO2022109300,
title = {Micro-behaviour with Reinforcement Knowledge-aware Reasoning for Explainable Recommendation},
journal = {Knowledge-Based Systems},
volume = {251},
pages = {109300},
year = {2022},
issn = {0950-7051},
doi = {https://doi.org/10.1016/j.knosys.2022.109300},
url = {https://www.sciencedirect.com/science/article/pii/S0950705122006529},
author = {Shaohua Tao and Runhe Qiu and Bo Xu and Yuan Ping}
}

@article{TAO2021107217,
title = {Multi-modal Knowledge-aware Reinforcement Learning Network for Explainable Recommendation},
journal = {Knowledge-Based Systems},
volume = {227},
pages = {107217},
year = {2021},
issn = {0950-7051},
doi = {https://doi.org/10.1016/j.knosys.2021.107217},
url = {https://www.sciencedirect.com/science/article/pii/S0950705121004792},
author = {Shaohua Tao and Runhe Qiu and Yuan Ping and Hui Ma}
}

@article{YANG2021106687,
title = {Accurate and Explainable Recommendation via Hierarchical Attention Network Oriented Towards Crowd Intelligence},
journal = {Knowledge-Based Systems},
volume = {213},
pages = {106687},
year = {2021},
issn = {0950-7051},
doi = {https://doi.org/10.1016/j.knosys.2020.106687},
url = {https://www.sciencedirect.com/science/article/pii/S0950705120308169},
author = {Chao Yang and Weixin Zhou and Zhiyu Wang and Bin Jiang and Dongsheng Li and Huawei Shen}
}

@article{LIU2020102099,
title = {Dynamic attention-based explainable recommendation with textual and visual fusion},
journal = {Information Processing \& Management},
volume = {57},
number = {6},
pages = {102099},
year = {2020},
issn = {0306-4573},
doi = {https://doi.org/10.1016/j.ipm.2019.102099},
url = {https://www.sciencedirect.com/science/article/pii/S0306457319301761},
author = {Peng Liu and Lemei Zhang and Jon Atle Gulla}
}

@article{LIANG202194,
title = {O3ERS: An explainable recommendation system with online learning, online recommendation, and online explanation},
journal = {Information Sciences},
volume = {562},
pages = {94-115},
year = {2021},
issn = {0020-0255},
doi = {https://doi.org/10.1016/j.ins.2020.12.070},
url = {https://www.sciencedirect.com/science/article/pii/S0020025520312366},
author = {Qianqiao Liang and Xiaolin Zheng and Yan Wang and Mengying Zhu}
}

@article{PAZRUZA2024102497,
title = {Sustainable transparency on recommender systems: Bayesian ranking of images for explainability},
journal = {Information Fusion},
volume = {111},
pages = {102497},
year = {2024},
issn = {1566-2535},
doi = {https://doi.org/10.1016/j.inffus.2024.102497},
url = {https://www.sciencedirect.com/science/article/pii/S1566253524002756},
author = {Jorge Paz-Ruza and Amparo Alonso-Betanzos and Bertha Guijarro-Berdiñas and Brais Cancela and Carlos Eiras-Franco}
}

@article{WEI2023202,
title = {ExpGCN: Review-aware Graph Convolution Network for explainable recommendation},
journal = {Neural Networks},
volume = {157},
pages = {202-215},
year = {2023},
issn = {0893-6080},
doi = {https://doi.org/10.1016/j.neunet.2022.10.014},
url = {https://www.sciencedirect.com/science/article/pii/S0893608022004087},
author = {Tianjun Wei and Tommy W.S. Chow and Jianghong Ma and Mingbo Zhao}
}

@article{KAUR2023100507,
title = {A deep learning knowledge graph neural network for recommender systems},
journal = {Machine Learning with Applications},
volume = {14},
pages = {100507},
year = {2023},
issn = {2666-8270},
doi = {https://doi.org/10.1016/j.mlwa.2023.100507},
url = {https://www.sciencedirect.com/science/article/pii/S2666827023000609},
author = {Gurinder Kaur and Fei Liu and Yi-Ping Phoebe Chen}
}

@article{YANG2020106194,
title = {HAGERec: Hierarchical Attention Graph Convolutional Network Incorporating Knowledge Graph for Explainable Recommendation},
journal = {Knowledge-Based Systems},
volume = {204},
pages = {106194},
year = {2020},
issn = {0950-7051},
doi = {https://doi.org/10.1016/j.knosys.2020.106194},
url = {https://www.sciencedirect.com/science/article/pii/S0950705120304196},
author = {Zuoxi Yang and Shoubin Dong}
}

@article{LI2024112042,
title = {An attention mechanism and residual network based knowledge graph-enhanced recommender system},
journal = {Knowledge-Based Systems},
volume = {299},
pages = {112042},
year = {2024},
issn = {0950-7051},
doi = {https://doi.org/10.1016/j.knosys.2024.112042},
url = {https://www.sciencedirect.com/science/article/pii/S0950705124006762},
author = {Weisheng Li and Hao Zhong and Junming Zhou and Chao Chang and Ronghua Lin and Yong Tang}
}

@article{WANG2020436,
title = {Learning user-item paths for explainable recommendation},
journal = {IFAC-PapersOnLine},
volume = {53},
number = {5},
pages = {436-440},
year = {2020},
note = {3rd IFAC Workshop on Cyber-Physical \& Human Systems CPHS 2020},
issn = {2405-8963},
doi = {https://doi.org/10.1016/j.ifacol.2021.04.119},
url = {https://www.sciencedirect.com/science/article/pii/S2405896321002305},
author = {Tongxuan Wang and Xiaolong Zheng and Saike He and Zhu Zhang and Desheng Dash Wu}
}

@article{GUO2021185,
title = {TAERT: Triple-Attentional Explainable Recommendation with Temporal Convolutional Network},
journal = {Information Sciences},
volume = {567},
pages = {185-200},
year = {2021},
issn = {0020-0255},
doi = {https://doi.org/10.1016/j.ins.2021.03.034},
url = {https://www.sciencedirect.com/science/article/pii/S0020025521002772},
author = {Siyuan Guo and Ying Wang and Hao Yuan and Zeyu Huang and Jianwei Chen and Xin Wang}
}

@article{MARKCHOM2023110258,
title = {Scalable and explainable visually-aware recommender systems},
journal = {Knowledge-Based Systems},
volume = {263},
pages = {110258},
year = {2023},
issn = {0950-7051},
doi = {https://doi.org/10.1016/j.knosys.2023.110258},
url = {https://www.sciencedirect.com/science/article/pii/S0950705123000084},
author = {Thanet Markchom and Huizhi Liang and James Ferryman}
}

@article{HAO2025113113,
title = {IReGNN: Implicit review-enhanced graph neural network for explainable recommendation},
journal = {Knowledge-Based Systems},
volume = {311},
pages = {113113},
year = {2025},
issn = {0950-7051},
doi = {https://doi.org/10.1016/j.knosys.2025.113113},
url = {https://www.sciencedirect.com/science/article/pii/S0950705125001601},
author = {Qingbo Hao and Chundong Wang and Yingyuan Xiao and Wenguang Zheng}
}

@article{LIU2025113217,
title = {Semantic relation-aware graph attention network with noise augmented layer-wise contrastive learning for recommendation},
journal = {Knowledge-Based Systems},
volume = {314},
pages = {113217},
year = {2025},
issn = {0950-7051},
doi = {https://doi.org/10.1016/j.knosys.2025.113217},
url = {https://www.sciencedirect.com/science/article/pii/S0950705125002643},
author = {Jianfang Liu and Wei Wang and Baolin Yi and Huanyu Zhang and Xiaoxuan Shen}
}

@article{WU2025129780,
title = {Cross-modal feature symbiosis for personalized meta-path generation in heterogeneous networks},
journal = {Neurocomputing},
volume = {633},
pages = {129780},
year = {2025},
issn = {0925-2312},
doi = {https://doi.org/10.1016/j.neucom.2025.129780},
url = {https://www.sciencedirect.com/science/article/pii/S0925231225004527},
author = {Xiaotong Wu and Liqing Qiu and Weidong Zhao}
}

@inproceedings{10.1145/3357384.3357925,
author = {Song, Weiping and Shi, Chence and Xiao, Zhiping and Duan, Zhijian and Xu, Yewen and Zhang, Ming and Tang, Jian},
title = {AutoInt: Automatic Feature Interaction Learning via Self-Attentive Neural Networks},
year = {2019},
isbn = {9781450369763},
publisher = {Association for Computing Machinery},
address = {New York, NY, USA},
url = {https://doi.org/10.1145/3357384.3357925},
doi = {10.1145/3357384.3357925},
booktitle = {Proceedings of the 28th ACM International Conference on Information and Knowledge Management},
pages = {1161–1170},
numpages = {10},
location = {Beijing, China},
series = {CIKM '19}
}

@inproceedings{10.1145/3219819.3219965,
author = {Hu, Binbin and Shi, Chuan and Zhao, Wayne Xin and Yu, Philip S.},
title = {Leveraging Meta-path based Context for Top- N Recommendation with A Neural Co-Attention Model},
year = {2018},
isbn = {9781450355520},
publisher = {Association for Computing Machinery},
address = {New York, NY, USA},
url = {https://doi.org/10.1145/3219819.3219965},
doi = {10.1145/3219819.3219965},
booktitle = {Proceedings of the 24th ACM SIGKDD International Conference on Knowledge Discovery \&amp; Data Mining},
pages = {1531–1540},
numpages = {10},
location = {London, United Kingdom},
series = {KDD '18}
}

@inproceedings{10.1145/3397271.3401137,
author = {Wang, Xiang and Jin, Hongye and Zhang, An and He, Xiangnan and Xu, Tong and Chua, Tat-Seng},
title = {Disentangled Graph Collaborative Filtering},
year = {2020},
isbn = {9781450380164},
publisher = {Association for Computing Machinery},
address = {New York, NY, USA},
url = {https://doi.org/10.1145/3397271.3401137},
doi = {10.1145/3397271.3401137},
booktitle = {Proceedings of the 43rd International ACM SIGIR Conference on Research and Development in Information Retrieval},
pages = {1001–1010},
numpages = {10},
location = {Virtual Event, China},
series = {SIGIR '20}
}

@inproceedings{10.1145/3343031.3351034,
author = {Wei, Yinwei and Wang, Xiang and Nie, Liqiang and He, Xiangnan and Hong, Richang and Chua, Tat-Seng},
title = {MMGCN: Multi-modal Graph Convolution Network for Personalized Recommendation of Micro-video},
year = {2019},
isbn = {9781450368896},
publisher = {Association for Computing Machinery},
address = {New York, NY, USA},
url = {https://doi.org/10.1145/3343031.3351034},
doi = {10.1145/3343031.3351034},
booktitle = {Proceedings of the 27th ACM International Conference on Multimedia},
pages = {1437–1445},
numpages = {9},
location = {Nice, France},
series = {MM '19}
}

@inproceedings{10.1145/2806416.2806504,
author = {He, Xiangnan and Chen, Tao and Kan, Min-Yen and Chen, Xiao},
title = {TriRank: Review-aware Explainable Recommendation by Modeling Aspects},
year = {2015},
isbn = {9781450337946},
publisher = {Association for Computing Machinery},
address = {New York, NY, USA},
url = {https://doi.org/10.1145/2806416.2806504},
doi = {10.1145/2806416.2806504},
booktitle = {Proceedings of the 24th ACM International on Conference on Information and Knowledge Management},
pages = {1661–1670},
numpages = {10},
location = {Melbourne, Australia},
series = {CIKM '15}
}

@inproceedings{10.1145/3159652.3159668,
author = {Chen, Xu and Xu, Hongteng and Zhang, Yongfeng and Tang, Jiaxi and Cao, Yixin and Qin, Zheng and Zha, Hongyuan},
title = {Sequential Recommendation with User Memory Networks},
year = {2018},
isbn = {9781450355810},
publisher = {Association for Computing Machinery},
address = {New York, NY, USA},
url = {https://doi.org/10.1145/3159652.3159668},
doi = {10.1145/3159652.3159668},
booktitle = {Proceedings of the Eleventh ACM International Conference on Web Search and Data Mining},
pages = {108–116},
numpages = {9},
location = {Marina Del Rey, CA, USA},
series = {WSDM '18}
}

@inproceedings{10.1145/3109859.3109890,
author = {Seo, Sungyong and Huang, Jing and Yang, Hao and Liu, Yan},
title = {Interpretable Convolutional Neural Networks with Dual Local and Global Attention for Review Rating Prediction},
year = {2017},
isbn = {9781450346528},
publisher = {Association for Computing Machinery},
address = {New York, NY, USA},
url = {https://doi.org/10.1145/3109859.3109890},
doi = {10.1145/3109859.3109890},
booktitle = {Proceedings of the Eleventh ACM Conference on Recommender Systems},
pages = {297–305},
numpages = {9},
location = {Como, Italy},
series = {RecSys '17}
}

@inproceedings{10.1145/3178876.3186154,
author = {Tay, Yi and Anh Tuan, Luu and Hui, Siu Cheung},
title = {Latent Relational Metric Learning via Memory-based Attention for Collaborative Ranking},
year = {2018},
isbn = {9781450356398},
publisher = {International World Wide Web Conferences Steering Committee},
address = {Republic and Canton of Geneva, CHE},
url = {https://doi.org/10.1145/3178876.3186154},
doi = {10.1145/3178876.3186154},
booktitle = {Proceedings of the 2018 World Wide Web Conference},
pages = {729–739},
numpages = {11},
location = {Lyon, France},
series = {WWW '18}
}

@inproceedings{10.1145/3631700.3665226,
author = {Afreen, Neda and Balloccu, Giacomo and Boratto, Ludovico and Fenu, Gianni and Malloci, Francesca Maridina and Marras, Mirko and Martis, Andrea Giovanni},
title = {Learner-centered Ontology for Explainable Educational Recommendation},
year = {2024},
isbn = {9798400704666},
publisher = {Association for Computing Machinery},
address = {New York, NY, USA},
url = {https://doi.org/10.1145/3631700.3665226},
doi = {10.1145/3631700.3665226},
booktitle = {Adjunct Proceedings of the 32nd ACM Conference on User Modeling, Adaptation and Personalization},
pages = {567–575},
numpages = {9},
location = {Cagliari, Italy},
series = {UMAP Adjunct '24}
}

@inproceedings{10.1145/3637528.3671781,
author = {Zhang, Jingsen and Tang, Jiakai and Chen, Xu and Yu, Wenhui and Hu, Lantao and Jiang, Peng and Li, Han},
title = {Natural Language Explainable Recommendation with Robustness Enhancement},
year = {2024},
isbn = {9798400704901},
publisher = {Association for Computing Machinery},
address = {New York, NY, USA},
url = {https://doi.org/10.1145/3637528.3671781},
doi = {10.1145/3637528.3671781},
booktitle = {Proceedings of the 30th ACM SIGKDD Conference on Knowledge Discovery and Data Mining},
pages = {4203–4212},
numpages = {10},
location = {Barcelona, Spain},
series = {KDD '24}
}

@inproceedings{10.1145/3543507.3583260,
author = {Zhang, Jingsen and Chen, Xu and Tang, Jiakai and Shao, Weiqi and Dai, Quanyu and Dong, Zhenhua and Zhang, Rui},
title = {Recommendation with Causality enhanced Natural Language Explanations},
year = {2023},
isbn = {9781450394161},
publisher = {Association for Computing Machinery},
address = {New York, NY, USA},
url = {https://doi.org/10.1145/3543507.3583260},
doi = {10.1145/3543507.3583260},
booktitle = {Proceedings of the ACM Web Conference 2023},
pages = {876–886},
numpages = {11},
location = {Austin, TX, USA},
series = {WWW '23}
}

@inproceedings{10.1145/3624918.3625331,
author = {Yu, Yi and Sugiyama, Kazunari and Jatowt, Adam},
title = {AdaReX: Cross-Domain, Adaptive, and Explainable Recommender System},
year = {2023},
isbn = {9798400704086},
publisher = {Association for Computing Machinery},
address = {New York, NY, USA},
url = {https://doi.org/10.1145/3624918.3625331},
doi = {10.1145/3624918.3625331},
booktitle = {Proceedings of the Annual International ACM SIGIR Conference on Research and Development in Information Retrieval in the Asia Pacific Region},
pages = {272–281},
numpages = {10},
location = {Beijing, China},
series = {SIGIR-AP '23}
}

@inproceedings{10.1145/3604915.3609491,
author = {B\"{o}lz, Felix and Nurbakova, Diana and Calabretto, Sylvie and Gerl, Armin and Brunie, Lionel and Kosch, Harald},
title = {HUMMUS: A Linked, Healthiness-Aware, User-centered and Argument-Enabling Recipe Data Set for Recommendation},
year = {2023},
isbn = {9798400702419},
publisher = {Association for Computing Machinery},
address = {New York, NY, USA},
url = {https://doi.org/10.1145/3604915.3609491},
doi = {10.1145/3604915.3609491},
booktitle = {Proceedings of the 17th ACM Conference on Recommender Systems},
pages = {1–11},
numpages = {11},
location = {Singapore, Singapore},
series = {RecSys '23}
}

@inproceedings{10.1145/3636555.3636898,
author = {Frej, Jibril and Shah, Neel and Knezevic, Marta and Nazaretsky, Tanya and K\"{a}ser, Tanja},
title = {Finding Paths for Explainable MOOC Recommendation: A Learner Perspective},
year = {2024},
isbn = {9798400716188},
publisher = {Association for Computing Machinery},
address = {New York, NY, USA},
url = {https://doi.org/10.1145/3636555.3636898},
doi = {10.1145/3636555.3636898},
booktitle = {Proceedings of the 14th Learning Analytics and Knowledge Conference},
pages = {426–437},
numpages = {12},
location = {Kyoto, Japan},
series = {LAK '24}
}

@inproceedings{10.1145/3616855.3635855,
author = {Liu, Xu and Yu, Tong and Xie, Kaige and Wu, Junda and Li, Shuai},
title = {Interact with the Explanations: Causal Debiased Explainable Recommendation System},
year = {2024},
isbn = {9798400703713},
publisher = {Association for Computing Machinery},
address = {New York, NY, USA},
url = {https://doi.org/10.1145/3616855.3635855},
doi = {10.1145/3616855.3635855},
booktitle = {Proceedings of the 17th ACM International Conference on Web Search and Data Mining},
pages = {472–481},
numpages = {10},
location = {Merida, Mexico},
series = {WSDM '24}
}

@book{hinkle2003applied,
  title={Applied statistics for the behavioral sciences},
  author={Hinkle, Dennis E and Wiersma, William and Jurs, Stephen G and others},
  volume={663},
  year={2003},
  publisher={Houghton Mifflin Boston}
}

@inproceedings{balog2020measuring,
    author = {Balog, Krisztian and Radlinski, Filip},
    title = {Measuring Recommendation Explanation Quality: The Conflicting Goals of Explanations},
    year = {2020},
    isbn = {9781450380164},
    publisher = {Association for Computing Machinery},
    address = {New York, NY, USA},
    url = {https://doi.org/10.1145/3397271.3401032},
    doi = {10.1145/3397271.3401032},
    booktitle = {Proceedings of the 43rd International ACM SIGIR Conference on Research and Development in Information Retrieval},
    pages = {329–338},
    numpages = {10},
    location = {Virtual Event, China},
    series = {SIGIR '20}
}

@Inbook{Wilcoxon1992,
author="Wilcoxon, Frank",
editor="Kotz, Samuel
and Johnson, Norman L.",
title="Individual Comparisons by Ranking Methods",
bookTitle="Breakthroughs in Statistics: Methodology and Distribution",
year="1992",
publisher="Springer New York",
address="New York, NY",
pages="196--202",
isbn="978-1-4612-4380-9",
doi="10.1007/978-1-4612-4380-9_16",
url="https://doi.org/10.1007/978-1-4612-4380-9_16"
}

@inproceedings{10.1145/3490099.3511104,
author = {Marinho, Leandro Balby and da Costa, J\'{u}lio Barreto Guedes and Parra, Denis and Santos, Rodrygo L. T.},
title = {Similarity-Based Explanations meet Matrix Factorization via Structure-Preserving Embeddings},
year = {2022},
isbn = {9781450391443},
publisher = {Association for Computing Machinery},
address = {New York, NY, USA},
url = {https://doi.org/10.1145/3490099.3511104},
doi = {10.1145/3490099.3511104},
booktitle = {Proceedings of the 27th International Conference on Intelligent User Interfaces},
pages = {782–793},
numpages = {12},
location = {Helsinki, Finland},
series = {IUI '22}
}

@article{papadimitriou2012generalized,
  title={A generalized taxonomy of explanations styles for traditional and social recommender systems},
  author={Papadimitriou, Alexis and Symeonidis, Panagiotis and Manolopoulos, Yannis},
  journal={Data mining and knowledge discovery},
  volume={24},
  pages={555--583},
  year={2012},
  publisher={Springer},
  doi={10.1007/s10618-011-0215-0},
  url = {https://doi.org/10.1007/s10618-011-0215-0},
}

@inproceedings{bilgic2005explaining,
  title={Explaining recommendations: Satisfaction vs. promotion},
  author={Bilgic, Mustafa and Mooney, Raymond J},
  booktitle={Beyond personalization workshop, IUI},
  volume={5},
  pages={153},
  year={2005}
}

@article{tintarev2015explaining,
  title={Explaining recommendations: Design and evaluation},
  author={Tintarev, Nava and Masthoff, Judith},
  journal={Recommender systems handbook},
  pages={353--382},
  year={2015},
  publisher={Springer}
}

@inproceedings{melchiorre2022protomf,
  title={ProtoMF: Prototype-based Matrix Factorization for Effective and Explainable Recommendations},
  author={Melchiorre, Alessandro B and Rekabsaz, Navid and Ganh{\"o}r, Christian and Schedl, Markus},
  booktitle={Proceedings of the 16th ACM Conference on Recommender Systems},
  pages={246--256},
  year={2022}
}

@inproceedings{zhao2020leveraging,
  title={Leveraging demonstrations for reinforcement recommendation reasoning over knowledge graphs},
  author={Zhao, Kangzhi and Wang, Xiting and Zhang, Yuren and Zhao, Li and Liu, Zheng and Xing, Chunxiao and Xie, Xing},
  booktitle={Proceedings of the 43rd international ACM SIGIR conference on research and development in information retrieval},
  pages={239--248},
  year={2020}
}

@article{balloccu2023reinforcement,
  title={Reinforcement recommendation reasoning through knowledge graphs for explanation path quality},
  author={Balloccu, Giacomo and Boratto, Ludovico and Fenu, Gianni and Marras, Mirko},
  journal={Knowledge-Based Systems},
  volume={260},
  pages={110098},
  year={2023},
  publisher={Elsevier}
}

@inproceedings{balloccu2022post,
  author = {Balloccu, Giacomo and Boratto, Ludovico and Fenu, Gianni and Marras, Mirko},
    title = {Post Processing Recommender Systems with Knowledge Graphs for Recency, Popularity, and Diversity of Explanations},
    year = {2022},
    isbn = {9781450387323},
    publisher = {Association for Computing Machinery},
    address = {New York, NY, USA},
    url = {https://doi.org/10.1145/3477495.3532041},
    doi = {10.1145/3477495.3532041},
    booktitle = {Proceedings of the 45th International ACM SIGIR Conference on Research and Development in Information Retrieval},
    pages = {646–656},
    numpages = {11},
    location = {Madrid, Spain},
    series = {SIGIR '22}
}

@inproceedings{tai2021user,
  title={User-centric path reasoning towards explainable recommendation},
  author={Tai, Chang-You and Huang, Liang-Ying and Huang, Chien-Kun and Ku, Lun-Wei},
  booktitle={Proceedings of the 44th International ACM SIGIR Conference on Research and Development in Information Retrieval},
  pages={879--889},
  year={2021}
}

@inproceedings{musto2019combining,
  title={Combining text summarization and aspect-based sentiment analysis of users' reviews to justify recommendations},
  author={Musto, Cataldo and Rossiello, Gaetano and de Gemmis, Marco and Lops, Pasquale and Semeraro, Giovanni},
  booktitle={Proceedings of the 13th ACM conference on recommender systems},
  pages={383--387},
  year={2019}
}

@inproceedings{hada2021rexplug,
  author = {Hada, Deepesh V. and M., Vijaikumar and Shevade, Shirish K.},
title = {ReXPlug: Explainable Recommendation using Plug-and-Play Language Model},
year = {2021},
isbn = {9781450380379},
publisher = {Association for Computing Machinery},
address = {New York, NY, USA},
url = {https://doi.org/10.1145/3404835.3462939},
doi = {10.1145/3404835.3462939},
booktitle = {Proceedings of the 44th International ACM SIGIR Conference on Research and Development in Information Retrieval},
pages = {81–91},
numpages = {11},
location = {Virtual Event, Canada},
series = {SIGIR '21}
}

@inproceedings{musto2016explod,
  author = {Musto, Cataldo and Narducci, Fedelucio and Lops, Pasquale and De Gemmis, Marco and Semeraro, Giovanni},
title = {ExpLOD: A Framework for Explaining Recommendations based on the Linked Open Data Cloud},
year = {2016},
isbn = {9781450340359},
publisher = {Association for Computing Machinery},
address = {New York, NY, USA},
url = {https://doi.org/10.1145/2959100.2959173},
doi = {10.1145/2959100.2959173},
booktitle = {Proceedings of the 10th ACM Conference on Recommender Systems},
pages = {151–154},
numpages = {4},
location = {Boston, Massachusetts, USA},
series = {RecSys '16}
}

@article{musto2019linked,
  title = {Linked open data-based explanations for transparent recommender systems},
journal = {International Journal of Human-Computer Studies},
volume = {121},
pages = {93-107},
year = {2019},
note = {Advances in Computer-Human Interaction for Recommender Systems},
issn = {1071-5819},
doi = {https://doi.org/10.1016/j.ijhcs.2018.03.003},
url = {https://www.sciencedirect.com/science/article/pii/S1071581918300946},
author = {Cataldo Musto and Fedelucio Narducci and Pasquale Lops and Marco {de Gemmis} and Giovanni Semeraro}
}

@article{du2022post,
  title = {Post-hoc recommendation explanations through an efficient exploitation of the DBpedia category hierarchy},
journal = {Knowledge-Based Systems},
volume = {245},
pages = {108560},
year = {2022},
issn = {0950-7051},
doi = {https://doi.org/10.1016/j.knosys.2022.108560},
url = {https://www.sciencedirect.com/science/article/pii/S0950705122002490},
author = {Yu Du and Sylvie Ranwez and Nicolas Sutton-Charani and Vincent Ranwez}
}

@inproceedings{tan2021contefactual,
author = {Tan, Juntao and Xu, Shuyuan and Ge, Yingqiang and Li, Yunqi and Chen, Xu and Zhang, Yongfeng},
title = {Counterfactual Explainable Recommendation},
year = {2021},
isbn = {9781450384469},
publisher = {Association for Computing Machinery},
address = {New York, NY, USA},
url = {https://doi.org/10.1145/3459637.3482420},
doi = {10.1145/3459637.3482420},
booktitle = {Proceedings of the 30th ACM International Conference on Information \& Knowledge Management},
pages = {1784–1793},
numpages = {10},
location = {Virtual Event, Queensland, Australia},
series = {CIKM '21}
}

@article{
    coba_confalonieri_zanker_2022, 
    title={RecoXplainer: A Library for Development and Offline Evaluation of Explainable Recommender Systems}, 
    volume={17}, 
    SSN={1556-603X}, 
    DOI={10.1109/mci.2021.3129958}, 
    number={1}, 
    journal={IEEE Computational Intelligence Magazine}, publisher={IEEE Computational Intelligence Magazine}, 
    author={Coba, Ludovik and Confalonieri, Roberto and Zanker, Markus}, year={2022}, pages={46–58} 
}

@article{adomavicius2011improving,
    author={Adomavicius, Gediminas and Kwon, YoungOk},
  journal={IEEE Transactions on Knowledge and Data Engineering}, 
  title={Improving Aggregate Recommendation Diversity Using Ranking-Based Techniques}, 
  year={2012},
  volume={24},
  number={5},
  pages={896-911},
  doi={10.1109/TKDE.2011.15}
}

@article{harper2015movielens,
  title={The movielens datasets: History and context},
  author={Harper, F Maxwell and Konstan, Joseph A},
  journal={Acm transactions on interactive intelligent systems (tiis)},
  volume={5},
  number={4},
  pages={1--19},
  year={2015},
  publisher={Acm New York, NY, USA}
}

@inproceedings{pillai2019comparing,
  title={Comparing DBpedia, Wikidata, and YAGO for web information retrieval},
  author={Pillai, Sini Govinda and Soon, Lay-Ki and Haw, Su-Cheng},
  booktitle={Intelligent and Interactive Computing: Proceedings of IIC 2018},
  pages={525--535},
  year={2019},
  organization={Springer}
}

@article{ferrari2021troubling,
  title={A troubling analysis of reproducibility and progress in recommender systems research},
  author={Ferrari Dacrema, Maurizio and Boglio, Simone and Cremonesi, Paolo and Jannach, Dietmar},
  journal={ACM Transactions on Information Systems (TOIS)},
  volume={39},
  number={2},
  pages={1--49},
  year={2021},
  publisher={ACM New York, NY, USA}
}

@inproceedings{da2018case,
  author = {da Costa, Arthur and Fressato, Eduardo and Neto, Fernando and Manzato, Marcelo and Campello, Ricardo},
title = {Case recommender: a flexible and extensible python framework for recommender systems},
year = {2018},
isbn = {9781450359016},
publisher = {Association for Computing Machinery},
address = {New York, NY, USA},
url = {https://doi.org/10.1145/3240323.3241611},
doi = {10.1145/3240323.3241611},
booktitle = {Proceedings of the 12th ACM Conference on Recommender Systems},
pages = {494–495},
numpages = {2},
location = {Vancouver, British Columbia, Canada},
series = {RecSys '18}
}

@inproceedings{He2017,
author = {He, Xiangnan and Liao, Lizi and Zhang, Hanwang and Nie, Liqiang and Hu, Xia and Chua, Tat-Seng},
title = {Neural Collaborative Filtering},
year = {2017},
isbn = {9781450349130},
publisher = {International World Wide Web Conferences Steering Committee},
address = {Republic and Canton of Geneva, CHE},
url = {https://doi.org/10.1145/3038912.3052569},
doi = {10.1145/3038912.3052569},
booktitle = {Proceedings of the 26th International Conference on World Wide Web},
pages = {173–182},
numpages = {10},
location = {Perth, Australia},
series = {WWW '17}
}

@inproceedings{Rendle2009,
author = {Rendle, Steffen and Freudenthaler, Christoph and Gantner, Zeno and Schmidt-Thieme, Lars},
title = {BPR: Bayesian Personalized Ranking from Implicit Feedback},
year = {2009},
isbn = {9780974903958},
publisher = {AUAI Press},
address = {Arlington, Virginia, USA},
booktitle = {Proceedings of the Twenty-Fifth Conference on Uncertainty in Artificial Intelligence},
pages = {452–461},
numpages = {10},
location = {Montreal, Quebec, Canada},
series = {UAI '09}
}

@inproceedings{Cremonesi2010,
author = {Cremonesi, Paolo and Koren, Yehuda and Turrin, Roberto},
title = {Performance of Recommender Algorithms on Top-n Recommendation Tasks},
year = {2010},
isbn = {9781605589060},
publisher = {Association for Computing Machinery},
address = {New York, NY, USA},
url = {https://doi.org/10.1145/1864708.1864721},
doi = {10.1145/1864708.1864721},
booktitle = {Proceedings of the Fourth ACM Conference on Recommender Systems},
pages = {39–46},
numpages = {8},
location = {Barcelona, Spain},
series = {RecSys '10}
}

@inproceedings{Resnick1994,
author = {Resnick, Paul and Iacovou, Neophytos and Suchak, Mitesh and Bergstrom, Peter and Riedl, John},
title = {GroupLens: An Open Architecture for Collaborative Filtering of Netnews},
year = {1994},
isbn = {0897916891},
publisher = {Association for Computing Machinery},
address = {New York, NY, USA},
url = {https://doi.org/10.1145/192844.192905},
doi = {10.1145/192844.192905},
booktitle = {Proceedings of the 1994 ACM Conference on Computer Supported Cooperative Work},
pages = {175–186},
numpages = {12},
location = {Chapel Hill, North Carolina, USA},
series = {CSCW '94}
}

@article{knijnenburg2015evaluating,
  title={Evaluating recommender systems with user experiments},
  author={Knijnenburg, Bart P and Willemsen, Martijn C},
  journal={Recommender systems handbook},
  pages={309--352},
  year={2015},
  publisher={Springer}
}

@inproceedings{rashid2002getting,
  title={Getting to know you: learning new user preferences in recommender systems},
  author={Rashid, Al Mamunur and Albert, Istvan and Cosley, Dan and Lam, Shyong K and McNee, Sean M and Konstan, Joseph A and Riedl, John},
  booktitle={Proceedings of the 7th international conference on Intelligent user interfaces},
  pages={127--134},
  year={2002}
}

@article{zhang2020explainable,
title={Explainable Recommendation: A Survey and New Perspectives},
volume={14},
ISSN={1554-0669}, 
DOI={10.1561/1500000066}, 
number={1}, 
journal={Foundations and Trends® in Information Retrieval}, 
publisher={Foundations and Trends® in Information Retrieval},
author={Zhang, Yongfeng and Chen, Xu}, 
year={2020}, 
pages={1–101} 
}

@inproceedings{kouki2019personalized,
author = {Kouki, Pigi and Schaffer, James and Pujara, Jay and O'Donovan, John and Getoor, Lise},
title = {Personalized explanations for hybrid recommender systems},
year = {2019},
isbn = {9781450362726},
publisher = {Association for Computing Machinery},
address = {New York, NY, USA},
url = {https://doi.org/10.1145/3301275.3302306},
doi = {10.1145/3301275.3302306},
booktitle = {Proceedings of the 24th International Conference on Intelligent User Interfaces},
pages = {379–390},
numpages = {12},
location = {Marina del Ray, California},
series = {IUI '19}
}

@article{10.1145/3716394,
author = {Wardatzky, Kathrin and Inel, Oana and Rossetto, Luca and Bernstein, Abraham},
title = {Whom do Explanations Serve? A Systematic Literature Survey of User Characteristics in Explainable Recommender Systems Evaluation},
year = {2025},
publisher = {Association for Computing Machinery},
address = {New York, NY, USA},
url = {https://doi.org/10.1145/3716394},
doi = {10.1145/3716394},
note = {Just Accepted},
journal = {ACM Trans. Recomm. Syst.},
month = feb
}

@inproceedings{10.1145/3627043.3659574,
author = {Manzoor, Ahtsham and Ziegler, Samuel C. and Garcia, Klaus Maria. Pirker and Jannach, Dietmar},
title = {ChatGPT as a Conversational Recommender System: A User-Centric Analysis},
year = {2024},
isbn = {9798400704338},
publisher = {Association for Computing Machinery},
address = {New York, NY, USA},
url = {https://doi.org/10.1145/3627043.3659574},
doi = {10.1145/3627043.3659574},
booktitle = {Proceedings of the 32nd ACM Conference on User Modeling, Adaptation and Personalization},
pages = {267–272},
numpages = {6},
location = {Cagliari, Italy},
series = {UMAP '24}
}

@inproceedings{10.1145/2959100.2959186,
author = {Jannach, Dietmar and Adomavicius, Gediminas},
title = {Recommendations with a Purpose},
year = {2016},
isbn = {9781450340359},
publisher = {Association for Computing Machinery},
address = {New York, NY, USA},
url = {https://doi.org/10.1145/2959100.2959186},
doi = {10.1145/2959100.2959186},
booktitle = {Proceedings of the 10th ACM Conference on Recommender Systems},
pages = {7–10},
numpages = {4},
location = {Boston, Massachusetts, USA},
series = {RecSys '16}
}

@article{nunes2017systematic,
title={A systematic review and taxonomy of explanations in decision support and recommender systems}, volume={27}, ISSN={0924-1868}, DOI={10.1007/s11257-017-9195-0}, number={3-5}, journal={User Modeling and User-Adapted Interaction}, publisher={User Modeling and User-Adapted Interaction}, author={Nunes, Ingrid and Jannach, Dietmar}, year={2017}, pages={393–444} }

@article{steck2021deep,
  title={Deep learning for recommender systems: A Netflix case study},
  author={Steck, Harald and Baltrunas, Linas and Elahi, Ehtsham and Liang, Dawen and Raimond, Yves and Basilico, Justin},
  journal={AI Magazine},
  volume={42},
  number={3},
  pages={7--18},
  year={2021}
}

@misc{alvino2015learning,
  title={Learning a Personalized Homepage-Netflix TechBlog},
  author={Alvino, C and Basilico, J},
  year={2015},
  publisher={Medium. https://netflixtechblog. com/learning-a-personalized-homepage~…}
}

@inproceedings{Cantador2011,
      author = {Cantador, Iv\'{a}n and Brusilovsky, Peter and Kuflik, Tsvi},
      title = {2nd Workshop on Information Heterogeneity and Fusion in Recommender Systems (HetRec 2011)},
      booktitle = {Proceedings of the 5th ACM conference on Recommender systems},
      series = {RecSys 2011},
      year = {2011},
      location = {Chicago, IL, USA},
      publisher = {ACM},
      address = {New York, NY, USA},
}

@inproceedings{zanon2024model,
  title={Model-Agnostic Knowledge Graph Embedding Explanations for Recommender Systems},
  author={Zanon, Andr{\'e} Levi and da Rocha, Leonardo Chaves Dutra and Manzato, Marcelo Garcia},
  booktitle={World Conference on Explainable Artificial Intelligence},
  pages={3--27},
  year={2024},
  organization={Springer}
}

@inproceedings{10.1145/2939672.2939778,
author = {Ribeiro, Marco Tulio and Singh, Sameer and Guestrin, Carlos},
title = {"Why Should I Trust You?": Explaining the Predictions of Any Classifier},
year = {2016},
isbn = {9781450342322},
publisher = {Association for Computing Machinery},
address = {New York, NY, USA},
url = {https://doi.org/10.1145/2939672.2939778},
doi = {10.1145/2939672.2939778},
booktitle = {Proceedings of the 22nd ACM SIGKDD International Conference on Knowledge Discovery and Data Mining},
pages = {1135–1144},
numpages = {10},
location = {San Francisco, California, USA},
series = {KDD '16}
}

@inproceedings{NIPS2017_8a20a862,
 author = {Lundberg, Scott M and Lee, Su-In},
 booktitle = {Advances in Neural Information Processing Systems},
 editor = {I. Guyon and U. Von Luxburg and S. Bengio and H. Wallach and R. Fergus and S. Vishwanathan and R. Garnett},
 pages = {},
 publisher = {Curran Associates, Inc.},
 title = {A Unified Approach to Interpreting Model Predictions},
 url = {https://proceedings.neurips.cc/paper_files/paper/2017/file/8a20a8621978632d76c43dfd28b67767-Paper.pdf},
 volume = {30},
 year = {2017}
}

@article{kitchenham2009systematic,
  title={Systematic literature reviews in software engineering--a systematic literature review},
  author={Kitchenham, Barbara and Brereton, O Pearl and Budgen, David and Turner, Mark and Bailey, John and Linkman, Stephen},
  journal={Information and software technology},
  volume={51},
  number={1},
  pages={7--15},
  year={2009},
  publisher={Elsevier}
}

@article{zhang2024large,
  title={Large Language Models as Evaluators for Recommendation Explanations},
  author={Zhang, Xiaoyu and Li, Yishan and Wang, Jiayin and Sun, Bowen and Ma, Weizhi and Sun, Peijie and Zhang, Min},
  journal={arXiv preprint arXiv:2406.03248},
  year={2024}
}

@inproceedings{Wen2022,
author = {Wen, Bingbing and Feng, Yunhe and Zhang, Yongfeng and Shah, Chirag},
title = {ExpScore: Learning Metrics for Recommendation Explanation},
year = {2022},
isbn = {9781450390965},
publisher = {Association for Computing Machinery},
address = {New York, NY, USA},
url = {https://doi.org/10.1145/3485447.3512269},
doi = {10.1145/3485447.3512269},
booktitle = {Proceedings of the ACM Web Conference 2022},
pages = {3740–3744},
numpages = {5},
location = {Virtual Event, Lyon, France},
series = {WWW '22}
}

@Article{ZANON2022109333,
  author    = {André Levi Zanon and Leonardo Chaves Dutra da Rocha and Marcelo Garcia Manzato},
  journal   = {Knowledge-Based Systems},
  title     = {Balancing the trade-off between accuracy and diversity in recommender systems with personalized explanations based on Linked Open Data},
  year      = {2022},
  issn      = {0950-7051},
  pages     = {109333},
  volume    = {252},
  doi       = {https://doi.org/10.1016/j.knosys.2022.109333},
  publisher = {Elsevier},
  url       = {https://www.sciencedirect.com/science/article/pii/S0950705122006682},
}

@Article{AI2025129692,
  author   = {Jun Ai and Haolin Li and Zhan Su and Fengyu Zhao},
  journal  = {Neurocomputing},
  title    = {An explainable recommendation algorithm based on content summarization and linear attention},
  year     = {2025},
  issn     = {0925-2312},
  pages    = {129692},
  volume   = {630},
  doi      = {https://doi.org/10.1016/j.neucom.2025.129692},
  url      = {https://www.sciencedirect.com/science/article/pii/S0925231225003649},
}

@InProceedings{10.1145/3269206.3271739,
  author    = {Wang, Hongwei and Zhang, Fuzheng and Wang, Jialin and Zhao, Miao and Li, Wenjie and Xie, Xing and Guo, Minyi},
  booktitle = {Proceedings of the 27th ACM International Conference on Information and Knowledge Management},
  title     = {RippleNet: Propagating User Preferences on the Knowledge Graph for Recommender Systems},
  year      = {2018},
  address   = {New York, NY, USA},
  pages     = {417–426},
  publisher = {Association for Computing Machinery},
  series    = {CIKM '18},
  doi       = {10.1145/3269206.3271739},
  isbn      = {9781450360142},
  location  = {Torino, Italy},
  numpages  = {10},
  url       = {https://doi.org/10.1145/3269206.3271739},
}

@inproceedings{10.1145/3178876.3186070,
author = {Chen, Chong and Zhang, Min and Liu, Yiqun and Ma, Shaoping},
title = {Neural Attentional Rating Regression with Review-level Explanations},
year = {2018},
isbn = {9781450356398},
publisher = {International World Wide Web Conferences Steering Committee},
address = {Republic and Canton of Geneva, CHE},
url = {https://doi.org/10.1145/3178876.3186070},
doi = {10.1145/3178876.3186070},
booktitle = {Proceedings of the 2018 World Wide Web Conference},
pages = {1583–1592},
numpages = {10},
location = {Lyon, France},
series = {WWW '18}
}

@inproceedings{10.1145/3178876.3186145,
author = {Cheng, Zhiyong and Ding, Ying and Zhu, Lei and Kankanhalli, Mohan},
title = {Aspect-Aware Latent Factor Model: Rating Prediction with Ratings and Reviews},
year = {2018},
isbn = {9781450356398},
publisher = {International World Wide Web Conferences Steering Committee},
address = {Republic and Canton of Geneva, CHE},
url = {https://doi.org/10.1145/3178876.3186145},
doi = {10.1145/3178876.3186145},
booktitle = {Proceedings of the 2018 World Wide Web Conference},
pages = {639–648},
numpages = {10},
location = {Lyon, France},
series = {WWW '18}
}

@inproceedings{10.1145/3308558.3313705,
author = {Cao, Yixin and Wang, Xiang and He, Xiangnan and Hu, Zikun and Chua, Tat-Seng},
title = {Unifying Knowledge Graph Learning and Recommendation: Towards a Better Understanding of User Preferences},
year = {2019},
isbn = {9781450366748},
publisher = {Association for Computing Machinery},
address = {New York, NY, USA},
url = {https://doi.org/10.1145/3308558.3313705},
doi = {10.1145/3308558.3313705},
booktitle = {The World Wide Web Conference},
pages = {151–161},
numpages = {11},
location = {San Francisco, CA, USA},
series = {WWW '19}
}

@inproceedings{Steck2019,
author = {Steck, Harald},
title = {Embarrassingly Shallow Autoencoders for Sparse Data},
year = {2019},
isbn = {9781450366748},
publisher = {Association for Computing Machinery},
address = {New York, NY, USA},
url = {https://doi.org/10.1145/3308558.3313710},
doi = {10.1145/3308558.3313710},
booktitle = {The World Wide Web Conference},
pages = {3251–3257},
numpages = {7},
location = {San Francisco, CA, USA},
series = {WWW '19}
}

@InProceedings{10.1145/3331184.3331203,
  author    = {Xian, Yikun and Fu, Zuohui and Muthukrishnan, S. and de Melo, Gerard and Zhang, Yongfeng},
  booktitle = {Proceedings of the 42nd International ACM SIGIR Conference on Research and Development in Information Retrieval},
  title     = {Reinforcement Knowledge Graph Reasoning for Explainable Recommendation},
  year      = {2019},
  address   = {New York, NY, USA},
  pages     = {285–294},
  publisher = {Association for Computing Machinery},
  series    = {SIGIR'19},
  doi       = {10.1145/3331184.3331203},
  isbn      = {9781450361729},
  location  = {Paris, France},
  numpages  = {10},
  url       = {https://doi.org/10.1145/3331184.3331203},
}

@inproceedings{10.1145/3442381.3450133,
author = {Wang, Xiang and Huang, Tinglin and Wang, Dingxian and Yuan, Yancheng and Liu, Zhenguang and He, Xiangnan and Chua, Tat-Seng},
title = {Learning Intents behind Interactions with Knowledge Graph for Recommendation},
year = {2021},
isbn = {9781450383127},
publisher = {Association for Computing Machinery},
address = {New York, NY, USA},
url = {https://doi.org/10.1145/3442381.3450133},
doi = {10.1145/3442381.3450133},
booktitle = {Proceedings of the Web Conference 2021},
pages = {878–887},
numpages = {10},
location = {Ljubljana, Slovenia},
series = {WWW '21}
}

@inproceedings{10.1145/3442381.3449788,
author = {Zheng, Yu and Gao, Chen and Li, Xiang and He, Xiangnan and Li, Yong and Jin, Depeng},
title = {Disentangling User Interest and Conformity for Recommendation with Causal Embedding},
year = {2021},
isbn = {9781450383127},
publisher = {Association for Computing Machinery},
address = {New York, NY, USA},
url = {https://doi.org/10.1145/3442381.3449788},
doi = {10.1145/3442381.3449788},
booktitle = {Proceedings of the Web Conference 2021},
pages = {2980–2991},
numpages = {12},
location = {Ljubljana, Slovenia},
series = {WWW '21}
}

@inproceedings{10.1145/3485447.3511937,
author = {Geng, Shijie and Fu, Zuohui and Tan, Juntao and Ge, Yingqiang and de Melo, Gerard and Zhang, Yongfeng},
title = {Path Language Modeling over Knowledge Graphsfor Explainable Recommendation},
year = {2022},
isbn = {9781450390965},
publisher = {Association for Computing Machinery},
address = {New York, NY, USA},
url = {https://doi.org/10.1145/3485447.3511937},
doi = {10.1145/3485447.3511937},
booktitle = {Proceedings of the ACM Web Conference 2022},
pages = {946–955},
numpages = {10},
location = {Virtual Event, Lyon, France},
series = {WWW '22}
}

@inproceedings{10.1145/3485447.3512029,
author = {Pan, Sicheng and Li, Dongsheng and Gu, Hansu and Lu, Tun and Luo, Xufang and Gu, Ning},
title = {Accurate and Explainable Recommendation via Review Rationalization},
year = {2022},
isbn = {9781450390965},
publisher = {Association for Computing Machinery},
address = {New York, NY, USA},
url = {https://doi.org/10.1145/3485447.3512029},
doi = {10.1145/3485447.3512029},
booktitle = {Proceedings of the ACM Web Conference 2022},
pages = {3092–3101},
numpages = {10},
location = {Virtual Event, Lyon, France},
series = {WWW '22}
}

@inproceedings{10.1145/3539618.3591776,
author = {Shuai, Jie and Wu, Le and Zhang, Kun and Sun, Peijie and Hong, Richang and Wang, Meng},
title = {Topic-enhanced Graph Neural Networks for Extraction-based Explainable Recommendation},
year = {2023},
isbn = {9781450394086},
publisher = {Association for Computing Machinery},
address = {New York, NY, USA},
url = {https://doi.org/10.1145/3539618.3591776},
doi = {10.1145/3539618.3591776},
booktitle = {Proceedings of the 46th International ACM SIGIR Conference on Research and Development in Information Retrieval},
pages = {1188–1197},
numpages = {10},
location = {Taipei, Taiwan},
series = {SIGIR '23}
}

@article{rana_daddio_manzato_bridge_2022, title={Extended recommendation-by-explanation}, volume={32}, ISSN={0924-1868}, DOI={10.1007/s11257-021-09317-4}, number={1-2}, journal={User Modeling and User-Adapted Interaction}, publisher={User Modeling and User-Adapted Interaction}, author={Rana, Arpit and D’Addio, Rafael M. and Manzato, Marcelo G. and Bridge, Derek}, year={2022}, pages={91–131} }

@article{haque_islam_mikalef_2025, title={To Explain or Not To Explain: An Empirical Investigation of AI-based Recommendations on Social Media Platforms}, volume={35}, ISSN={1019-6781}, DOI={10.1007/s12525-024-00741-z}, number={1}, journal={Electronic Markets}, publisher={Electronic Markets}, author={Haque, Akm Bahalul and Islam, Najmul and Mikalef, Patrick}, year={2025} }

@article{ranjbar_momtazi_homayoonpour_2024, title={Explaining recommendation system using counterfactual textual explanations}, volume={113}, ISSN={0885-6125}, DOI={10.1007/s10994-023-06390-1}, number={4}, journal={Machine Learning}, publisher={Machine Learning}, author={Ranjbar, Niloofar and Momtazi, Saeedeh and Homayoonpour, Mohammadmehdi}, year={2024}, pages={1989–2012} }

@inbook{caro-martinez_jorro-aragoneses_diaz-agudo_recio-garcia_2024, title={Graph-Based Interface for Explanations by Examples in Recommender Systems: A User Study}, ISSN={1865-0929}, DOI={10.1007/978-3-031-63797-1_2}, booktitle={Communications in Computer and Information Science}, publisher={Communications in Computer and Information Science}, author={Caro-Martínez, Marta and Jorro-Aragoneses, José L. and Díaz-Agudo, Belén and Recio-García, Juan A.}, year={2024}, pages={28–41} }

@article{de_campos_fernandez_luna_huete_2024, title={An explainable content-based approach for recommender systems: a case study in journal recommendation for paper submission}, volume={34}, ISSN={0924-1868}, DOI={10.1007/s11257-024-09400-6}, number={4}, journal={User Modeling and User-Adapted Interaction}, publisher={User Modeling and User-Adapted Interaction}, author={De Campos, Luis M. and Fernández-Luna, Juan M. and Huete, Juan F.}, year={2024}, pages={1431–1465} }

@article{xie_wang_xu_chen_zheng_tang_2024, title={A Review-Level Sentiment Information Enhanced Multitask Learning Approach for Explainable Recommendation}, volume={11}, ISSN={2329-924X}, DOI={10.1109/tcss.2024.3376728}, number={5}, journal={IEEE Transactions on Computational Social Systems}, publisher={IEEE Transactions on Computational Social Systems}, author={Xie, Fenfang and Wang, Yuansheng and Xu, Kun and Chen, Liang and Zheng, Zibin and Tang, Mingdong}, year={2024}, pages={5925–5934} }

@inproceedings{le_abel_gouspillou_2023, title={Combining Embedding-Based and Semantic-Based Models for Post-Hoc Explanations in Recommender Systems}, DOI={10.1109/smc53992.2023.10394410}, author={Le, Ngoc Luyen and Abel, Marie-Hélène and Gouspillou, Philippe}, journal={IEEE International Conference on Systems, Man and Cybernetics}, year={2023}, pages={4619–4624} }

@inproceedings{bastola_shakya_2024, title={Knowledge-Enriched Graph Convolution Network for Hybrid Explainable Recommendation from Review Texts and Reasoning Path}, DOI={10.1109/icict60155.2024.10544384}, author={Bastola, Rama and Shakya, Subarna}, year={2024}, pages={590–599}, journal={International Conference on Inventive Computation Technologies}}

@article{hu_liu_miao_lin_miao_2022, title={Aspect-guided Syntax Graph Learning for Explainable Recommendation}, ISSN={1041-4347}, DOI={10.1109/tkde.2022.3221847}, journal={IEEE Transactions on Knowledge and Data Engineering}, publisher={IEEE Transactions on Knowledge and Data Engineering}, author={Hu, Yidan and Liu, Yong and Miao, Chunyan and Lin, Gongqi and Miao, Yuan}, year={2022}, pages={1–14} }

@inproceedings{zhan_li_li_liu_gupta_kot_2023, title={Towards Explainable Recommendation Via Bert-Guided Explanation Generator}, DOI={10.1109/icassp49357.2023.10096389}, author={Zhan, Huijing and Li, Ling and Li, Shaohua and Liu, Weide and Gupta, Manas and Kot, Alex C.}, year={2023}, pages={1–5}, journal={International Conference on Acoustics, Speech, and Signal Processing}, publisher={International Conference on Acoustics, Speech, and Signal Processing}}

@article{balloccu_boratto_fenu_marras_2022, title={XRecSys: A framework for path reasoning quality in explainable recommendation}, volume={14}, ISSN={2665-9638}, DOI={10.1016/j.simpa.2022.100404}, journal={Software Impacts}, publisher={Software Impacts}, author={Balloccu, Giacomo and Boratto, Ludovico and Fenu, Gianni and Marras, Mirko}, year={2022}, pages={100404} }

@article{balloccu_boratto_fenu_marras_2023, title={Reinforcement recommendation reasoning through knowledge graphs for explanation path quality}, volume={260}, ISSN={0950-7051}, DOI={10.1016/j.knosys.2022.110098}, journal={Knowledge-Based Systems}, publisher={Knowledge-Based Systems}, author={Balloccu, Giacomo and Boratto, Ludovico and Fenu, Gianni and Marras, Mirko}, year={2023}, pages={110098} }

@inproceedings{10.1145/2766462.2767755,
author = {McAuley, Julian and Targett, Christopher and Shi, Qinfeng and van den Hengel, Anton},
title = {Image-Based Recommendations on Styles and Substitutes},
year = {2015},
isbn = {9781450336215},
publisher = {Association for Computing Machinery},
address = {New York, NY, USA},
url = {https://doi.org/10.1145/2766462.2767755},
doi = {10.1145/2766462.2767755},
booktitle = {Proceedings of the 38th International ACM SIGIR Conference on Research and Development in Information Retrieval},
pages = {43–52},
numpages = {10},
location = {Santiago, Chile},
series = {SIGIR '15}
}

@inproceedings{10.1145/3340531.3411992,
author = {Li, Lei and Zhang, Yongfeng and Chen, Li},
title = {Generate Neural Template Explanations for Recommendation},
year = {2020},
isbn = {9781450368599},
publisher = {Association for Computing Machinery},
address = {New York, NY, USA},
url = {https://doi.org/10.1145/3340531.3411992},
doi = {10.1145/3340531.3411992},
booktitle = {Proceedings of the 29th ACM International Conference on Information \& Knowledge Management},
pages = {755–764},
numpages = {10},
location = {Virtual Event, Ireland},
series = {CIKM '20}
}

@inproceedings{10.1145/3292500.3330989,
author = {Wang, Xiang and He, Xiangnan and Cao, Yixin and Liu, Meng and Chua, Tat-Seng},
title = {KGAT: Knowledge Graph Attention Network for Recommendation},
year = {2019},
isbn = {9781450362016},
publisher = {Association for Computing Machinery},
address = {New York, NY, USA},
url = {https://doi.org/10.1145/3292500.3330989},
doi = {10.1145/3292500.3330989},
booktitle = {Proceedings of the 25th ACM SIGKDD International Conference on Knowledge Discovery \& Data Mining},
pages = {950–958},
numpages = {9},
location = {Anchorage, AK, USA},
series = {KDD '19}
}

@inproceedings{10.1145/3351095.3372852,
author = {Zhang, Yunfeng and Liao, Q. Vera and Bellamy, Rachel K. E.},
title = {Effect of confidence and explanation on accuracy and trust calibration in AI-assisted decision making},
year = {2020},
isbn = {9781450369367},
publisher = {Association for Computing Machinery},
address = {New York, NY, USA},
url = {https://doi.org/10.1145/3351095.3372852},
doi = {10.1145/3351095.3372852},
booktitle = {Proceedings of the 2020 Conference on Fairness, Accountability, and Transparency},
pages = {295–305},
numpages = {11},
location = {Barcelona, Spain},
series = {FAT* '20}
}

@inproceedings{10.1145/3485447.3512031,
author = {Yang, Aobo and Wang, Nan and Cai, Renqin and Deng, Hongbo and Wang, Hongning},
title = {Comparative Explanations of Recommendations},
year = {2022},
isbn = {9781450390965},
publisher = {Association for Computing Machinery},
address = {New York, NY, USA},
url = {https://doi.org/10.1145/3485447.3512031},
doi = {10.1145/3485447.3512031},
booktitle = {Proceedings of the ACM Web Conference 2022},
pages = {3113–3123},
numpages = {11},
location = {Virtual Event, Lyon, France},
series = {WWW '22}
}

@Article{WU2024111133,
  author   = {Huiqiong Wu and Guibing Guo and Enneng Yang and Yudong Luo and Yabo Chu and Linying Jiang and Xingwei Wang},
  journal  = {Knowledge-Based Systems},
  title    = {PESI: Personalized Explanation recommendation with Sentiment Inconsistency between ratings and reviews},
  year     = {2024},
  issn     = {0950-7051},
  pages    = {111133},
  volume   = {283},
  doi      = {https://doi.org/10.1016/j.knosys.2023.111133},
  url      = {https://www.sciencedirect.com/science/article/pii/S0950705123008833},
}

@Article{XIE2021235,
  author   = {Jin Xie and Fuxi Zhu and Xuefei Li and Sheng Huang and Shichao Liu},
  journal  = {Neurocomputing},
  title    = {Attentive preference personalized recommendation with sentence-level explanations},
  year     = {2021},
  issn     = {0925-2312},
  pages    = {235-247},
  volume   = {426},
  doi      = {https://doi.org/10.1016/j.neucom.2020.10.041},
  url      = {https://www.sciencedirect.com/science/article/pii/S0925231220315903},
}

@Article{LI2025110542,
  author   = {Ying Li and Ming Li and Jin Ding and Yixue Bai},
  journal  = {Engineering Applications of Artificial Intelligence},
  title    = {Two-layer knowledge graph transformer network-based question and answer explainable recommendation},
  year     = {2025},
  issn     = {0952-1976},
  pages    = {110542},
  volume   = {149},
  doi      = {https://doi.org/10.1016/j.engappai.2025.110542},
  url      = {https://www.sciencedirect.com/science/article/pii/S0952197625005421},
}

@article{krishnadisagreement,
  title={The Disagreement Problem in Explainable Machine Learning: A Practitioner’s Perspective},
  author={Krishna, Satyapriya and Han, Tessa and Gu, Alex and Wu, Steven and Jabbari, Shahin and Lakkaraju, Himabindu},
  journal={Transactions on Machine Learning Research}
}

@article{xu2023reusable,
author = {Xu, Zhichao and Zeng, Hansi and Tan, Juntao and Fu, Zuohui and Zhang, Yongfeng and Ai, Qingyao},
title = {A Reusable Model-agnostic Framework for Faithfully Explainable Recommendation and System Scrutability},
year = {2023},
issue_date = {January 2024},
publisher = {Association for Computing Machinery},
address = {New York, NY, USA},
volume = {42},
number = {1},
issn = {1046-8188},
url = {https://doi.org/10.1145/3605357},
doi = {10.1145/3605357},
journal = {ACM Trans. Inf. Syst.},
month = aug,
articleno = {29},
numpages = {29}
}

@article{10.1145/3583558,
author = {Nauta, Meike and Trienes, Jan and Pathak, Shreyasi and Nguyen, Elisa and Peters, Michelle and Schmitt, Yasmin and Schl\"{o}tterer, J\"{o}rg and van Keulen, Maurice and Seifert, Christin},
title = {From Anecdotal Evidence to Quantitative Evaluation Methods: A Systematic Review on Evaluating Explainable AI},
year = {2023},
issue_date = {December 2023},
publisher = {Association for Computing Machinery},
address = {New York, NY, USA},
volume = {55},
number = {13s},
issn = {0360-0300},
url = {https://doi.org/10.1145/3583558},
doi = {10.1145/3583558},
journal = {ACM Comput. Surv.},
month = jul,
articleno = {295},
numpages = {42}
}

@article{schünemann_moja_2015, title={Reviews: Rapid! Rapid! Rapid! …and systematic}, volume={4}, ISSN={2046-4053}, url={https://dx.doi.org/10.1186/2046-4053-4-4}, DOI={10.1186/2046-4053-4-4}, number={1}, journal={Systematic Reviews}, publisher={Systematic Reviews}, author={Schünemann, Holger J and Moja, Lorenzo}, year={2015}, pages={4} }

@article{smela_toumi_świerk_francois_biernikiewicz_clay_boyer_2023, title={Rapid literature review: definition and methodology}, volume={11}, ISSN={2001-6689}, DOI={10.1080/20016689.2023.2241234}, number={1}, journal={Journal of Market Access \& Health Policy}, publisher={Journal of Market Access \& Health Policy}, author={Smela, Beata and Toumi, Mondher and Świerk, Karolina and Francois, Clement and Biernikiewicz, Małgorzata and Clay, Emilie and Boyer, Laurent}, year={2023} }

@inproceedings{10.1145/3640457.3688137,
author = {Petruzzelli, Alessandro and Musto, Cataldo and Laraspata, Lucrezia and Rinaldi, Ivan and de Gemmis, Marco and Lops, Pasquale and Semeraro, Giovanni},
title = {Instructing and Prompting Large Language Models for Explainable Cross-domain Recommendations},
year = {2024},
isbn = {9798400705052},
publisher = {Association for Computing Machinery},
address = {New York, NY, USA},
url = {https://doi.org/10.1145/3640457.3688137},
doi = {10.1145/3640457.3688137},
booktitle = {Proceedings of the 18th ACM Conference on Recommender Systems},
pages = {298–308},
numpages = {11},
location = {Bari, Italy},
series = {RecSys '24}
}

@article{faul_erdfelder_buchner_lang_2009, title={Statistical power analyses using G*Power 3.1: Tests for correlation and regression analyses}, volume={41}, ISSN={1554-351X}, DOI={10.3758/brm.41.4.1149}, number={4}, journal={Behavior Research Methods}, publisher={Behavior Research Methods}, author={Faul, Franz and Erdfelder, Edgar and Buchner, Axel and Lang, Albert-Georg}, year={2009}, pages={1149–1160} }

@article{BARREDOARRIETA202082,
title = {Explainable Artificial Intelligence (XAI): Concepts, taxonomies, opportunities and challenges toward responsible AI},
journal = {Information Fusion},
volume = {58},
pages = {82-115},
year = {2020},
issn = {1566-2535},
doi = {https://doi.org/10.1016/j.inffus.2019.12.012},
url = {https://www.sciencedirect.com/science/article/pii/S1566253519308103},
author = {Alejandro {Barredo Arrieta} and Natalia Díaz-Rodríguez and Javier {Del Ser} and Adrien Bennetot and Siham Tabik and Alberto Barbado and Salvador Garcia and Sergio Gil-Lopez and Daniel Molina and Richard Benjamins and Raja Chatila and Francisco Herrera}
}

@inproceedings{10.1145/3705328.3748028,
author = {Mohammadi, Amir Reza and Peintner, Andreas and M\"{u}ller, Michael and Zangerle, Eva},
title = {Beyond Top-1: Addressing Inconsistencies in Evaluating Counterfactual Explanations for Recommender Systems},
year = {2025},
isbn = {9798400713644},
publisher = {Association for Computing Machinery},
address = {New York, NY, USA},
url = {https://doi.org/10.1145/3705328.3748028},
doi = {10.1145/3705328.3748028},
booktitle = {Proceedings of the Nineteenth ACM Conference on Recommender Systems},
pages = {515–520},
numpages = {6},
location = {
},
series = {RecSys '25}
}

@misc{wang2024nextgenerationllmbasedrecommendersystems,
      title={Towards Next-Generation LLM-based Recommender Systems: A Survey and Beyond}, 
      author={Qi Wang and Jindong Li and Shiqi Wang and Qianli Xing and Runliang Niu and He Kong and Rui Li and Guodong Long and Yi Chang and Chengqi Zhang},
      year={2024},
      eprint={2410.19744},
      archivePrefix={arXiv},
      primaryClass={cs.IR},
      url={https://arxiv.org/abs/2410.19744},
      doi = {10.48550/arXiv.2410.19744}
}


\appendix

\newpage
\section{Sources of Papers on the Rapid Literature Review}
\label{apen:sources}

\begin{figure}[!ht]
    \centering
    \includegraphics[width=\textwidth]{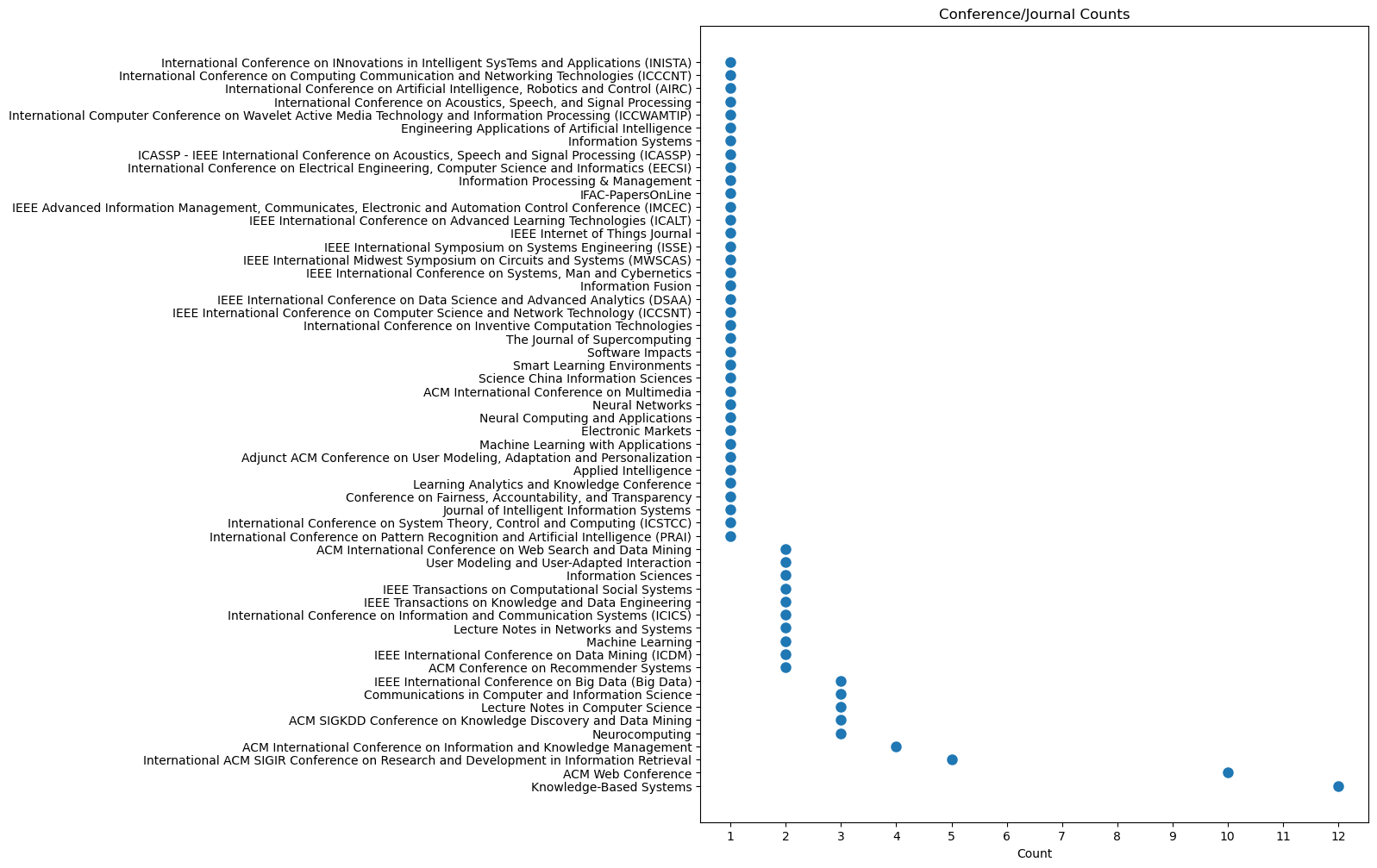}
    \caption{Distribution of the papers found by our rapid literature review by journal or conference.}
    \label{fig:sources}
\end{figure}

\newpage
\section{Rapid Literature Review Paper Categorization}
\label{apen:lit_review}

\begin{table}[htbp]
\caption{Categorization of the papers found by our rapid literature review. \textbf{Style} and \textbf{Method} columns categorize the information used to generate explanations and if whether explanations were generated agnostic to the \ac{RS} or intrinsically to the model. \textbf{Offline} and \textbf{Online} are binary and represent whether evaluations were measured by offline and\slash or online experiments, respectively. The \textbf{Offline Metric} column displays the offline metric used when an offline evaluation was conducted, in addition to the number of users from the dataset that were used to evaluate explanations in column \textbf{\# of Users}. Similarity, the \textbf{Online Metric} represent how the conducted user study captured user perception.\looseness=-1}
\label{tab:lit_review}
\end{table}
\begin{xltabular}{\textwidth}{l|l|l|l|l|l|l|l}
\toprule
\textbf{Citation} & \textbf{Style} & \textbf{Method} & \textbf{Offline} & \textbf{Offline Metric} & \textbf{\# of Users} & \textbf{Online} & \textbf{Online Metric} \\
\midrule \midrule
\cite{takami_flanagan_dai_ogata_2023} & Personality & Agnostic & No & - & - & Yes & A\slash B \\
\cite{lin_zhang_lin_zeng_zhou_wu_2024} & Content & Intrinsic & Yes & Explainable Items & All dataset & No & - \\
\cite{zanon_da_rocha_manzato_2024} & Content & Agnostic & Yes & Path Metrics & All dataset & No & - \\
\cite{samih_ghadi_fennan_2023} & Content & Agnostic & Yes & Anecdotal & Examples & No & - \\
\cite{li_liu_zhang_kou_liu_qu_2025} & Review & Agnostic & Yes & Precision/Recall & Examples & No & - \\
\cite{wen_liu_jing_yu_2024} & User-based & Agnostic & Yes & Counterfactual & All dataset & No & - \\
\cite{long_jin_2024} & Review & Agnostic & Yes & BLEU/ROUGE & All dataset & No & - \\
\cite{jendal_le_lauw_lissandrini_dolog_hose_2024} & Hybrid & Agnostic & Yes & BLEU/ROUGE & All dataset & No & - \\
\cite{zhong_negre_2022} & Review & Intrinsic & No & - & - & Yes & User Trial \\
\cite{alizadeh_noughabi_behkamal_zarrinkalam_kahani_2024} & Hybrid & Agnostic & Yes & Path Metrics & All dataset & No & - \\
\cite{zhang_zhu_wang_2023} & User-based & Intrinsic & Yes & Explainable Items & All dataset & No & - \\
\cite{zheng_chen_cao_peng_huang_2024} & Content & Intrinsic & Yes & Anecdotal & Examples & No & - \\
\cite{wang_xie_ding_chen_xiang_2025} & Content & Intrinsic & Yes & Anecdotal & Examples & No & - \\
\cite{sang_yang_zhang_liao_2025} & Content & Intrinsic & Yes & Anecdotal & Examples & No & - \\
\cite{9811151} & Review & Intrinsic & No & - & - & No & - \\
\cite{8622439} & Review & Agnostic & Yes & Anecdotal & Examples & No & - \\
\cite{10741116} & Parameters & Agnostic & Yes & Precision/Recall & All dataset & No & - \\
\cite{9607106} & Content & Intrinsic & Yes & Anecdotal & Examples & No & - \\
\cite{9260076} & User-based & Agnostic & Yes & Explainable Items & All dataset & No & - \\
\cite{10776491} & Content & Agnostic & Yes & Anecdotal & Examples & No & - \\
\cite{9005590} & Hybrid & Intrinsic & Yes & Anecdotal & Examples & No & - \\
\cite{10260804} & Content & Intrinsic & No & - & - & Yes & User Trial \\
\cite{10683822} & Content & Agnostic & Yes & Anecdotal & Examples & No & - \\
\cite{9079084} & Review & Intrinsic & Yes & Explainable Items & All dataset & No & - \\
\cite{9836983} & User-based & Agnostic & Yes & Anecdotal & Examples & No & - \\
\cite{10873804} & Review & Intrinsic & Yes & Anecdotal & Examples & No & - \\
\cite{8594883} & Review & Agnostic & Yes & Correlation & All dataset & Yes & A\slash B \\
\cite{10048787} & User-based & Intrinsic & Yes & Anecdotal & Examples & Yes & - \\
\cite{10446052} & Review & Agnostic & Yes & BLEU/ROUGE & All dataset & No & - \\
\cite{10827288} & Parameters & Agnostic & Yes & Anecdotal & Examples & No & - \\
\cite{9482221} & Review & Intrinsic & Yes & BLEU/ROUGE & All dataset & No & - \\
\cite{10658914} & LLM & Agnostic & Yes & Anecdotal & Examples & No & - \\
\cite{10334552} & Content & Intrinsic & Yes & Path Metrics & All dataset & No & - \\
\cite{10825771} & Parameters & Agnostic & Yes & Anecdotal & All dataset & No & - \\
\cite{10308154} & Parameters & Agnostic & Yes & Anecdotal & All dataset & No & - \\
\cite{10742303} & Content & Intrinsic & Yes & Anecdotal & Examples & No & - \\
\cite{10623784} & Content & Agnostic & Yes & Anecdotal & Examples & No & - \\
\cite{10884422} & Review & Intrinsic & Yes & Precision/Recall & All dataset & No & - \\
\cite{SHIMIZU2022107970} & Content & Intrinsic & Yes & Anecdotal & Examples & No & - \\
\cite{BRUNOT2022102021} & Parameters & Agnostic & Yes & Precision/Recall & All dataset & No & - \\
\cite{TAO2022109300} & Content & Intrinsic & Yes & Anecdotal & Examples & No & - \\
\cite{TAO2021107217} & Content & Intrinsic & Yes & Anecdotal & Examples & No & - \\
\cite{YANG2021106687} & Review & Intrinsic & Yes & Precision/Recall & All dataset & No & - \\
\cite{LIU2020102099} & Hybrid & Intrinsic & Yes & Anecdotal & Examples & Yes & User Trial \\
\cite{LIANG202194} & Content & Agnostic & Yes & Anecdotal & Examples & No & - \\
\cite{ZANON2022109333} & Content & Agnostic & No & - & - & No & - \\
\cite{PAZRUZA2024102497} & Hybrid & Intrinsic & No & - & - & No & - \\
\cite{AI2025129692} & Review & Intrinsic & Yes & BLEU/ROUGE & All dataset & No & - \\
\cite{WEI2023202} & Hybrid & Intrinsic & No & - & - & No & - \\
\cite{KAUR2023100507} & Content & Intrinsic & No & - & - & No & - \\
\cite{YANG2020106194} & Content & Intrinsic & Yes & Anecdotal & Examples & No & - \\
\cite{LI2024112042} & Content & Intrinsic & No & - & - & No & - \\
\cite{WANG2020436} & Content & Intrinsic & No & - & - & No & - \\
\cite{GUO2021185} & Review & Intrinsic & Yes & Anecdotal & Examples & No & - \\
\cite{MARKCHOM2023110258} & Hybrid & Intrinsic & Yes & Explainable Items & All dataset & No & - \\
\cite{HAO2025113113} & Review & Intrinsic & Yes & BLEU/ROUGE & All dataset & No & - \\
\cite{LIU2025113217} & Content & Intrinsic & No & - & - & No & - \\
\cite{WU2025129780} & Content & Intrinsic & Yes & Anecdotal & Examples & No & - \\
\cite{10.1145/3269206.3271739} & Content & Intrinsic & Yes & Anecdotal & Examples & No & - \\
\cite{10.1145/3357384.3357925} & Content & Intrinsic & Yes & Correlation & All dataset & No & - \\
\cite{10.1145/3219819.3219965} & Content & Intrinsic & Yes & Anecdotal & Examples & No & - \\
\cite{10.1145/3178876.3186070} & Review & Intrinsic & Yes & Precision/Recall & All dataset & No & - \\
\cite{10.1145/3308558.3313705} & Content & Intrinsic & Yes & Anecdotal & Examples & No & - \\
\cite{10.1145/3397271.3401137} & Review & Intrinsic & Yes & Anecdotal & Examples & No & - \\
\cite{10.1145/3343031.3351034} & Hybrid & Intrinsic & Yes & Anecdotal & Examples & No & - \\
\cite{10.1145/2806416.2806504} & Review & Intrinsic & Yes & Anecdotal & Examples & No & - \\
\cite{10.1145/3159652.3159668} & User-based & Intrinsic & Yes & Anecdotal & Examples & No & - \\
\cite{10.1145/3109859.3109890} & Review & Intrinsic & Yes & Anecdotal & Examples & No & - \\
\cite{10.1145/3331184.3331203} & Content & Intrinsic & Yes & Anecdotal & Examples & No & - \\
\cite{10.1145/3442381.3450133} & Content & Intrinsic & Yes & Anecdotal & Examples & No & - \\
\cite{10.1145/3178876.3186145} & Review & Intrinsic & Yes & Anecdotal & Examples & No & - \\
\cite{10.1145/3178876.3186154} & User-based & Intrinsic & No & - & - & No & - \\
\cite{10.1145/3442381.3449788} & User-based & Intrinsic & Yes & Correlation & All dataset & No & - \\
\cite{10.1145/3631700.3665226} & Content & Intrinsic & No & - & - & Yes & User Trial \\
\cite{10.1145/3637528.3671781} & Review & Agnostic & Yes & BLEU/ROUGE & All dataset & No & - \\
\cite{10.1145/3543507.3583260} & Review & Agnostic & Yes & BLEU/ROUGE & All dataset & No & - \\
\cite{10.1145/3624918.3625331} & Review & Agnostic & Yes & BLEU/ROUGE & All dataset & No & - \\
\cite{10.1145/3485447.3511937} & Content & Intrinsic & Yes & Anecdotal & Examples & No & - \\
\cite{10.1145/3485447.3512029} & Review & Intrinsic & Yes & Anecdotal & Examples & No & - \\
\cite{10.1145/3604915.3609491} & Content & Agnostic & No & - & - & - & - \\
\cite{10.1145/3636555.3636898} & Content & Intrinsic & No & - & - & Yes & User Trial \\
\cite{10.1145/3616855.3635855} & User-based & Agnostic & No & - & - & Yes & A\slash B \\
\cite{10.1145/3539618.3591776} & Review & Intrinsic & Yes & BLEU/ROUGE & All dataset & No & - \\
\cite{rana_daddio_manzato_bridge_2022} & Review & Intrinsic & No & - & - & Yes & A\slash B \\
\cite{haque_islam_mikalef_2025} & User-based & Agnostic & No & - & - & Yes & User Trial \\
\cite{ranjbar_momtazi_homayoonpour_2024} & Review & Intrinsic & Yes & Counterfactual & All dataset & No & - \\
\cite{caro-martinez_jorro-aragoneses_diaz-agudo_recio-garcia_2024} & Content & Agnostic & No & - & - & Yes & A\slash B \\
\cite{de_campos_fernandez_luna_huete_2024} & Content & Agnostic & No & - & - & Yes & User Trial \\
\cite{xie_wang_xu_chen_zheng_tang_2024} & Review & Intrinsic & Yes & BLEU/ROUGE & All dataset & No & - \\
\cite{le_abel_gouspillou_2023} & Content & Intrinsic & Yes & Anecdotal & Examples & No & - \\
\cite{bastola_shakya_2024} & Hybrid & Intrinsic & Yes & BLEU/ROUGE & All dataset & No & - \\
\cite{hu_liu_miao_lin_miao_2022} & Review & Intrinsic & Yes & BLEU/ROUGE & All dataset & No & - \\
\cite{zhan_li_li_liu_gupta_kot_2023} & Review & Intrinsic & Yes & BLEU/ROUGE & All dataset & No & - \\
\cite{LI2025110542} & Content & Intrinsic & Yes & Anecdotal & All dataset & No & - \\
\cite{XIE2021235} & Review & Intrinsic & Yes & Anecdotal & Examples & No & - \\
\cite{balloccu_boratto_fenu_marras_2022} & Content & Agnostic & No & Path Metrics & - & No & - \\
\cite{balloccu_boratto_fenu_marras_2023} & Content & Agnostic & Yes & Path Metrics & All dataset & No & - \\
\cite{WU2024111133} & Review & Intrinsic & Yes & BLEU/ROUGE & All dataset & No & - \\
\cite{10.1145/2766462.2767755} & Hybrid & Intrinsic & Yes & Anecdotal & Examples & No & - \\
\cite{10.1145/3340531.3411992} & Review & Intrinsic & Yes & BLEU/ROUGE & All dataset & No & - \\
\cite{10.1145/3292500.3330989} & Content & Intrinsic & Yes & Anecdotal & Examples & No & - \\
\cite{10.1145/3351095.3372852} & Parameters & Agnostic & No & - & - & Yes & A\slash B \\
\cite{10.1145/3485447.3512031} & Review & Intrinsic & Yes & BLEU/ROUGE & All dataset & No & - \\
\bottomrule
\end{xltabular}

\newpage
\section{Edge Types Distribution on Artists KG}
\label{apen:type_movies}

\begin{figure}[!ht]
    \centering
    \includegraphics[scale=.8]{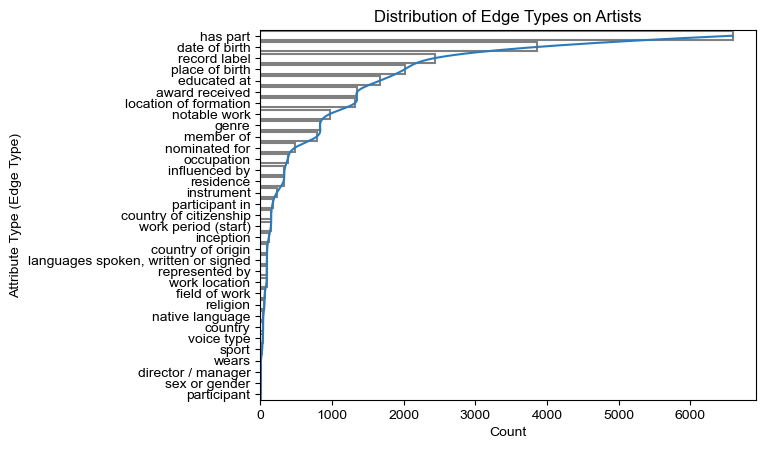}
    \caption{Distribution of Edge Types References from Item and Attribute Nodes on the Artists Wikidata KG}
    \label{fig:edge_type_artists}
\end{figure}

\newpage
\section{Genre distribution on Artists KG}
\label{apen:genre_distribution}

\begin{figure}[!ht]
    \centering
    \includegraphics[scale=0.7]{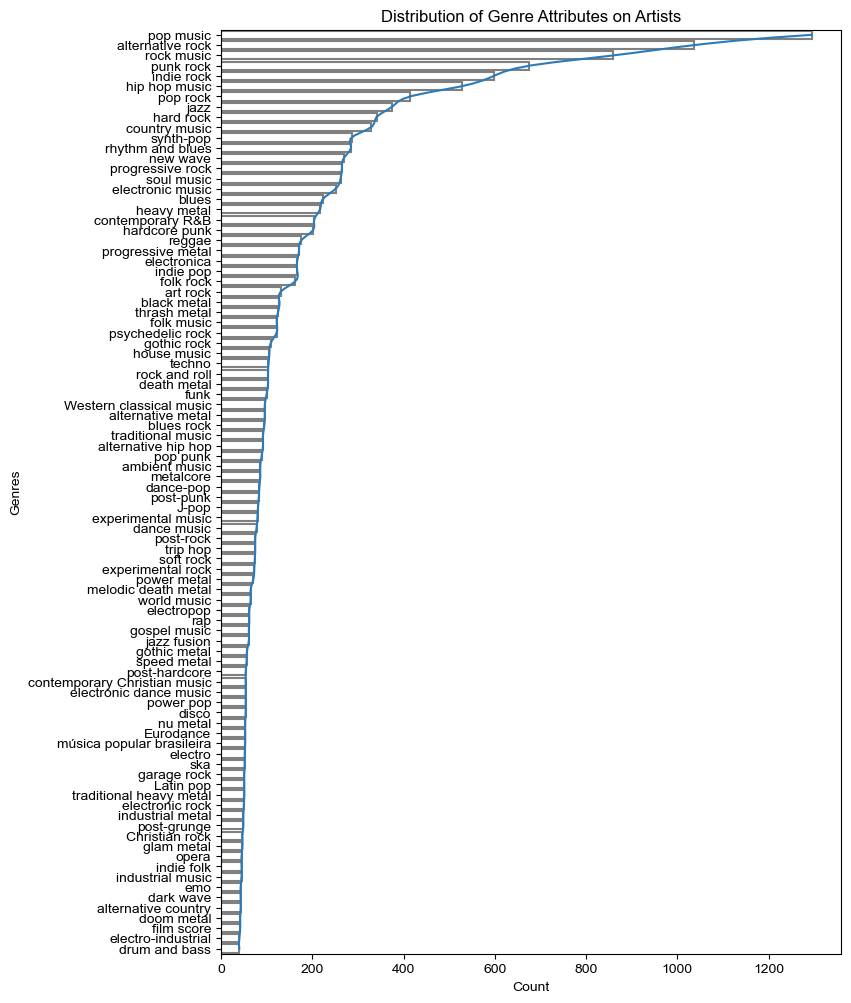}
    \caption{Distribution of Genre Attributes References from Item Nodes on the Artists Wikidata KG}
    \label{fig:genre_artists}
\end{figure}

\newpage
\section{User Profile Construction Screen}
\label{apen:profile}

\begin{figure}[h]
    \centering
    \includegraphics[width=\textwidth]{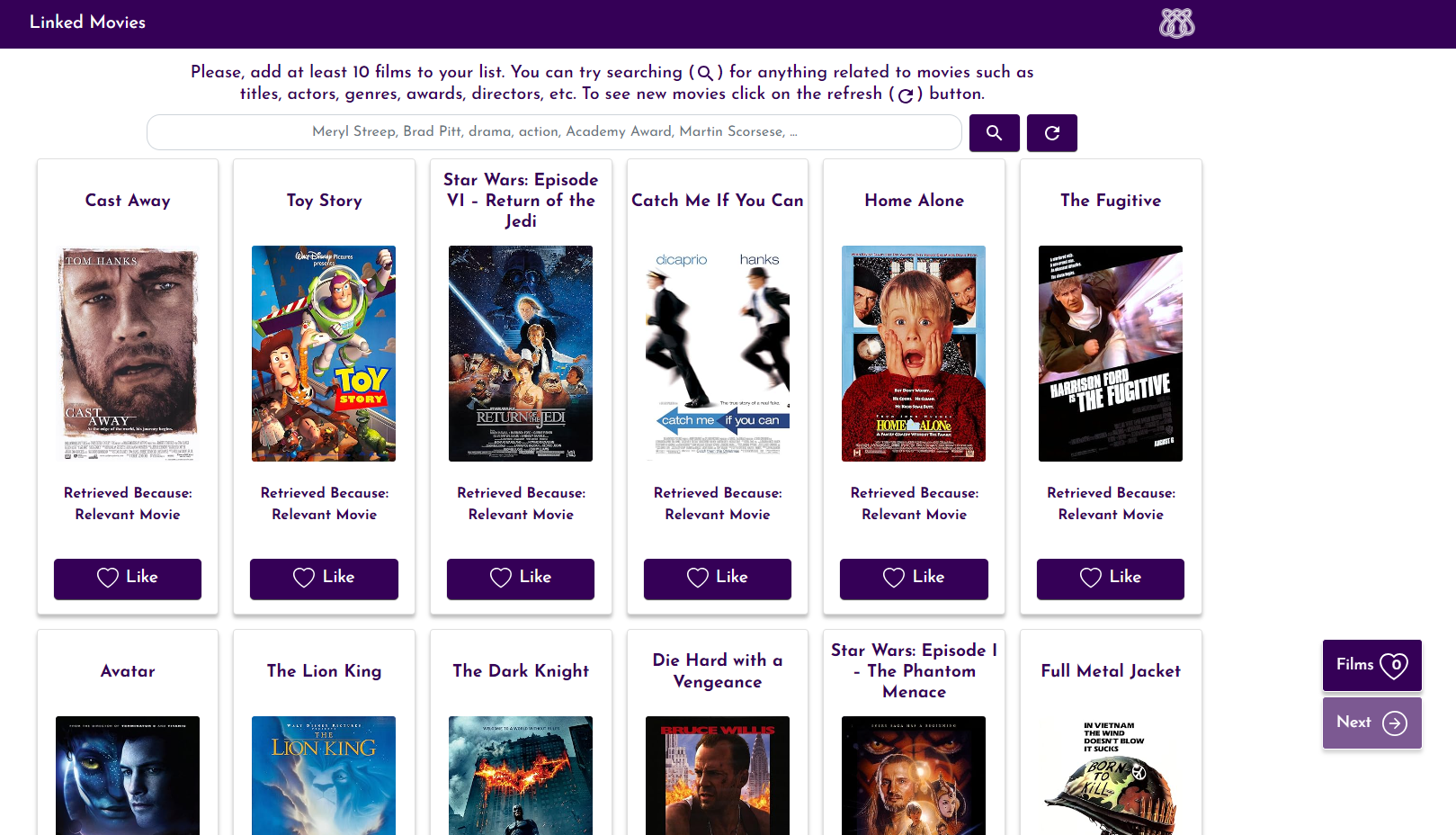}
    \caption{Screen for the user to create the profile}
    \label{fig:user_pro}
\end{figure}

\newpage
\section{Evaluation Screen}
\label{apen:expl}

\begin{figure}[h]
    \centering
    \includegraphics[width=\textwidth]{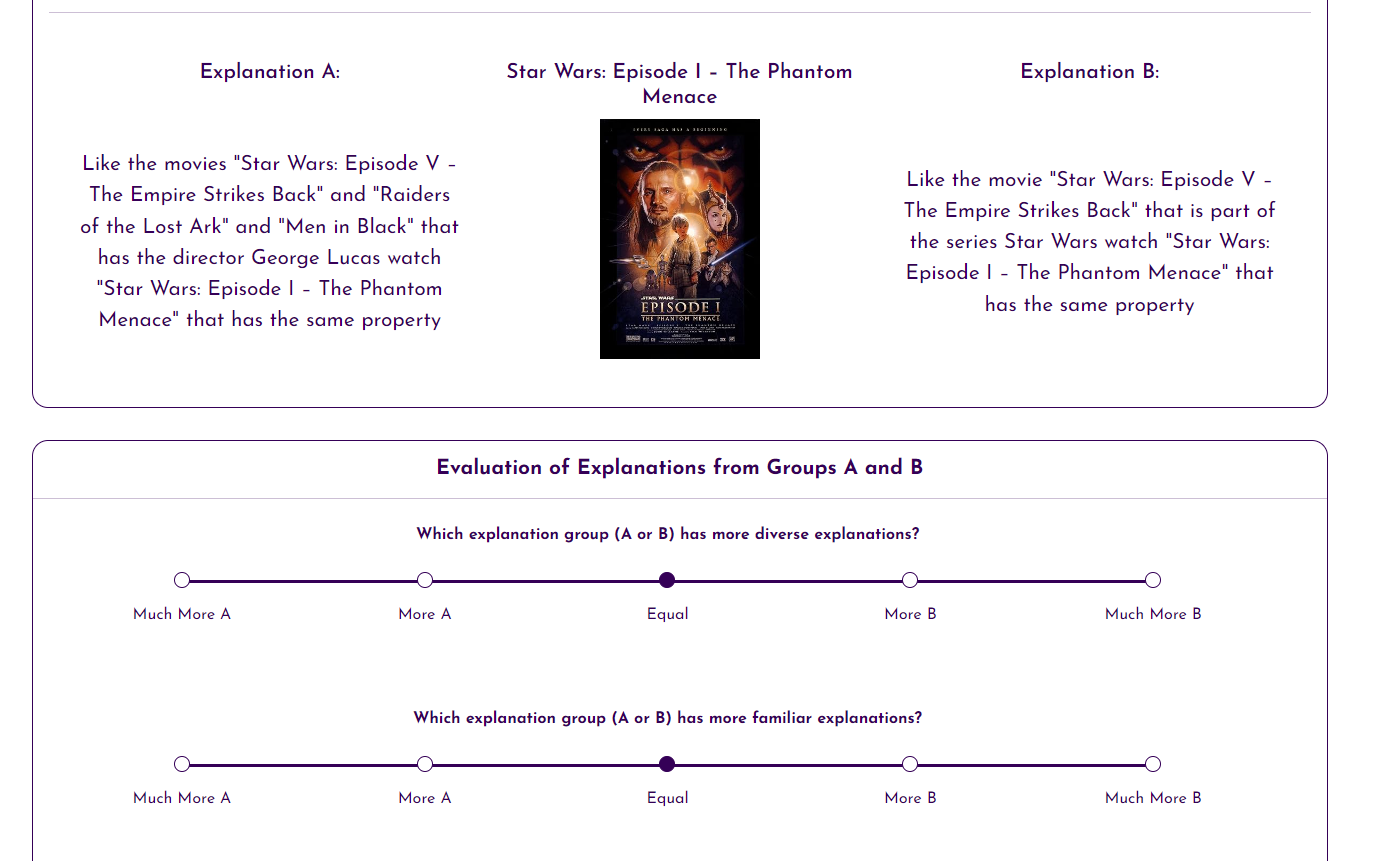}
    \caption{Example of a user's screen with recommendations and a set of questions to be answered}
    \label{fig:user_rec}
\end{figure}

\newpage
\section{LastFM Offline Explanation Metrics Results}
\label{apen:lastfm}

Unlike the \ac{MovieLens} dataset, which logs interactions as user, item, and timestamp triples, the \ac{LastFM} dataset comprises triples of user, artist, and weight. Here, the weight reflects the frequency of a user's listening to an artist. Therefore, in the \ac{LastFM} dataset, the LIR metric is linked to how often a user listens to an artist instead of the timestamp of the interaction.\looseness=-1

\begin{table*}[!htbp]
\caption{Offline results for the metrics for the top-1 recommendation for the \ac{LastFM} dataset. \textbf{Bold} results with \greentriangleup are the best values considering the three explanation algorithms for a recommendation algorithm.\looseness=-1}
\begin{tabular}{cc|ccc||ccc|}
\cline{3-8}
\multicolumn{1}{l}{}                            & \multicolumn{1}{l|}{} & \multicolumn{3}{c||}{Item Metrics}                                                         & \multicolumn{3}{c|}{Attribute Metrics}                                        \\ \cline{3-8} 
\textbf{}                                       & \textbf{}             & \multicolumn{1}{c|}{MID}            & \multicolumn{1}{c|}{TID}           & LIR            & \multicolumn{1}{c|}{ETD} & \multicolumn{1}{c|}{TPD}          & SEP            \\ \hline
\multicolumn{1}{|c|}{\multirow{3}{*}{MostPop}}  & ExpLOD                & \multicolumn{1}{c|}{\textbf{2,943}  \greentriangleup} & \multicolumn{1}{c|}{\textbf{1427}  \greentriangleup} & \textbf{0,020}  \greentriangleup & \multicolumn{1}{c|}{1}   & \multicolumn{1}{c|}{44}           & 0,788          \\ \cline{2-8} 
\multicolumn{1}{|c|}{}                          & ExpLOD v2             & \multicolumn{1}{c|}{2,740}          & \multicolumn{1}{c|}{1103}          & \textbf{0,020}  \greentriangleup & \multicolumn{1}{c|}{1}   & \multicolumn{1}{c|}{40}           & \textbf{0,843}  \greentriangleup \\ \cline{2-8} 
\multicolumn{1}{|c|}{}                          & PEM                   & \multicolumn{1}{c|}{1,773}          & \multicolumn{1}{c|}{371}           & 0,015          & \multicolumn{1}{c|}{1}   & \multicolumn{1}{c|}{\textbf{88}  \greentriangleup}  & 0,100          \\ \hline \hline
\multicolumn{1}{|c|}{\multirow{3}{*}{UserKNN}}  & ExpLOD                & \multicolumn{1}{c|}{2,969}          & \multicolumn{1}{c|}{1389}          & 0,023          & \multicolumn{1}{c|}{1}   & \multicolumn{1}{c|}{137}          & 0,585          \\ \cline{2-8} 
\multicolumn{1}{|c|}{}                          & ExpLOD v2             & \multicolumn{1}{c|}{2,905}          & \multicolumn{1}{c|}{1212}          & \textbf{0,022}  \greentriangleup & \multicolumn{1}{c|}{1}   & \multicolumn{1}{c|}{109}          & \textbf{0,622}  \greentriangleup \\ \cline{2-8} 
\multicolumn{1}{|c|}{}                          & PEM                   & \multicolumn{1}{c|}{\textbf{2,167}  \greentriangleup} & \multicolumn{1}{c|}{807}           & 0,015          & \multicolumn{1}{c|}{1}   & \multicolumn{1}{c|}{\textbf{304}  \greentriangleup} & 0,132          \\ \hline \hline
\multicolumn{1}{|c|}{\multirow{3}{*}{PageRank}} & ExpLOD                & \multicolumn{1}{c|}{\textbf{2,962}  \greentriangleup} & \multicolumn{1}{c|}{\textbf{1407}  \greentriangleup} & 0,021          & \multicolumn{1}{c|}{1}   & \multicolumn{1}{c|}{97}           & 0,635          \\ \cline{2-8} 
\multicolumn{1}{|c|}{}                          & ExpLOD v2             & \multicolumn{1}{c|}{2,838}          & \multicolumn{1}{c|}{1111}          & \textbf{0,021}  \greentriangleup & \multicolumn{1}{c|}{1}   & \multicolumn{1}{c|}{79}           & \textbf{0,729}  \greentriangleup \\ \cline{2-8} 
\multicolumn{1}{|c|}{}                          & PEM                   & \multicolumn{1}{c|}{1,995}          & \multicolumn{1}{c|}{565}           & 0,015          & \multicolumn{1}{c|}{1}   & \multicolumn{1}{c|}{\textbf{189}  \greentriangleup} & 0,093          \\ \hline \hline
\multicolumn{1}{|c|}{\multirow{3}{*}{BPRMF}}    & ExpLOD                & \multicolumn{1}{c|}{2,946}          & \multicolumn{1}{c|}{\textbf{1438}  \greentriangleup} & 0,022          & \multicolumn{1}{c|}{1}   & \multicolumn{1}{c|}{145}          & 0,622          \\ \cline{2-8} 
\multicolumn{1}{|c|}{}                          & ExpLOD v2             & \multicolumn{1}{c|}{2,852}          & \multicolumn{1}{c|}{1273}          & \textbf{0,017}  \greentriangleup & \multicolumn{1}{c|}{1}   & \multicolumn{1}{c|}{125}          & \textbf{0,629}  \greentriangleup \\ \cline{2-8} 
\multicolumn{1}{|c|}{}                          & PEM                   & \multicolumn{1}{c|}{\textbf{2,122}  \greentriangleup} & \multicolumn{1}{c|}{923}           & 0,015          & \multicolumn{1}{c|}{1}   & \multicolumn{1}{c|}{\textbf{349}  \greentriangleup} & 0,157          \\ \hline \hline
\multicolumn{1}{|c|}{\multirow{3}{*}{EASE}}     & ExpLOD                & \multicolumn{1}{c|}{2,972}          & \multicolumn{1}{c|}{\textbf{1386}  \greentriangleup} & 0,026          & \multicolumn{1}{c|}{1}   & \multicolumn{1}{c|}{125}          & 0,588          \\ \cline{2-8} 
\multicolumn{1}{|c|}{}                          & ExpLOD v2             & \multicolumn{1}{c|}{2,915}          & \multicolumn{1}{c|}{1195}          & \textbf{0,021}  \greentriangleup & \multicolumn{1}{c|}{1}   & \multicolumn{1}{c|}{103}          & \textbf{0,644}  \greentriangleup \\ \cline{2-8} 
\multicolumn{1}{|c|}{}                          & PEM                   & \multicolumn{1}{c|}{\textbf{2,181}  \greentriangleup} & \multicolumn{1}{c|}{811}           & 0,017          & \multicolumn{1}{c|}{1}   & \multicolumn{1}{c|}{\textbf{289}  \greentriangleup} & 0,128          \\ \hline \hline
\multicolumn{1}{|c|}{\multirow{3}{*}{NCF}}      & ExpLOD                & \multicolumn{1}{c|}{2,875}          & \multicolumn{1}{c|}{\textbf{1530}  \greentriangleup} & \textbf{0,017}  \greentriangleup & \multicolumn{1}{c|}{1}   & \multicolumn{1}{c|}{173}          & \textbf{0,626}  \greentriangleup \\ \cline{2-8} 
\multicolumn{1}{|c|}{}                          & ExpLOD v2             & \multicolumn{1}{c|}{2,811}          & \multicolumn{1}{c|}{1450}          & 0,020          & \multicolumn{1}{c|}{1}   & \multicolumn{1}{c|}{173}          & 0,560          \\ \cline{2-8} 
\multicolumn{1}{|c|}{}                          & PEM                   & \multicolumn{1}{c|}{\textbf{2,195}  \greentriangleup} & \multicolumn{1}{c|}{1211}          & 0,013          & \multicolumn{1}{c|}{1}   & \multicolumn{1}{c|}{\textbf{577}  \greentriangleup} & 0,190          \\ \hline
\end{tabular}
\label{tab:offresultslastfm_n1}
\end{table*}

\newpage 
\begin{table*}[!htbp]
\caption{Offline results for the metrics for the top-5 recommendations for the \ac{LastFM} dataset. \textbf{Bold} results with \greentriangleup are the best values considering the three explanation algorithms for a recommendation algorithm. \looseness=-1}
\begin{tabular}{cc|ccc||ccc|}
\cline{3-8}
\multicolumn{1}{l}{}                            & \multicolumn{1}{l|}{} & \multicolumn{3}{c||}{Item Metrics}                                                         & \multicolumn{3}{c|}{Attribute Metrics}                                                     \\ \cline{3-8} 
                                                &                       & \multicolumn{1}{c|}{MID}            & \multicolumn{1}{c|}{TID}           & LIR            & \multicolumn{1}{c|}{ETD}             & \multicolumn{1}{c|}{TPD}           & SEP            \\ \hline
\multicolumn{1}{|c|}{\multirow{3}{*}{MostPop}}  & ExpLOD                & \multicolumn{1}{c|}{\textbf{7,806}  \greentriangleup} & \multicolumn{1}{c|}{\textbf{2681}  \greentriangleup} & \textbf{0,018}  \greentriangleup & \multicolumn{1}{c|}{0,702}           & \multicolumn{1}{c|}{118}           & 0,710          \\ \cline{2-8} 
\multicolumn{1}{|c|}{}                          & ExpLOD v2             & \multicolumn{1}{c|}{5,375}          & \multicolumn{1}{c|}{1795}          & 0,019          & \multicolumn{1}{c|}{0,493}           & \multicolumn{1}{c|}{64}            & \textbf{0,754}  \greentriangleup \\ \cline{2-8} 
\multicolumn{1}{|c|}{}                          & PEM                   & \multicolumn{1}{c|}{5,5243}         & \multicolumn{1}{c|}{1344}          & 0,0143         & \multicolumn{1}{c|}{\textbf{0,9212}  \greentriangleup} & \multicolumn{1}{c|}{\textbf{236}  \greentriangleup}  & 0,1214         \\ \hline \hline
\multicolumn{1}{|c|}{\multirow{3}{*}{UserKNN}}  & ExpLOD                & \multicolumn{1}{c|}{6,326}          & \multicolumn{1}{c|}{2482}          & 0,018          & \multicolumn{1}{c|}{0,534}           & \multicolumn{1}{c|}{238}           & \textbf{0,529}  \greentriangleup \\ \cline{2-8} 
\multicolumn{1}{|c|}{}                          & ExpLOD v2             & \multicolumn{1}{c|}{6,333}          & \multicolumn{1}{c|}{2179}          & \textbf{0,019}  \greentriangleup & \multicolumn{1}{c|}{0,535}           & \multicolumn{1}{c|}{173}           & 0,281          \\ \cline{2-8} 
\multicolumn{1}{|c|}{}                          & PEM                   & \multicolumn{1}{c|}{\textbf{7,303}} & \multicolumn{1}{c|}{1939}          & 0,016          & \multicolumn{1}{c|}{\textbf{0,911}}  & \multicolumn{1}{c|}{\textbf{653}  \greentriangleup}  & 0,142          \\ \hline \hline
\multicolumn{1}{|c|}{\multirow{3}{*}{PageRank}} & ExpLOD                & \multicolumn{1}{c|}{\textbf{7,108}  \greentriangleup} & \multicolumn{1}{c|}{\textbf{2674}  \greentriangleup} & 0,019          & \multicolumn{1}{c|}{0,610}           & \multicolumn{1}{c|}{224}           & 0,650          \\ \cline{2-8} 
\multicolumn{1}{|c|}{}                          & ExpLOD v2             & \multicolumn{1}{c|}{6,134}          & \multicolumn{1}{c|}{2205}          & \textbf{0,021}  \greentriangleup & \multicolumn{1}{c|}{0,545}           & \multicolumn{1}{c|}{150}           & \textbf{0,716}  \greentriangleup \\ \cline{2-8} 
\multicolumn{1}{|c|}{}                          & PEM                   & \multicolumn{1}{c|}{6,526}          & \multicolumn{1}{c|}{1662}          & 0,013          & \multicolumn{1}{c|}{\textbf{0,944}  \greentriangleup}  & \multicolumn{1}{c|}{\textbf{477}  \greentriangleup}  & 0,121          \\ \hline \hline
\multicolumn{1}{|c|}{\multirow{3}{*}{BPRMF}}    & ExpLOD                & \multicolumn{1}{c|}{7,168}          & \multicolumn{1}{c|}{\textbf{2727}  \greentriangleup} & 0,019          & \multicolumn{1}{c|}{0,615}           & \multicolumn{1}{c|}{230}           & 0,561          \\ \cline{2-8} 
\multicolumn{1}{|c|}{}                          & ExpLOD v2             & \multicolumn{1}{c|}{7,050}          & \multicolumn{1}{c|}{2434}          & \textbf{0,020}  \greentriangleup & \multicolumn{1}{c|}{0,620}           & \multicolumn{1}{c|}{180}           & \textbf{0,630}  \greentriangleup \\ \cline{2-8} 
\multicolumn{1}{|c|}{}                          & PEM                   & \multicolumn{1}{c|}{\textbf{7,298}  \greentriangleup} & \multicolumn{1}{c|}{2133}          & 0,016          & \multicolumn{1}{c|}{\textbf{0,945}  \greentriangleup}  & \multicolumn{1}{c|}{\textbf{682}  \greentriangleup}  & 0,176          \\ \hline \hline
\multicolumn{1}{|c|}{\multirow{3}{*}{EASE}}     & ExpLOD                & \multicolumn{1}{c|}{6,479}          & \multicolumn{1}{c|}{\textbf{2445}  \greentriangleup} & 0,019          & \multicolumn{1}{c|}{0,547}           & \multicolumn{1}{c|}{238}           & \textbf{0,531}  \greentriangleup \\ \cline{2-8} 
\multicolumn{1}{|c|}{}                          & ExpLOD v2             & \multicolumn{1}{c|}{6,553}          & \multicolumn{1}{c|}{2130}          & \textbf{0,019}  \greentriangleup & \multicolumn{1}{c|}{0,558}           & \multicolumn{1}{c|}{174}           & 0,286          \\ \cline{2-8} 
\multicolumn{1}{|c|}{}                          & PEM                   & \multicolumn{1}{c|}{\textbf{7,457}  \greentriangleup} & \multicolumn{1}{c|}{1881}          & 0,015          & \multicolumn{1}{c|}{\textbf{0,925}  \greentriangleup}  & \multicolumn{1}{c|}{\textbf{650}  \greentriangleup}  & 0,147          \\ \hline \hline
\multicolumn{1}{|c|}{\multirow{3}{*}{NCF}}      & ExpLOD                & \multicolumn{1}{c|}{8,708}          & \multicolumn{1}{c|}{\textbf{3436}  \greentriangleup} & \textbf{0,018}  \greentriangleup & \multicolumn{1}{c|}{0,777}           & \multicolumn{1}{c|}{303}           & \textbf{0,590}  \greentriangleup \\ \cline{2-8} 
\multicolumn{1}{|c|}{}                          & ExpLOD v2             & \multicolumn{1}{c|}{8,733}          & \multicolumn{1}{c|}{3298}          & 0,016          & \multicolumn{1}{c|}{0,787}           & \multicolumn{1}{c|}{302}           & 0,551          \\ \cline{2-8} 
\multicolumn{1}{|c|}{}                          & PEM                   & \multicolumn{1}{c|}{\textbf{8,751}  \greentriangleup} & \multicolumn{1}{c|}{3089}          & 0,016          & \multicolumn{1}{c|}{\textbf{0,959}  \greentriangleup}  & \multicolumn{1}{c|}{\textbf{1157}  \greentriangleup} & 0,275          \\ \hline
\end{tabular}
\label{tab:offresultslastfm_n5}
\end{table*}

\newpage
\section{Recommender Systems Ranking Metrics}
\label{apen:ranking}

\begin{table*}[!htbp]
\caption{Mean 10-fold ranking metrics for each recommendation algorithm on the \ac{LastFM} dataset. \textbf{Bold} values with \greentriangleup are the best for a metric. NCF algorithm does not have beyond accuracy metrics because a leave-one-out evaluation was used as in the original paper \cite{He2017}.\looseness=-1}
\begin{tabular}{|c|c|c|c|c|c|c|c|}
\hline
Metric                    & K  & MostPop & BPRMF           & PageRank & UserKNN & EASE            & NCF    \\ \hline
\multirow{4}{*}{NDCG}     & 1  & 0,0706  & 0,1404          & 0,1550   & 0,2252  & \textbf{0,2407} \greentriangleup & 0,1672 \\ \cline{2-8} 
                          & 3  & 0,0990  & 0,1885          & 0,1989   & 0,2847  & \textbf{0,2988} \greentriangleup & 0,2350 \\ \cline{2-8} 
                          & 5  & 0,1062  & 0,1986          & 0,2078   & 0,2942  & \textbf{0,3090} \greentriangleup & 0,2544 \\ \cline{2-8} 
                          & 10 & 0,1100  & 0,2014          & 0,2098   & 0,2903  & \textbf{0,3033} \greentriangleup & 0,2695 \\ \hline
\multirow{4}{*}{MAP}      & 1  & 0,0706  & 0,1404          & 0,1550   & 0,2252  & \textbf{0,2407} \greentriangleup & 0,1672 \\ \cline{2-8} 
                          & 3  & 0,1261  & 0,2343          & 0,2411   & 0,3412  & \textbf{0,3546} \greentriangleup & 0,3006 \\ \cline{2-8} 
                          & 5  & 0,1417  & 0,2571          & 0,2615   & 0,3654  & \textbf{0,3807} \greentriangleup & 0,3404 \\ \cline{2-8} 
                          & 10 & 0,1600  & 0,2783          & 0,2807   & 0,3821  & \textbf{0,3965} \greentriangleup & 0,3793 \\ \hline
\multirow{3}{*}{AGG-DIV}  & 1  & 9,2     & \textbf{184,6} \greentriangleup  & 60,5     & 155,5   & 132,8           & -      \\ \cline{2-8} 
                          & 3  & 15,1    & \textbf{314,6} \greentriangleup  & 125,1    & 294,6   & 266,6           & -      \\ \cline{2-8} 
                          & 5  & 20,9    & \textbf{402,6} \greentriangleup  & 179,7    & 400     & 375,4           & -      \\ \hline
\multirow{3}{*}{Entropy}  & 1  & 0,4095  & \textbf{1,7667} \greentriangleup & 0,9659   & 1,7390  & 1,6393          & -      \\ \cline{2-8} 
                          & 3  & 0,8013  & \textbf{1,9158} \greentriangleup & 1,2481   & 1,9392  & 1,8695          & -      \\ \cline{2-8} 
                          & 5  & 0,9799  & \textbf{1,9971} \greentriangleup & 1,3946   & 2,0531  & 1,9898          & -      \\ \hline
\multirow{3}{*}{Gini}     & 1  & 0,9997  & \textbf{0,9954} \greentriangleup & 0,9991   & 0,9960  & 0,9968          & -      \\ \cline{2-8} 
                          & 3  & 0,9994  & \textbf{0,9938} \greentriangleup & 0,9986   & 0,9937  & 0,9968          & -      \\ \cline{2-8} 
                          & 5  & 0,9992  & \textbf{0,9926} \greentriangleup & 0,9981   & 0,9919  & 0,9930          & -      \\ \hline
\multirow{3}{*}{Coverage} & 1  & 0,0008  & \textbf{0,0167} \greentriangleup & 0,0055   & 0,0141  & 0,0120          & -      \\ \cline{2-8} 
                          & 3  & 0,0014  & \textbf{0,0167} \greentriangleup & 0,0113   & 0,0267  & 0,0241          & -      \\ \cline{2-8} 
                          & 5  & 0,0019  & \textbf{0,0364} \greentriangleup & 0,0163   & 0,0362  & 0,0340          & -      \\ \hline
\end{tabular}
\end{table*}

\begin{table*}[!htbp]
\label{}
\caption{Mean 10-fold ranking metrics for each recommendation algorithm on the \ac{MovieLens} dataset. \textbf{Bold} values with \greentriangleup are the best for a metric. NCF algorithm does not have beyond accuracy metrics because a leave-one-out evaluation was used as in the original paper \cite{He2017}.\looseness=-1}
\begin{tabular}{|c|c|c|c|c|c|c|c|}
\hline
Metric                    & K  & MostPop & BPR-MF          & PageRank & UserKNN & EASE            & NCF    \\ \hline
\multirow{4}{*}{NDCG}     & 1  & 0,1690  & 0,2059          & 0,2021   & 0,2901  & \textbf{0,3582} \greentriangleup & 0,2306 \\ \cline{2-8} 
                          & 3  & 0,2677  & 0,3507          & 0,3511   & 0,4443  & \textbf{0,5118} \greentriangleup & 0,3996 \\ \cline{2-8} 
                          & 5  & 0,2928  & 0,3789          & 0,3776   & 0,4634  & \textbf{0,5229} \greentriangleup & 0,4463 \\ \cline{2-8} 
                          & 10 & 0,3084  & 0,3919          & 0,3919   & 0,4659  & \textbf{0,5209} \greentriangleup & 0,4957 \\ \hline
\multirow{4}{*}{MAP}      & 1  & 0,1690  & 0,2059          & 0,2021   & 0,2901  & \textbf{0,3582} \greentriangleup & 0,2306 \\ \cline{2-8} 
                          & 3  & 0,2201  & 0,2814          & 0,2800   & 0,3703  & \textbf{0,4374} \greentriangleup & 0,3165 \\ \cline{2-8} 
                          & 5  & 0,2309  & 0,2941          & 0,2929   & 0,3770  & \textbf{0,4384} \greentriangleup & 0,3392 \\ \cline{2-8} 
                          & 10 & 0,2278  & 0,2867          & 0,2879   & 0,3604  & \textbf{0,4153} \greentriangleup & 0,3584 \\ \hline
\multirow{3}{*}{AGG-DIV}  & 1  & 16,6    & \textbf{185,6} \greentriangleup  & 44,8     & 110,1   & 123,9           & -      \\ \cline{2-8} 
                          & 3  & 32,8    & \textbf{343,3}  \greentriangleup & 89,8     & 187     & 236,2           & -      \\ \cline{2-8} 
                          & 5  & 47,2    & \textbf{446,1} \greentriangleup  & 126,6    & 248     & 314,8           & -      \\ \hline
\multirow{3}{*}{Entropy}  & 1  & 0,6987  & \textbf{1,9851} \greentriangleup & 0,8659   & 1,7723  & 1,7783          & -      \\ \cline{2-8} 
                          & 3  & 1,0248  & \textbf{2,1690} \greentriangleup & 1,2282   & 1,9355  & 1,9838          & -      \\ \cline{2-8} 
                          & 5  & 1,1881  & \textbf{2,2571} \greentriangleup & 1,4076   & 2,0266  & 2,0943          & -      \\ \hline
\multirow{3}{*}{Gini}     & 1  & 0,9995  & \textbf{0,9908} \greentriangleup & 0,9991   & 0,9948  & 0,9946          & -      \\ \cline{2-8} 
                          & 3  & 0,9990  & \textbf{0,9861} \greentriangleup & 0,9983   & 0,9926  & 0,9915          & -      \\ \cline{2-8} 
                          & 5  & 0,9986  & \textbf{0,9833} \greentriangleup & 0,9976   & 0,9910  & 0,9892          & -      \\ \hline
\multirow{3}{*}{Coverage} & 1  & 0,0018  & \textbf{0,0202} \greentriangleup & 0,0049   & 0,0120  & 0,0135          & -      \\ \cline{2-8} 
                          & 3  & 0,0036  & \textbf{0,0374} \greentriangleup & 0,0098   & 0,0204  & 0,0257          & -      \\ \cline{2-8} 
                          & 5  & 0,0051  & \textbf{0,0486} \greentriangleup & 0,0138   & 0,0270  & 0,0343          & -      \\ \hline
\end{tabular}
\end{table*}

\clearpage
\section{Full Correlation Matrices}
\label{apen:correlations}

\begin{figure}[!ht]
    \centering
    \includegraphics[width=.9\textwidth]{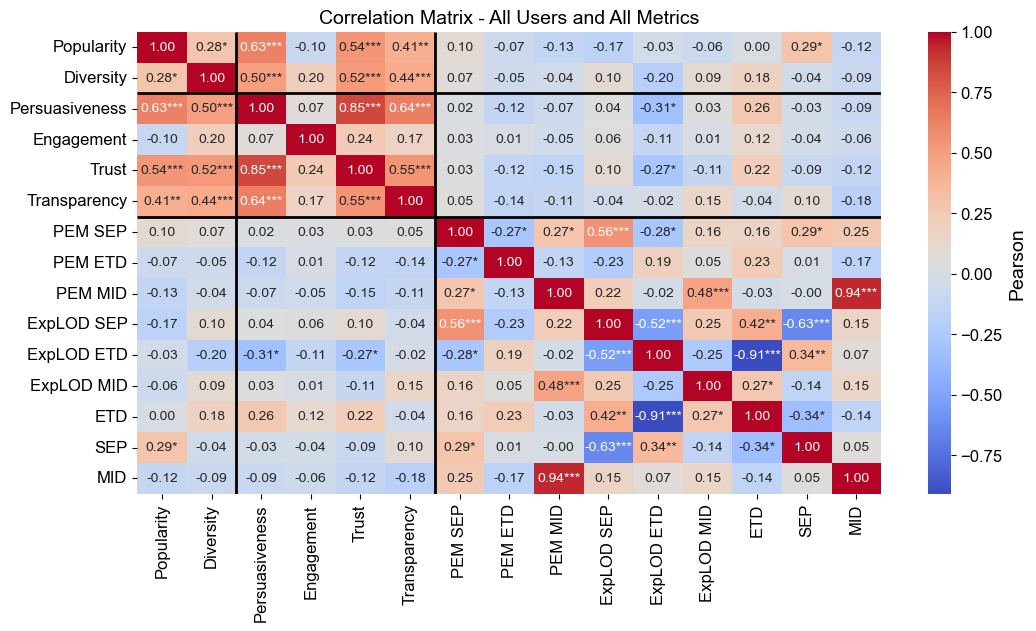}
    \caption{Correlation analysis between differences in user ratings for the \ac{PEM} and ExpLOD v2 algorithms, considering online explanation goals and offline metrics. The ETD, SEP, and MID columns denote the correlations for the differences between the corresponding \ac{PEM} and ExpLOD v2 offline metrics. Lines and columns 7–12 also reports the individual offline explanation metrics for \ac{PEM} and ExpLOD v2, using their names as prefixes. The symbol * indicates $p < 0.05$; ** indicates $p < 0.01$; and *** indicates $p < 0.001$.\looseness=-1}
    \label{fig:corr_all_full}
\end{figure}

\begin{figure}[!ht]
    \centering
    \includegraphics[width=.9\textwidth]{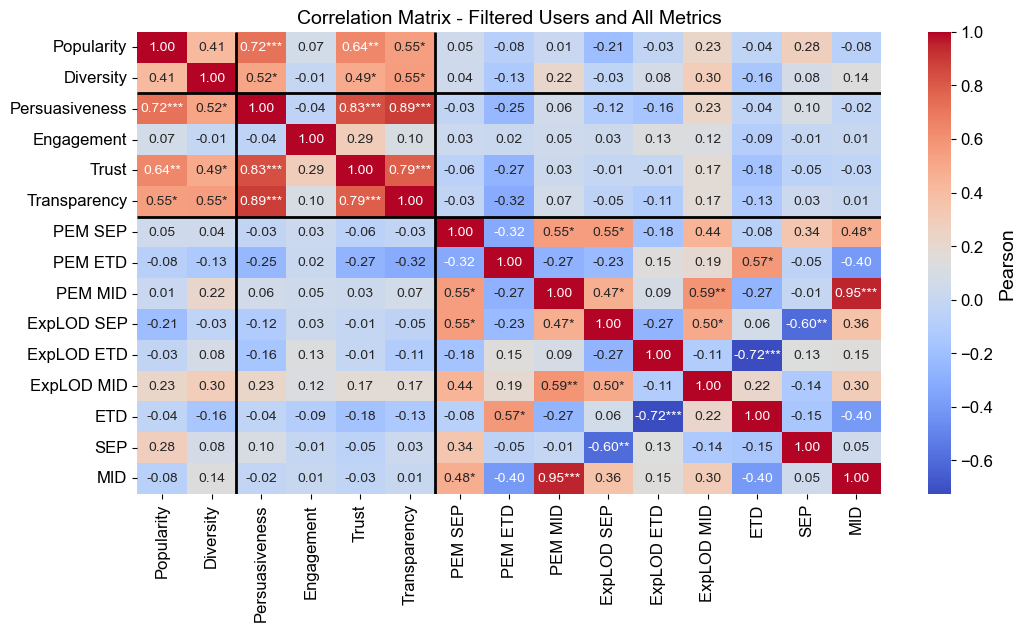}
    \caption{Correlation analysis between differences in user ratings for the \ac{PEM} and ExpLOD v2 algorithms, considering online explanation goals and offline metrics when they display diverse explanation attributes. The ETD, SEP, and MID columns denote the correlations for the differences between the corresponding \ac{PEM} and ExpLOD v2 offline metrics. Lines and columns 7–12 also reports the individual offline explanation metrics for \ac{PEM} and ExpLOD v2, using their names as prefixes. The symbol * indicates $p < 0.05$; ** indicates $p < 0.01$; and *** indicates $p < 0.001$.\looseness=-1}
    \label{fig:corr_f_full}
\end{figure}

\end{document}